\documentclass[aps,prb,twocolumn,superscriptaddress,floatfix,citeautoscript,hyperref=pdftex]{revtex4-1}
\usepackage{times}
\usepackage{graphicx}
\usepackage{dcolumn}
\usepackage{bm}
\usepackage{color}
\usepackage{amsmath}
\usepackage{amssymb}
\usepackage{braket}
\newcommand {\beq} {\begin{equation}}
\newcommand {\eeq} {\end{equation}}
\newcommand {\bqa} {\begin{eqnarray}}
\newcommand {\eqa} {\end{eqnarray}}

\newsavebox{\mybox}

\usepackage[colorlinks=true]{hyperref}

\begin{document}

\title{Fractionalized pair density wave in the pseudo-gap phase of cuprate superconductors}

\author{D. Chakraborty}

\affiliation{Institut de Physique Th\'eorique, Universit\'e Paris-Saclay, CEA, CNRS, F-91191 Gif-sur-Yvette, France.}

\author{M. Grandadam}

\affiliation{Institut de Physique Th\'eorique, Universit\'e Paris-Saclay, CEA, CNRS, F-91191 Gif-sur-Yvette, France.}

\author{M. H. Hamidian}

\affiliation{Department of Physics, Harvard University, Cambridge, MA 02138, USA.}

\author{J. C. S. Davis}

\affiliation{Department of Physics, University College Cork, Cork T12R5C, Ireland.}

\affiliation{Clarendon Laboratory, University of Oxford, Parks Road, Oxford, OX1 3PU, UK.}

\author{Y. Sidis}

\affiliation{Laboratoire L\'eon Brillouin, CEA-CNRS, Universit\'e Paris-Saclay, CEA Saclay, Gif-sur-Yvette 91191, France.}

\author{C. P\'epin}

\affiliation{Institut de Physique Th\'eorique, Universit\'e Paris-Saclay, CEA, CNRS, F-91191 Gif-sur-Yvette, France.}

\begin{abstract}

The mysterious pseudo-gap (PG) phase of cuprate superconductors has been the subject of intense investigation over the last thirty years, but without a clear agreement about its origin. Owing to a recent observation in Raman spectroscopy, of a precursor in the charge channel, on top of the well known fact of a precursor in the superconducting channel, we present here a novel idea: the PG is formed through a Higgs mechanism, where two kinds of preformed pairs, in the particle-particle and particle-hole channels, become entangled through a freezing of their global phase. Remarkably, this entanglement is equivalent to fractionalizing a Cooper pair density wave (PDW) into its elementary parts; the particle-hole pair, giving rise to both density modulations and current modulations, and the particle-particle counterpart, leading to the formation of Cooper pairs. From this perspective, the ``fractionalized PDW'' becomes the central object around the formation of the pseudo-gap. The ``locking'' of phases between the charge and superconducting modes gives a unique explanation for the unusual global phase coherence of short-range charge modulations, observed below $T_{c}$ on phase sensitive scanning tunneling microscopy (STM). A simple microscopic model enables us to estimate the mean-field values of the precursor gaps in each channel and the PG energy scale, and to compare them to the values observed in Raman scattering spectroscopy. We also discuss the possibility of a multiplicity of orders in the PG phase and give an overview of the phase diagram.

\end{abstract}

\maketitle

\section{Introduction}\label{sec:Intro}

\subsection{General introduction}\label{sec:GenIntro}

The PG `phase' in the cuprate superconductors remains one of the most enduring mysteries of condensed matter physics. It was first observed as loss of density of states at intermediate oxygen doping \cite{Alloul89,Warren89} $0.08<p<0.20$, where part of the Fermi surface is gapped in the anti-nodal (AN) region ( $\left(0,\pm\pi\right)$ and $\left(\pm\pi,0\right)$ ) of the Brillouin zone, leading to the formation of Fermi arcs (see e.g. Ref. [\onlinecite{Norman03,Lee06}]). The partial gapping of the Fermi surface is very puzzling, because it breaks the Luttinger theorem which counts the number of electrons in a reconfiguration of the Fermi surface. To account for this very unusual observation, several approaches have been put forward. The first one focuses on the proximity to the Mott transition, and states that due to the strong Coulomb interaction ($U\simeq1$eV), the electron fractionalizes into elementary parts \cite{Coleman84,Baskaran88,Nagaosa90,Lee92,Senthil:2000eb}, for example spinons and holons, which accounts for the formation of the pseudo-gap. This line of thought was developed over the years with one famous candidate: the formation of spin singlet through the Resonating-Valence-Bond (RVB) state \cite{Anderson87}. A second line of thought remarked that, in the vicinity of a localization transition, the phase of all fields fluctuates wildly \cite{EmeryVJ:1995dr}. Scenarios with phase fluctuations and preformed but incoherent Cooper pairs were thus proposed \cite{Norman:1998va,Chien2009}. These scenarios were very strong in describing the fluctuations above $T_{c}$. For example, the unchanged AN spectroscopic gap across $T_{c}$, up to $T^{*}$, has been understood as the presence of preformed pairs -or superconducting fluctuations, above $T_{c}$ \cite{Campuzano98,Campuzano:1996fb}. Unfortunately, preformed Cooper pairs could only be observed up to a small temperature above $T_c$ and not up to $T^*$ \cite{RullierAlbenque:2011ji,Bergeal:2008gf,CyrChoiniere:2018ed}.
\\
\\
In this paper, we give a second life to the preformed pair scenario with a new idea. We propose that the PG phase is comprised of two kinds of competing preformed pairs: particle-particle (p-p) and particle-hole (p-h), having very distinct symmetries, but which get entangled at $T^{*}$ through a freezing of their global phase. In our theory, $T^{*}$ is a true phase transition temperature with a broken $U(1)\times U(1)$ gauge symmetry, the second $U(1)$ gauge symmetry getting broken at $T_{c}$. One U(1) correspond to the electromagnetic charge symmetry and the other U(1) is associated to the gradient of the local phase of the p-h pairs. Since p-h pairs are neutral to the electromagnetic field, the second U(1) is identified as a neutral gauge field, minimization of which generates a constraint between the two pairs. Remarkably, the emergence of a neutral gauge field can be seen from another novel parallel point of view where a PDW order parameter fractionalizes into elementary p-p and p-h pairs with a constraint between them and thus entangling them. Within this parallel viewpoint, the `fractionalized' PDW can be seen as a fundamental object lying at the origin of the formation of the PG phase. A gauge field associated to a neutral field resulting in a constraint is also considered in the case of electron's fractionalization where the constraint is that of no double occupancy of electrons on lattice sites. Here, instead of fractionalizing electrons, we fractionalize an order parameter, PDW, the one which is fragile and difficult to stabilize in most theoretical approaches.

Each of the two preformed pairs leads to the formation of a `primary' state at low temperatures, but our theoretical formulation is generic and could accommodate for other `primary' states, like anti-ferromagnetic stripes, thus opening space for the solution of various debates in the PG puzzle.

\subsection{Prescription of two kinds of preformed pairs and fractionalized PDW}\label{sec:2pairs}

One of the recent developments in the physics of cuprate superconductors is the ubiquitous observation of Charge Density Modulations (CDM) in the underdoped regime. It was first observed by STM in the superconducting phase, as modulations inside the vortex core \cite{Hoffman02,Matsuba07,Yoshizawa13,Machida16,Hamidian15a}. Observation of quantum oscillations\cite{Doiron-Leyraud07,Sebastian12}, X-ray\cite{Chang12,Blanco-Canosa13,Blackburn13a,Ghiringhelli12,Gerber:2015gx,Chang16} and NMR measurements \cite{Wu11,Wu13a,Wu:2015bt,Julien15} have completed the picture. The 2D CDM have a predominantly d-wave symmetric form factor\cite{Hamidian15a,Comin15} and lives on the Cu-Cu bonds in a one band picture. 

Very recently, a new feature emerges in Raman spectroscopy which reports for the first time a precursor gap in the charge (particle-hole) channel, observed as a peak in the $B_{2g}$ response of the cuprate compound HgBa$_{2}$Ca$_{2}$Cu$_{3}$O$_{8+\delta}$ (Hg-1223) \cite{Loret19}. Similar feature is also observed in other cuprates like HgBa$_{2}$CuO$_{4+\delta}$ (Hg-1201) \cite{Li13,Loret19} and YBa$_{2}$Cu$_{3}$O$_{6+x}$ (YBCO) \cite{Loret19}. It is shown in Ref. [\onlinecite{Loret19}] that the spectral gap associated to the charge order is of the same order of magnitude as the superconducting gap, and that both gaps behave in a similar way with doping, following $T^{*}$ rather than $T_{c}$. This very intriguing experiment is calling for a reconsideration of the scenario of preformed pairs, but with two kinds of preformed pairs in competition, in the p-p and p-h channels. Due to the near degeneracy, the system hesitates energetically between forming p-p and p-h pairs. This fact motivates our ansatz of an entangled state of p-p and p-h pairs for the PG state and a constraint between the two relates their energies to the PG energy scale.

Moreover, a PDW order has been recently observed below $T_c$ in the halo \cite{Edkins18} surrounding the vortex core in the cuprate compound Bi$_{2}$Sr$_{2}$CaCu$_{2}$O$_{8+\delta}$ (BSCCO). This PDW occurs with modulations both at the same wave vector as the charge modulations and at half of its value. A PDW in the same compound was also observed in the absence of magnetic field using a superconducting STM tip \cite{Hamidian16}. The zero field PDW was only observed with the same vector as charge modulations. These observations inspired several theoretical works \cite{Wang18,Dai18,Norman18} indicating their importance in the PG phase, but with no consensus on whether the fundamental state is a PDW or a charge ordered state. In our formulation, we provide, for the first time, a viewpoint based on `fractionalized' PDW which is the fundamental object. In the PG phase, the PDW is fractionalized to p-p pairs and p-h pairs. Only at low temperatures these two pairs recombine and a PDW can be observed. Through this new perspective, we reconcile the debate of the nature of the fundamental state.
\\
\\
On a different side, there are growing experimental indications that the PG phase sustain a `true' symmetry broken state. Resonant ultrasound spectroscopy \cite{Shekhter13} reports a thermodynamic phase transition at $T^{*}$ associated with the emergence of the PG phase. STM \cite{Lawler10,Kohsaka07}, anomalous Nernst effect \cite{Daou10}, torque-magnetometry \cite{Sato2017} and polarized neutron diffraction \cite{Mangin_Thro17} measurements all indicate that the four-fold ($C_{4}$) rotational symmetry is broken at $T^{*}$. In addition, polarized elastic neutron scattering \cite{Fauque06,Bourges18} and optical second harmonic generation \cite{Zhao2016} measurements suggest that the time reversal symmetry and parity can be further broken at $T^{*}$. All of these indicate that there can be intra-unit cell $Q=0$ (translational symmetry preserving) orders developing at $T^{*}$. None of these $Q=0$ orders can explain the opening of a gap in the AN region in the fermionic spectrum or the existence of a finite $Q$ charge order at lower temperatures.

\begin{figure}[t]
{\includegraphics[width=0.8\linewidth]{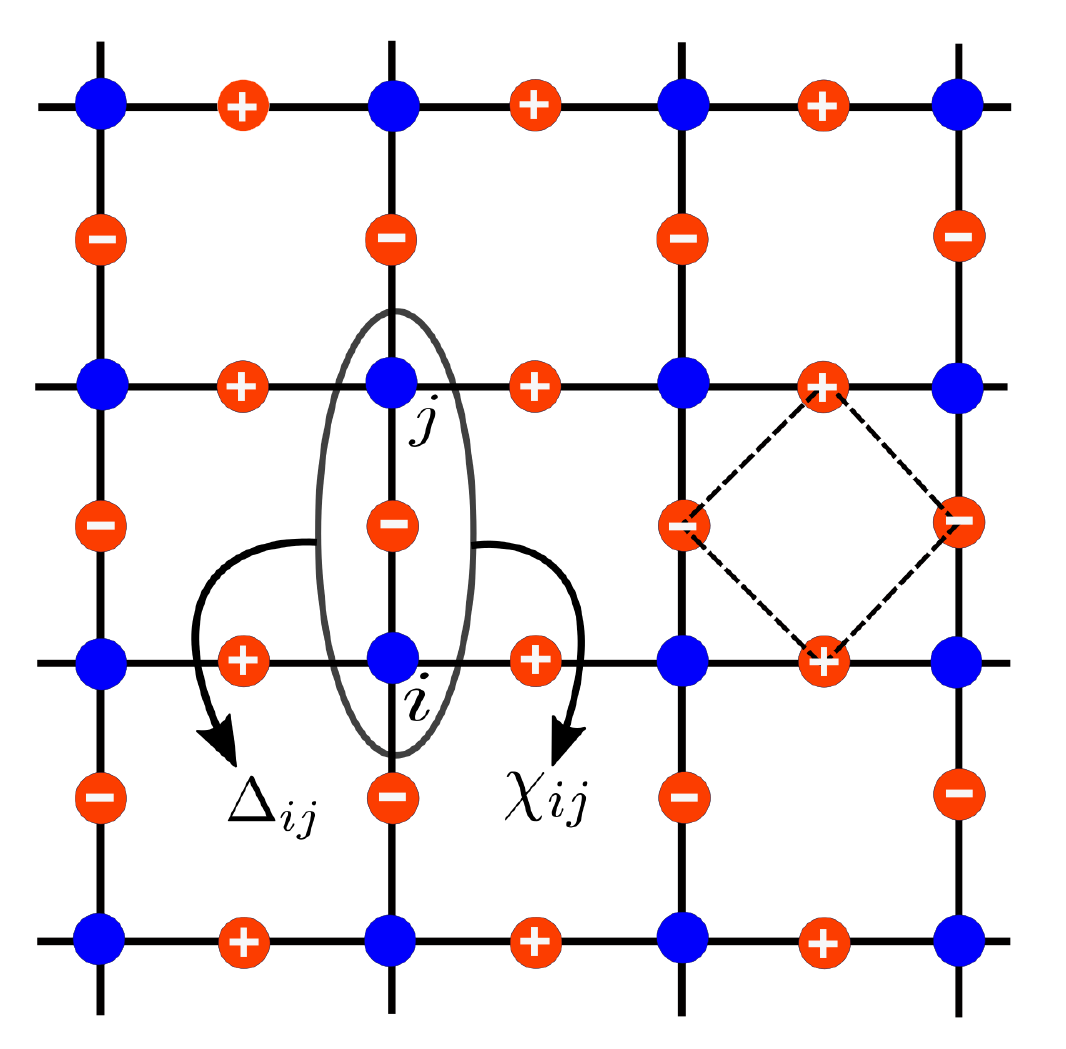}
} 
\caption{Schematic representation of the square planar Cu lattice with both $\Delta_{ij}$ and $\chi_{ij}$ living on nearest neighbor bonds $\langle ij \rangle$ (Cu-Cu bonds). The site $j$ can be either of the four nearest neighbors of $i$ whose coordinates are $r_{j}=r_{i}+ \delta$  with $\delta= \pm \hat{u}_x~ \mbox{or}~ \pm \hat{u}_y$ where $\hat{u}$ is the lattice translational operator. The operator $\hat{d}$ gives the d-wave character with $\hat{d}=1$ for $\delta= \pm \hat{u}_x$ and $\hat{d}=-1$ for $\delta= \pm \hat{u}_y$. We have constructed the continuum field theory in Sec.~\ref{sec:Higgs} using a coordinate system on the midpoints of the bonds, $r=(r_i+r_j)/2$. The midpoints are shown as orange dots with `+' and `-' indicating the d-wave character of the bonds. These midpoints constitute a tilted square lattice (shown with dotted lines) with an `anti-ferromagnet' like arrangement.}
\label{Fig:bond} 
\end{figure}

The PG phase thus behaves like a `Frankenstein' creature showing numerous different puzzling properties which seems apparently disconnected to each other. A theory that coherently connects all of these phenomenological features is a need of the hour. In this endeavor, we postulate that the PG state is an `entangled' state of preformed Cooper pairs $\left(\Delta_{ij}\equiv\hat{d}\sum_{\sigma}\sigma c_{j{-\sigma}}c_{i\sigma}\right)$ and preformed bond-excitonic pairs (particle-hole pairs) \footnote{The term `excitons' has been extensively used in the history of condensed matter physics in a different context. Nevertheless, we will use the term bond-excitonic order for identifying a particle-hole order parameter on bonds of a lattice with a modulation wave vector $Q$.} $\left(\chi_{ij}\equiv\hat{d}\sum_{\sigma}c_{i\sigma}^{\dagger}c_{j\sigma}e^{iQ.(r_{i}+r_{j})/2}\right)$, both of which live on nearest neighbor bonds $\langle ij\rangle$ (see Fig.~\ref{Fig:bond}) of the square planar Cu lattice and $\hat{d}$ gives a d-wave structure factor. The formulation of this paper does not need any specific form of the modulation wave vector $Q$. The preformed p-p pairs on bonds will correspond to the d-wave superconducting (d-SC) order and the preformed p-h pairs on bonds will correspond to the modulating d-wave bond-excitonic (d-BDW) order. Since the d-BDW order parameter is complex, the real part can lead to the d-wave charge density wave (d-CDW) and the imaginary part can lead to the d-wave current density wave (d-currentDW), as in previous studies \cite{Wang14}. The PG state is described by a spinor or a doublet with two constituent states - d-SC and d-BDW. From an alternative viewpoint, these two constituent states emerge from the fractionalization of a PDW field. The PDW field can be observed when it recombines due to condensation in either d-SC or d-BDW at low temperatures. Fluctuations in the PDW field can further give rise to `auxiliary' $Q=0$ orders. Thus on the one hand our proposal has the prospects of generating finite $Q$ orders like d-CDW and PDW at relatively low temperatures, it can also induce the `auxiliary' magneto electric loop current order parameter at $Q=0$ which can account for the breaking of both time reversal symmetry and parity in the PG phase. 

\subsection{Theoretical perspective}\label{sec:theorypers}

\subsubsection{Higgs mechanism and fractionalization of a PDW}\label{sec:higgsandpreformed}

In field theory, the ``Higgs mechanism'' is typically associated to the freezing of a phase resulting in a broken gauge invariance. The vector potential corresponding
to the gradient of this phase hence gets expelled from the system; it gets massive \cite{altlandBook}. A prototype example of Higgs mechanism in condensed matter physics is superconductivity \cite{Anderson:1963vi}, where the phase of the superconducting order parameter is frozen. Here, the $T^{*}$ line is ascribed to a specific Higgs mechanism which freezes the global phase of the two kinds of preformed pairs. The gauge field corresponding to the global U(1) phase acquires a mass $E^{*}=\sqrt{\left|\chi_{ij}\right|^{2}+\left|\Delta_{ij}\right|^{2}}$ which is identified with the spectroscopic PG energy scale $E^{*}$. It is shown that the freezing of the global phase entangles the two kinds of preformed pairs at $T^{*}$. Due to the composite nature of the gauge field, the electromagnetic (EM) field does not get expelled at $T^{*}$: there is no Meissner effect.

A state with two preformed pairs corresponding to the d-SC order and the d-BDW order can be described by a U(1) $\times$ U(1) gauge theory. One U(1) corresponds to the usual charge symmetry (usually broken by superconducting ground state) and the other is related to the local phase $\theta_{\chi}$ of the d-BDW. Since the d-BDW is neutral to electromagnetism, the gradient of $\theta_{\chi}$ is a neutral gauge field $\alpha_{\mu}=\partial_{\mu} \theta_{\chi}$, minimization of which generates the constraint $\chi_{ij}^{2}+\Delta_{ij}^{2}=(E^*)^2$ ($E^{*}=1$ being a high energy scale). The theory described in this paper is a unique proposal, which differs from the existing gauge theories \cite{Baskaran88,Lee92} in one essential way. Our description does not involve any fractionalization of the electron's degrees of freedom, but rather, we fractionalize an ``order parameter" \cite{Sachdev19,Komijani18}. The neutral gauge field can be thought as the fractionalization of a PDW into two elementary parts: particle-particle and particle-hole on a bond $\Delta_{\text{PDW}}=\Delta_{ij}\chi_{ij}^*$, associated to the constraint $\chi_{ij}^{2}+\Delta_{ij}^{2}=1$.

\subsubsection{PG energy scale}\label{sec:PGE}

Historically, it has been argued that the spectroscopic signatures of the PG revealed two energy scales \cite{Lee06}. One corresponds to a kink along with a depletion close to the Fermi level in spectroscopic probes like angle resolved photo emission spectroscopy (ARPES) \cite{Ding:1998iv,Damascelli03,Vishik:2010fn}, scanning tunneling spectroscopy \cite{Kugler01,Fischer07} and Raman spectroscopy \cite{Sacuto:2013ez,Devereaux07,LeTacon06}. The other, higher energy scale, is associated to the downturn in the Knight-shift measured from NMR experiments \cite{Alloul89} and the higher energy hump in ARPES or Raman spectroscopy. The argument of two distinct PG energy scales, typical of strong coupling approaches, interprets the higher scale as responsible for spin singlet formation (a typical example is the RVB state) and the lower scale arising due to the superconducting fluctuations. Importantly in our paper, the PG is associated to only one energy scale $E^*$ arising from the same constraint obtained both from the Higgs perspective and from the parallel perspective of fractionalized PDW. $E^*$ is visible as coherence peak in STM or ARPES and pair-breaking peak in Raman spectroscopy at roughly the same energy. This single energy scale acquires its definition for temperatures below $T^{*}$. In our view, the higher energy hump seen in ARPES or Raman spectroscopy is not an independent energy scale and is possibly related to the coupling of fermions to a collective mode \cite{Norman97,Onufrieva99,Eschrig:2000bf,Eschrig:2006ky,Chubukov06}.

\subsubsection{Entanglement of two kinds of preformed pairs}\label{sec:entanglement}

The U(1) $\times$ U(1) gauge theory can be reformulated for fields on bonds $\left(i,j\right)$, in terms of a global phase \footnote{We use the term `global' phase to represent the common phase between the Cooper pairs and the bond-excitonic pairs. This should not be confused with a phase which is same at all lattice sites.} and a relative phase of the two preformed pairs without any loss of generality. In the case of superconductivity, the ground state is given as $\ket{SC}=\Delta_{ij}\ket{0}$ where $\ket{0}$ is the vacuum state. This ground state breaks the charge U(1) gauge invariance and the gauge field acquires a mass $\sqrt{\braket{SC|SC}}=\Delta$ where $\Delta$ is the uniform superconducting gap. In contrast,  the ground state corresponding to the PG phase is an entangled state given as $\ket{PG}=\left(\Delta_{ij}+\chi_{ij}\right)\ket{0}$, which is a quantum superposition of d-SC and d-BDW orders (a `super-solid' phase). At $T^{*}$, $\braket{PG|PG}$ gets condensed to a non-zero value with a broken gauge symmetry, henceforth the gauge field acquires a mass $E^{*}=\sqrt{\braket{PG|PG}}=\sqrt{\left|\chi_{ij}\right|^{2}+\left|\Delta_{ij}\right|^{2}}$, which characterizes the PG energy scale. In order to minimize the energy, the system chooses to condense in this entangled state instead of condensing separately in either of the pairs (also see Sec.~\ref{sec:mfgapeq}). Our ansatz is equivalent to fractionalizing a PDW in the PG phase. This special Higgs mechanism induces a strong competition between the two preformed pairs. As a result, the amplitudes of the d-SC and d-BDW orders fluctuate wildly just below $T^{*}$ with no uniform components. Thus, neither of the three orders, d-SC or d-BDW or PDW, condense at this temperature showing that the translational symmetry or the charge symmetry is not broken at $T^{*}$. 

\begin{figure}[t]
\includegraphics[width=0.9\linewidth]{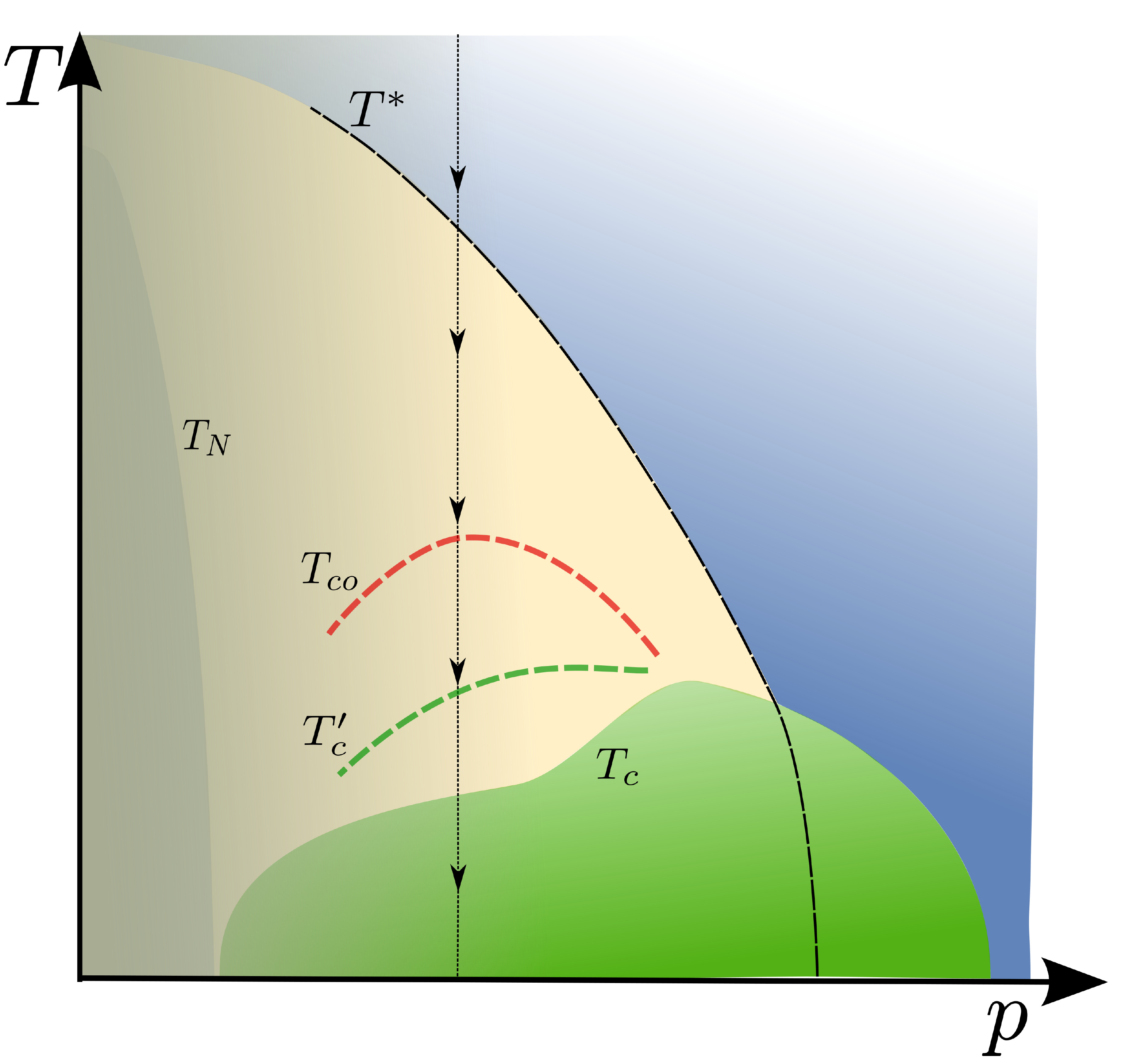} \caption{Schematic temperature (T)- hole doping ($p$) phase diagram for a cuprate superconductor. The vertical dotted black line demonstrates an adiabatic decrease in temperature from a representative high temperature ($T>T^*$) point in the phase diagram. As explained in the text, the system hits the first Higgs mechanism freezing the global phase of the p-p and p-h preformed pairs entangling them at $T^*$. This induces a constraint between the amplitudes of the two order parameters. The fluctuations of the relative phase and the two amplitudes can be described by an O(3) non linear $\sigma$-model. Lower temperature crossover lines $T_{co}$ and $T^{\prime}_c$ correspond to the mean field lines where the amplitudes of the two preformed pairs get condensed giving a uniform component to each. A second Higgs mechanism occurs at $T_c$, where the relative phase also gets quenched. We also note that the theory described in this paper is strictly valid for dopings $p>6\%$. Especially, we do not intend to explain the Neel temperature ($T_N$) demarcating the anti-ferromagnetic phase.  For lower dopings ($p<6\%$), there are other effects like competing magnetic orders or modifications in the effective action owing to the strong electronic correlations \cite{Ferraz11}. We neglect these effects in the current picture.}
\label{fig:scdiagram} 
\end{figure}

\subsubsection{Phase diagram}\label{sec:pdintro}

With this prelude, we describe the phase diagram of the underdoped cuprates. A first Higgs mechanism at $T^{*}$ freezes the global phase of the two preformed pairs. The PG state below $T^*$ is thus a state with entangled p-p and p-h pairs with no long-range order. The concept of two kinds of preformed pairs makes the amplitude and the phase fluctuations of the d-SC and d-BDW orders distinct. As a result, this opens up possibilities of different temperature lines existing in the rich phase diagram of underdoped cuprates, as depicted in Fig.~\ref{fig:scdiagram}. Lower temperature crossover lines $T_{co}$ and $T_{c}^{\prime}$ correspond to the mean field lines of the p-h and p-p pairs respectively, where the amplitudes of the d-BDW and d-SC orders condense to give uniform components in the same spirit as that of Bose condensation of preformed pairs (for details see Sec.~\ref{sec:fluc}). At $T_{co}$, the short-range d-CDW can be observed in X-ray, STM or NMR measurements due to the pinning of the phase of the d-BDW order. An NMR perspective on pinning of the charge order in YBCO and its similarity with pinning in layered metals is given in Ref.~\onlinecite{Wu:2015bt}.  Since $T_{co}$ and $T_{c}^{\prime}$ are mean-field lines, their relative position in the phase diagram depends crucially on the details of the microscopic models. Here, we consider $T_{co}>T_{c}^{\prime}$. A possible justification comes from the microscopic model (Eq.~\eqref{eq:11b}) chosen in this study. A large off-site density-density interaction in this model can lead to an enhanced $T_{co}$. The mean-field precursor gaps of both the d-SC and d-BDW orders become well defined below $T_{c}^{\prime}$. But the relative phase still fluctuates and thus there is no phase coherence in d-SC or d-BDW orders. $T_{c}^{\prime}$ marks the onset of the pairing fluctuations as observed in Nernst effect \cite{CyrChoiniere:2018ed}, transport studies \cite{RullierAlbenque:2011ji} and Josephson SQUID experiments \cite{Bergeal:2008gf}. The relative phase of the two orders gets frozen at a lower temperature $T_{c}$, where the phase coherence sets in for both the d-SC and d-BDW orders with a formation of a `super-solid' like phase. Some signatures of a `super-solid' like phase can be seen by the observation of the charge order in X-ray \cite{Blackburn13a,Tabis14}, STM \cite{Wise08,HowaldKapitulnik03} and NMR \cite{Wu:2015bt} measurements even in the superconducting state at zero magnetic field for temperatures below $T_{c}$ down to $T=0$. The correlation length of the charge order is not expected to increase for $T<T_{c}$ due to a strong competition with d-SC \cite{Hayward14,Caplan15}. Instead, the correlation length features a maximum at $T_{c}$ \cite{Comin15a} showing an intimate connection between the d-SC and d-BDW orders. We remark that if the pinning of the d-BDW order is too strong, no superconductivity can emerge below $T_{c}$. Our formalism thus implies that the pinning is present but weaker than the Higgs mechanism giving rise to a bulk superconductor at $T_{c}$. Lastly, as already noted, since the d-BDW is a complex field, preemptive orders breaking discrete symmetries like parity, time reversal or lattice rotation, usually discussed in the context of $\mathbf{Q}=0$ orders such as electronic nematicity or loop current state, at higher temperature have to be present, in the same line of thought as in previous studies \cite{Wang14,Wang15b,Wang15a}.

The phase diagram can also viewed from the perspective of fractionalization of the PDW field. As mentioned earlier, the entanglement of p-p and p-h pairs at $T^*$ is equivalent to fractionalizing a PDW $\Delta_{\text{PDW}}=\Delta_{ij}\chi_{ij}^*$ into elementary p-p and p-h pairs. The PDW reconfines locally when either of the two elementary constituents condenses. Similar confinement transition occurs in the theories of electron's fractionalization where electron reconfines when either of the elementary constituents `spinons' and `holons' condense. At $T=T_{co}$, the PDW field reconfines locally due to the condensation of the d-BDW field amplitude. The system will show a short-range PDW state. For $T<T_{co}$, the theory allows for two possible PDW fields: $\tilde{\Delta}_{\text{PDW}}=\Delta_{ij}\chi_{ij}$ involving the global phase of the p-p and p-h pairs and $\Delta_{\text{PDW}}$ involving the relative phase. While $\tilde{\Delta}_{\text{PDW}}$ acquires global phase coherence at $T_{c}^{\prime}$, $\Delta_{\text{PDW}}$ obtains global phase coherence only at $T_c$.

A true long-range charge order, PDW or `super-solid' is never established in the absence of magnetic field due to the omnipresence of disorder in cuprates. Disorder acts on the charge order as a `random-field' \cite{DelMaestro06}. Following Imry-Ma criterion \cite{Imry75}, any strength of `random-field' disorder disrupts the long-range coherence in charge order in dimensions $d \le 4$. This is not the case for the superconducting order as disorder does not directly couple to the superconducting order parameter as `random-fields'. Thus for $T<T_c$, the superconducting order shows a true long-range nature in $d=3$ or a quasi long-range nature in $d=2$. But, a 3D charge order acquires a true long-range nature only at high magnetic fields when it shows uniaxial behavior (breaking a nematic discrete symmetry \cite{Nie13}) or in the additional presence of chain-disorder \cite{Caplan17}. As a consequence, the PDW order, which is a bilinear combination of the charge order and the superconducting order, can show long-range features only at zero temperature or at high magnetic field.

\subsubsection{Connection with theories on preformed Cooper pairs}\label{sec:PPcon}

In the past, the preformed Cooper pairs \cite{Kanigel:2008wm} were explored in details in scenarios where the phase of the Cooper pairs \cite{EmeryVJ:1995dr,Franz:1998et,Banerjee:2011bz,Banerjee:2011cu} fluctuates. A distinction has to be made between fluctuating scenarios\cite{Kivelson:1998ir,Benfatto:2000gy}, where the focus is on the strength of the fluctuations, and preformed pair scenarios \cite{Chien2009,Bozovic:2004cp,Boyack:2014fl,Boyack:2017gb} where the emphasis is put on strong short-range Cooper pairs which lead to models analogous to the Bose Einstein Condensation (BEC) phenomenon. For cuprates, it is indeed natural to assume that the size of the preformed pair, if they exist, is very short, of a few lattice sizes, giving credit to models which treat them as hard core bosons. In this preformed pairs scenario, the PG can be related to a precursor gap of fluctuating preformed Cooper pairs (p-p pairs), which acquire a phase coherence only below $T_c$ \cite{Kanigel:2008wm,Chien2009}. This approach goes well with the experimental observation that the `coherence peak' position in ARPES does not change when the temperature is reduced across $T_c$. However presence of superconducting fluctuations up to $T^*$ was largely debated. Experimental observations of Nernst effect, transport studies and Josephson SQUID measurements identified the region of superconducting fluctuations only up to a small temperature above $T_c$. This issue can be resolved by invoking a partner competitor like p-h pairs, which reduces the temperature window of superconducting fluctuations near $T_c$ \cite{Wachtel:2014ke}. Furthermore, recent observation of preformed p-p pairs up to $T^*$ in pump probe experiments \cite{Rajasekaran18,Hu14,Fausti11} revived the idea of fluctuating preformed pairs.

In our work, we extend to two kinds of competing p-p and p-h preformed pairs keeping various phenomenological advantages of the preformed p-p scenario. For example: (a) In both the approaches, the presence of p-p pairs will result into the Fermi arcs as observed in ARPES. Superconductivity is a whole Fermi surface instability. At temperatures higher than $T_c$, the nodal quasi particles will be more prone to fluctuations leaving the antinodal gap unperturbed \cite{Campuzano98}. As a result, the Fermi surface will be gapped in the anti-nodal region. (b) In analogy with the preformed p-p scenario, in our approach, both $T_c$ and $T_{co}$ is expected to show a dome shaped doping dependence whereas $T^*$ decrease monotonically with doping. This is because of an additional source of fluctuations at low doping due to closeness to the Mott transition, so that phase fluctuations are too strong to stabilize d-SC and the coherent puddles of charge modulations. (c) Owing to the notion of preformed pairs, the PG $T^*$ line will show universal \cite{Alloul14} features independent of disorder or magnetic field in contrast to both d-CDW and d-SC ground states which are affected by non-magnetic impurities like Zn \cite{Nachumi96,Alloul09} and pressure \cite{Cyr-Choiniere18,Souliou18,Vinograd19}. This is very similar to the idea of persistent gap (because of preformed p-p pairs) in s-wave superconductors in the presence of strong disorder \cite{Ghosal98} or magnetic field \cite{Ganguly17}.

\subsubsection{Connection with emergent SU(2) theories}\label{sec:SU2con}

In order to describe the PG phase of the underdoped cuprates, there were earlier proposals based on emergent symmetries between the d-SC order and a non-superconducting `partner'. Some of these include an SO(5) symmetry with antiferromagnetism \cite{Zhang97} and an SU(2) symmetry with d-density wave \cite{Chakravarty01} or $\pi$-flux state \cite{Lee:1998cr}. More recently, theories with emergent SU(2) symmetry \cite{Metlitski10b,Efetov13,Kloss:2016hu,Montiel16,Morice18a} are explored where the non-superconducting `partner' correspond to charge order. The SU(2) symmetry admits only a few exact realizations in condensed matter physics. In the case of the attractive Hubbard model at half-filling, the symmetry is exactly realized in the ground state with a commensurate modulation wave vector $(\pi,\pi)$ \cite{Micnas:1990ee}. The eight hot spots model provides as well an exact realization of the SU(2) symmetry between the d-SC order and the d-BDW order, with an incommensurate d-BDW modulation wave vector relating two adjacent hot spots on the diagonal \cite{Efetov13}. Although the SU(2) emergent symmetry provides a strong phenomenology for underdoped cuprates, a major drawback is that its exact realization in ground states is fragile \cite{Wang18a} with respect to the variation of tunable parameters like doping or the curvature of the Fermi surface at the hot spots \cite{Fradkin15}. The present formulation solves this issue by providing a robust mechanism for the opening of a gap. The Higgs mechanism at $T^{*}$ leads to a constraint between the amplitudes of the two order parameters ($\left|\chi_{ij}\right|^{2}+\left|\Delta_{ij}\right|^{2}={E^{*}}^{2}$) where the relative phase as well as the two amplitudes fluctuate. In contrast, a similar constraint is an outcome of the SU(2) symmetry in the emergent symmetry theories. In the eight hot spots model, for example, the symmetry is imposed at the hot spots via an exact superposition of the two order parameters satisfying the constraint. In the present model though, the constraint is imposed at each bond, such that the two order parameters fight for phase space in momentum space, but also gain freedom to gap out a larger part of the Fermi surface. In spite of having very similar phenomenologies (like the constraint between the two order parameters), the two models differ in that we get the $T^{*}$ line as a true phase transition, associated with the breaking of a U(1) gauge symmetry. 

In both approaches though, the fluctuations below $T^{*}$ are described by a Non Linear $\sigma$-Model (NL$\sigma$M): the O(4) NL$\sigma$M for the eight hot spots model and the O(3) NL$\sigma$M or equivalently the SU(2) chiral model with the fluctuation space reduced to an $S_{2}$ sphere. These fluctuations can be further recast into the $CP^{1}$ model and remain protected by the Higgs mechanism in a wide range of doping. If we try to accommodate multiple `partners' in the theory, the fluctuations can be recast into a $CP^{n}$ model or an SU($n$+1) chiral model with $n$+1 complex fields satisfying the constraint. The $CP^{n}$ model and the SU($n$+1) chiral model are topologically equivalent for any general $n$ \cite{Perelomov81} (also see Sec.~\ref{sec:warmup}). However, the equivalence of a $CP^{n}$ model with an $O(n+2)$ NL$\sigma$M is only valid for $n=1$. For example, $O(n+2)$ NL$\sigma$M does not have topological properties for $n\geq 2$ whereas the $CP^{n}$ model is topologically non-trivial for all $n$. Thus, an extension of our formalism to the SO(5) model is not possible.

Though $T^{*}$ corresponds to a true phase transition in our theory, the presence of two dimensional fluctuations and topological fluctuations coming from the O(3) NL$\sigma$M will obscure the observation of singularities in thermodynamic probes.

\subsection{Organization of the paper}\label{sec:organiz}

We organize the paper in the following way. In the first part of the paper (Sec.~\ref{sec:Higgs}), we formulate the Higgs mechanism for a generic spinor comprising of two complex order parameters with a U(1) $\times$ U(1) gauge structure (also see Table.~\ref{tab:table1} in Appendix \ref{subsec:Higgsanalogy}). The Higgs mechanism freezes the global U(1) phase of the spinor below a temperature $T^{*}$. This freezing of the global phase can be interpreted as the Hopf fibration of an $S_{3}$ sphere to an $S_{2}$ sphere (Sec.~\ref{sec:subhopf}). As a result of this reduction to the $S_{2}$ sphere, we accommodate topological structures like skyrmions of pseudo-spin operators. With a special choice of the spinor where the individual components correspond to the d-SC and d-BDW order parameters, we show how the Higgs mechanism influences the response of these individual orders to an external EM field (Sec.~\ref{sec:London}). This Higgs phenomenon leaves the conventional London equations describing Meissner effect invariant. We further demonstrate that the structure of the fluctuations below $T^{*}$ can be explained using an O(3) NL$\sigma$M (Sec.~\ref{sec:fluc}).

In the second part of the paper (Sec.~\ref{sec:cuprate}), we illustrate the U(1) $\times$ U(1) gauge theory in the context of cuprate superconductors and relate $T^{*}$ with the pseudo-gap temperature. In this case, the spinor comprises of a d-SC order parameter and a d-BDW order parameter with an entanglement of the global phase of the two at $T^*$ (Sec.~\ref{sec:cupratefieldcon}). We show that the phase entanglement can be interpreted as fractionalizing a PDW order parameter with a neutral gauge field (Sec.~\ref{sec:fracpdw}). Beside giving a simple account for the understanding of cuprate phase diagram in the underdoped region, the theoretical framework discussed in this paper can explain many unique signatures seen in existing experiments on different cuprates. In Sec.~\ref{sec:micro}, \ref{sec:STM} and \ref{sec:orders}, we focus on the following experimental features:

\begin{enumerate}

\item \textit{Precursor gaps of preformed pairs (observed as pair-breaking peaks in Raman Spectroscopy)}: Recent electronic Raman spectroscopy \cite{Loret19} for the first time identified a precursor gap in the p-h charge channel, characterized as a peak in the $B_{2g}$ response of a prototype cuprate. In the context of this paper, these measurements highlight two key features: a) The near degeneracy of the associated energy scales of p-h and p-p pair breaking peaks, and b) the same doping dependence of both these peaks as that of $T^*$. In Sec.~\ref{sec:micro}, using a simplified microscopic model, we estimate the mean-field values of the precursor gaps. We also give a mean-field estimate of the PG energy scale and show that there is only one energy scale characterizing the PG phase. Calculating the momentum dependence, we obtain a gap repartition in the Brillouin zone by two kinds of preformed pairs. The doping dependence of these gaps in different parts of the Brillouin zone have a close resemblance to what is observed in Raman spectroscopy. 

\item \textit{d-CDW spatial phase coherence (observed in STM):} Another distinctive feature of our formalism is that both the relative phase and the global phase of the d-SC order parameter and the d-BDW order parameter is fixed below $T_{c}$. Thus $\chi_{ij}$ and $\Delta_{ij}$ both acquire a spatial phase coherence. In the presence of a magnetic field, the d-SC amplitude is suppressed near a vortex core. As a result, owing to the constraint between the two order parameters, the d-BDW order becomes more recognizable near a vortex core than it is away from the core. This was illustrated by the observation of the enhanced d-CDW (real part of the d-BDW order) near the vortex cores in STM \cite{Hoffman02,Matsuba07,Yoshizawa13,Machida16}. Remarkably, the d-CDW puddles formed near vortex cores exhibit a strong spatial phase coherence \cite{Hamidian15}, substantiating the theory proposed in this paper (also see Sec.~\ref{sec:STM}).

\item \textit{Multiple orders in the PG phase (PDW and loop currents):} The Higgs mechanism at $T^{*}$ results in the freezing of the global phase of the spinor. This leads to an emergence of a long-range phase coherent PDW order below $T_{c}^{\prime}$. Fluctuations in the PDW order for $T>T_{c}^{\prime}$ give the possibility of an auxiliary `loop current' \cite{Fauque06} order in the PG phase, breaking discrete symmetries like time reversal symmetry and parity (see Sec.~\ref{sec:orders}).

\end{enumerate} 

Finally in Sec.~\ref{sec:conclu}, we conclude by summarizing our main outcomes and placing our results in the context of the existing literature in cuprates.

\section{The U(1) $\times$ U(1) gauge theory}\label{sec:Higgs}

Now, we come to the theoretical formulation of our paper, i.e., to describe the Higgs mechanism for a spinor (or doublet). We start with an action acting on a doublet field, which shows U(1) $\times$ U(1) gauge invariance, and we freeze the corresponding overall U(1) phase through a Higgs phenomenon. The mass of the corresponding Higgs boson can be a good candidate for the estimation of the pseudo-gap energy scale of underdoped cuprates. 

\subsection{The model, gauge invariance and the entanglement scale}

We consider two complex fields $z_{1}$ and $z_{2}$ forming a spinor 
\begin{align}
\psi & =\left(\begin{array}{c}
z_{1}\\
z_{2}
\end{array}\right); & \psi^{\dagger}=\left(z_{1}^{*},z_{2}^{*}\right).\label{eq:1.0}
\end{align}
The corresponding action reads as 

{\small{
\begin{align}
{\cal S}_{a,b} & =\int d^{d}x\left[\frac{1}{2g}\left|D_{\mu}\psi\right|^{2}+V\left(\psi\right)+\frac{1}{4}F_{\mu\nu}F^{\mu\nu}+\frac{1}{4}\tilde{F}_{\mu\nu}\tilde{F}^{\mu\nu}\right],\label{eq:1}\\
\mbox{with } & D_{\mu}=\partial_{\mu}-ia_{\mu}-i\tau_{3}b_{\mu},\nonumber \\
 & F_{\mu\nu}=\partial_{\mu}a_{\nu}-\partial_{\nu}a_{\mu},\nonumber \\
\mbox{and } & \tilde{F}_{\mu\nu}=\partial_{\mu}b_{\nu}-\partial_{\nu}b_{\mu},\nonumber 
\end{align}}}
wherein, $\tau_{3}$ is Pauli matrix in the 2 $\times$2 spinorial
space, and $a_{\mu}$ and $b_{\mu}$ are gauge fields corresponding
respectively to the spinor's global phase $\theta$ and relative phase
$\varphi$. The $F_{\mu\nu}F^{\mu\nu}$ (or $\tilde{F}_{\mu\nu}\tilde{F}^{\mu\nu}$)
terms are there to get the most generic form of the action, but they
can be put to zero if one gauge field is neutral. While we carry out calculations in this section without specifying the form of $z_1$ and $z_2$ for generality, the physical meaning of the gauge fields for the case of cuprates is made apparent in Sec.~\ref{sec:cuprate} where we take a specific choice for $z_1$ and $z_2$ (Eq.~\eqref{eq:20}). The form of $\left|D_{\mu}\psi\right|^{2}$ in Eq.~\eqref{eq:1} is explicitly SU(2) symmetric and a motivation for this choice will be given in Sec.~\ref{sec:fracpdw1} (Eq.~\eqref{eq:fracPDWchiral1}). The potential $V(\psi)$ is chosen in such a way that the action in Eq.~\eqref{eq:1} is invariant under the two U(1) gauge transformations 
\begin{align}
\psi & \rightarrow e^{i\theta}\psi, & a_{\mu}\rightarrow a_{\mu}+\partial_{\mu}\theta,\nonumber \\
\psi & \rightarrow e^{i\tau_{3}\varphi}\psi, & b_{\mu}\rightarrow b_{\mu}+\partial_{\mu}\varphi,\label{eq:1a}
\end{align}
without necessarily imposing the SU(2) symmetry. Though the action in Eq.~\eqref{eq:1} resembles that of a Weinberg-Salam model of electroweak interaction, there is an important difference. The Weinberg-Salam model possesses a U(1) $\times$ SU(2) gauge structure. On the contrary, the presence of $V(\psi)$ term in Eq.~\eqref{eq:1} can explicitly break the SU(2) symmetry and the resultant gauge structure is U(1) $\times$ U(1). The U(1) $\times$ U(1) gauge theory is constructed such that one U(1) is related to the global phase $\theta$ of the spinor and the other
U(1) is connected to the relative phase $\varphi$ of the spinor.  The
phase $\theta$ is the same for both components of the spinor $\psi$.
The U(1) $\theta$-gauge invariance of the action in Eq.~\eqref{eq:1}
can be associated with a Higgs mechanism which freezes the common
U(1) phase $\theta$. First, solving for the minimization equations
$\delta{\cal S}_{a,b}/\delta a_{\mu}=0$, and $\delta{\cal S}_{a,b}/\delta b_{\mu}=0$, we
get respectively $a_{\mu}=\partial_{\mu}\theta$, and $b_{\mu}=\partial_{\mu}\varphi$. 

\begin{figure}
{\includegraphics[width=0.8\linewidth]{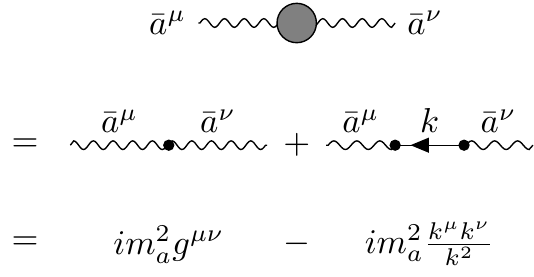} } \caption{\textcolor{black}{Diagrams describing the Higgs phenomenon for the
gauge fields $\bar{a}_{\mu}$ and the integration of the Goldstone mode
$\theta$. The integration over the field $\theta$ has made the gauge
field $\bar{a}_{\mu}$ massive and transverse to the direction of propagation
with $\bar{\mathbf{a}}^{\perp}=\bar{\mathbf{a}}-\mathbf{q}\left(\mathbf{q}\cdot\bar{\mathbf{a}}\right)/q^{2}$. For example, see Ref.~[\onlinecite{peskinBook}].}}
\label{Fig1a} 
\end{figure}

We explore below the possibility that the global phase of the spinor $\theta$ freezes at a typical energy scale whereas the relative phase $\varphi$ remains untouched. A good guess for such a scenario, is that the freezing of the phase $\theta$ corresponds to opening of a mass $E^{*}$ in the spinor field $\psi$, with 
\begin{align}
\sqrt{\left|z_{1}\right|^{2}+\left|z_{2}\right|^{2}} & :=\left|\psi_{0}\right|:=E^{*}.\label{eq:3}
\end{align}
This mass can be obtained from Eq.~\eqref{eq:1} with a specific choice of $V(\psi)$ depending only on the modulus $\psi^{\dagger}\psi$. Evaluation of this energy scale from a microscopic model is done in Sec.~\ref{sec:micro}. Integrating out the phase $\theta$ in Eq.~\eqref{eq:1} (details are given in Appendix~\ref{subsec:Pseudo-spin-Higgs-mechanism}) and differentiating with respect to $\mathbf{a_{q}}$ leads to
\begin{align}
\left(\frac{\left|\psi_{0}\right|}{2g}^{2}+q^{2}\right)\mathbf{a}_{\perp} & =-\left(\frac{\psi^{\dagger}\tau_{3}\psi}{2g}+q^{2}\right)\mathbf{b}_{\perp,}\label{eq:3b}
\end{align}
with $\mathbf{a}^{\perp}=\mathbf{a}-\mathbf{q}\left(\mathbf{q}\cdot\mathbf{a}\right)/q^{2}$ (idem for $\mathbf{b}_{\perp}$). The transverse gauge field $\mathbf{a}^{\perp}$
never becomes fully massive since on average $\left\langle \psi^{\dagger}\tau_{3}\psi\right\rangle =0$.
This feature removes all possibility of a Meissner effect at $T^{*}$.
We can say that the composite field $\bar{a}_{\mu}=a_{\mu}+\text{\ensuremath{\left(\psi^{\dagger}\tau_{3}\psi\right)}/\ensuremath{\left|\psi_{0}\right|}}^{2}b_{\mu}$
becomes massive , with a contribution to the action 
\begin{align}
\Delta{\cal S}_{a} & =\frac{1}{2}m_{a}^{2}\overline{a}_{\mu}\overline{a}^{\mu},\ \ m_{a}=\frac{1}{\sqrt{g}}E^{*}.\label{eq:3a}
\end{align}
This ``Higgs mechanism'' is pictured diagrammatically in Fig.\ref{Fig1a},
where it can be seen that after integration of the Goldstone boson
$\theta$, the condensate contribution to the polarization amplitude
gives a mass $m_{a}$ (Eq.~(\ref{eq:3a})) to the transverse propagator
$D_{\bar{a}_{\mu}}^{-1}=\left\langle {\cal T}\overline{a}_{\mu}\overline{a}_{\nu}\right\rangle ^{-1}=im_{a}^{2}\left(g^{\mu\nu}-\frac{k^{\mu}k^{\nu}}{k^{2}}\right)$.
$g^{\mu\nu}$ is the metric and $k^{\mu}$ is the four momentum.

\subsection{Analogy with the chiral model, Hopf fibration}\label{sec:subhopf}

The Higgs phenomenon exposed above has roots into the Hopf fibration of the sphere $S_{3}$ which can be factorized into $S_{2}$ by taking out a U(1) phase, $S_{3}\sim U(1)\times S_{2}$. In Eq.~\eqref{eq:3}, two complex order parameters are linked through a constraint, which makes an $S_{3}$ sphere. By factorizing a global phase as in Eq.~(\ref{eq:1a}),
the sphere $S_{3}$ reduces to $S_{2}$. The structure of the gauge field in Eq.~\eqref{eq:1} can be much more apparent in an analogous $CP^{1}$ representation of a chiral SU(2) model. A chiral SU(2) model is described by an action, 
\begin{align}
S=\frac{1}{2}\int d^{d}x & Tr[\partial_{\mu}\varphi^{\dagger}\partial_{\mu}\varphi], & \mbox{with } & \varphi_{ab}=\frac{\delta_{ab}}{2}-z_{a}z_{b}^{*},\label{eq:prel2}\\
\mbox{and } & \sum_{a=1}^{2}\left|z_{a}\right|^{2}=1,\nonumber 
\end{align}
where $\varphi$ is a matrix field belonging to the Lie algebra of
the group SU(2) and is parametrized using 2 complex numbers $z_{1}$
and $z_{2}$ with a constraint $\left|z_{1}\right|^{2}+\left|z_{2}\right|^{2}=1$. If the field $\varphi$ is a charge-2 boson, the action in Eq.~\eqref{eq:prel2} can be modified with $\partial_{\mu}\rightarrow\partial_{\mu}-2iA_{\mu}$, where $A_{\mu}$ is the EM vector potential. The action in Eq.~\eqref{eq:prel2} can be recast into a form \cite{Perelomov81} (more details given in Appendix \ref{sec:warmup}),
\begin{align}
S_{a} & =\int d^{d}x\left|D_{\mu}z\right|^{2}, & \mbox{with } & \sum_{\alpha=1}^{2}\left|z_{\alpha}\right|^{2}=1,\label{eq:prel4}\\
D_{\mu} & =\partial_{\mu}-ia_{\mu}, & a_{\mu}= & -i\sum_{a=1}^{2}{z_{a}}^{*}\partial_{\mu}z_{a},\label{eq:prel4.1}
\end{align}
where $z$ is a short hand notation for the doublet $z=\left(z_{1},z_{2}\right)$.
The action is called a $CP^{1}$ model. This $CP^{1}$ model has same
structure as the action in Eq.~\eqref{eq:1} but with $a_{\mu}=-i\sum_{a=1}^{2}{z_{a}}^{*}\partial_{\mu}z_{a}$.
The action in Eq.~\eqref{eq:prel4} is invariant under a global U(1)
change of phase in $z$ reflecting analogous gauge symmetry as in
Eq.~\eqref{eq:1a}. Similar gauge symmetry is frequently used in
the study of spin systems through, for example, a $CP^{1}$ representation
of the SU(2) spinor \cite{Fradkinbook}. The $CP^{1}$ model is equivalent
to the $O(3)$ NL$\sigma$M as defined later in the text in Eq.~\eqref{eq:9a}.
Hence in this case the gauge symmetry of the $CP^{1}$ model is also
associated with the Hopf fibration of $S_{3}\rightarrow S_{2}$, which
effectively transforms an $O(4)$ NL$\sigma$M into an $O(3)$ NL$\sigma$M.

At $T^{*}$ , through a freezing of the U(1) phase $\theta$, the
constraint in Eq.~\eqref{eq:3} is generated. $E^{*}$ in Eq.~(\ref{eq:3}),
which defines the scale of the constraint in Eq.~\eqref{eq:prel4},
can be identified as the energy scale of the PG, whereas $z_{1}$
and $z_{2}$ are respectively d-SC and d-BDW fields. Here, we have considered the spinor consisting of two components $z_1$ and $z_2$ where d-BDW field $z_2$ has only one modulation wave vector $\mathbf{Q}$. However, the formulation of this paper is quite generic and can be extended to incorporate $n+1$ complex fields for a generic $n$, as shown in Appendix \ref{sec:warmup}. For example, if the d-BDW has two modulation wave vectors $(Q_x,0)$ and $(0,Q_y)$, we can have a triplet instead of a doublet. The Higgs mechanism at $T^*$ will then induce a constraint between three complex order parameters. The corresponding action can be recast into a $CP^{2}$ representation of a chiral SU(3) model (see Appendix \ref{sec:su3cp2} for details).

\subsection{London-type Equations for the superconducting
transition}\label{sec:London}
 
In this section, starting
from Eq.~\eqref{eq:1} we look at what happens at lower temperatures
when the second Higgs mechanism, corresponding to the freezing of
the field $\varphi$ occurs. The covariant derivative writes, with
the gauge fields $a_{\mu}$ and $b_{\mu}$ defined above, as
\begin{align}
\left|D_{\mu}\psi\right|^{2}=\left|\partial_{\mu}\psi_{0}\right|^{2}+\psi^{\dagger}\left(\partial_{\mu}\theta+\tau_{3}\partial_{\mu}\varphi-a_{\mu}-\tau_{3}b_{\mu}\right)^{2}\psi.\label{eq:deriexp}
\end{align}
We formally integrate out the Goldstone modes $\theta$ and $\varphi$
one after another (see Appendix \ref{subsec:Derivation-of-the}).
As a result, we get an effective action 
\begin{align}
{\cal S}_{a,b}^{eff} & =\sum_{\mathbf{q}}[\frac{n_{s}^{+}}{2g}\left(\mathbf{a}_{\mathbf{q}}^{\perp}\cdot \mathbf{a}_{-\mathbf{q}}^{\perp}+\mathbf{b}_{\mathbf{q}}^{\perp}\cdot \mathbf{b}_{-\mathbf{q}}^{\perp}\right)\nonumber \\
&+\frac{n_{s}^{-}}{2g}\left(\mathbf{a}_{\mathbf{q}}^{\perp}\cdot \mathbf{b}_{-\mathbf{q}}^{\perp}+\mathbf{a}_{-\mathbf{q}}^{\perp}\cdot \mathbf{b}_{\mathbf{q}}^{\perp}\right)\nonumber \\
 & +\frac{q^{2}}{2}\mathbf{a}_{\mathbf{q}}^{\perp}\cdot \mathbf{a}_{-\mathbf{q}}^{\perp}+\frac{q^{2}}{2}\mathbf{b}_{\mathbf{q}}^{\perp}\cdot \mathbf{b}_{-\mathbf{q}}^{\perp}]+V\left(\psi_0\right),\label{eq:6}\\
\mbox{with } & n_{s}^{+}=\psi^{\dagger}\psi=\left|z_{1}\right|^{2}+\left|z_{2}\right|^{2},\label{eq:nspdef} \\
\mbox{and } & n_{s}^{-}=\psi^{\dagger}\tau_{3}\psi=\left|z_{1}\right|^{2}-\left|z_{2}\right|^{2},\label{eq:nsmdef} 
\end{align}
where a shorthand notation has been taken
for $\mathbf{a}^{\perp}=\mathbf{a}-\mathbf{q}\left(\mathbf{q}\cdot\mathbf{a}\right)/q^{2}$, which is the
transverse component of the gauge field (idem for $\mathbf{b}^{\perp}$).
The $q^{2}$-terms in Eq. (\ref{eq:6}) come from the gauge field strength
$F_{\mu\nu}F^{\mu\nu}$ and $\tilde{F}_{\mu\nu}\tilde{F}^{\mu\nu}$
in Eq.(\ref{eq:1}).

With the help of Eq.~(\ref{eq:6}) and Eq.~(\ref{eq:3b}),
we can describe the generic phase diagram of underdoped cuprates.
Let's take a point at a high temperature as pictured in Fig.~\ref{fig:scdiagram}
and adiabatically decrease the temperature. At $T^{*}$, the system
hits the first Higgs transition, which freezes the phase $\theta$
and the gauge field $\mathbf{a}^{\perp}$ becomes massive, with a mass proportional
to $\sqrt{n_{s}^{+}}=E^{*}$ (see Eq.~\eqref{eq:3a}). The relative phase
$\varphi$ as well as the amplitudes of the fields $\left|z_{1}\right|$
and $\left|z_{2}\right|$ are still fluctuating at $T^{*}$, in such
a way that the line is deprived of the typical thermodynamic sharpness
which usually accompanies the formation of a Higgs phase. The amplitude
of the doublet field $\psi$ condenses at this temperature with no
condensation (no long-range component) in individual field amplitudes
$\left|z_{1}\right|$ and $\left|z_{2}\right|$. So at $T^{*}$, $n_{s}^{-}=0$
from Eq.~\eqref{eq:nsmdef}. The amplitudes of the individual components
$z_{1}$ and $z_{2}$ get condensed, or attain mean-field values,
at lower temperatures $T_{co}$ (for $\left|z_{2}\right|$) and $T_{c}^{\prime}$
(for $\left|z_{1}\right|$). At a lower temperature $T_{c}$ (see
Fig.\ref{fig:scdiagram}), a second Higgs mechanism takes place, where
the remaining phase $\varphi$ freezes. As a result, both $z_{1}$
and $z_{2}$ acquire global phase coherence and the system gets into
a `super-solid' like phase. At this transition, both vector potentials
$\mathbf{a}^{\perp}$ and $\mathbf{b}^{\perp}$ gets associated to a phase stiffness.
Differentiating Eq.~\eqref{eq:6} with respect to $\mathbf{a}^{\perp}$ and $\mathbf{b}^{\perp}$, we get

\begin{align}
\left(\frac{2\left|z_{1}\right|^{2}}{g}+q^{2}\right)(\mathbf{a}_{\mathbf{q}}^{\perp}+\mathbf{b}_{\mathbf{q}}^{\perp}) & =0,\label{eq:6da}\\
\left(\frac{2\left|z_{2}\right|^{2}}{g}+q^{2}\right)(\mathbf{a}_{\mathbf{q}}^{\perp}-\mathbf{b}_{\mathbf{q}}^{\perp}) & =0,\label{eq:6db}
\end{align}

or equivalently 
\begin{align}
\left(\frac{2\left|z_{1}\right|^{2}\left|z_{2}\right|^{2}}{gn_{s}^{+}}+q^{2}\right)\mathbf{a}_{\mathbf{q}}^{\perp} & =0,\label{eq:6f}\\
\left(\frac{2\left|z_{1}\right|^{2}\left|z_{2}\right|^{2}}{gn_{s}^{+}}+q^{2}\right)\mathbf{b}_{\mathbf{q}}^{\perp} & =0,\nonumber 
\end{align}
which defines second London equation. 

In the case for cuprate superconductors, the spinor is identified as,
\begin{align}
\psi &=\left(\begin{array}{c}
z_{1}\\
z_{2}
\end{array}\right) =e^{i\theta}e^{i\tau_{3}\varphi}\psi_{0}, & \mbox{{with}\ } & \psi_{0}=\left(\begin{array}{c}
\tilde{z}_{1}\\
\tilde{z}_{2}
\end{array}\right),\label{eq:5a}
\end{align}
where $\tilde{z}_1=\hat{d} \left| z_1 \right|$ and $\tilde{z}_2=\hat{d} \left| z_2 \right| e^{iQ.r}$ with $\hat{d}$ being the d-wave form factor ($\hat{d}=\pm 1$ depending on the direction of the bond) and $Q$ is the modulation wave vector of the d-BDW field $z_2$. It should be emphasized that the amplitude $\left|z_2\right|$ depends on the choice of the modulation wave vector $Q$ and should be written as $\left|z_2^{Q}\right|$. For simplicity of notations, we have not used the superscript $Q$ in $\left|z_2\right|$. We consider d-wave form factor in both $\tilde{z}_1$ and $\tilde{z}_2$ and consider modulations only in $\tilde{z}_2$. Note that the constraint is still given by Eq.~\eqref{eq:3}. The two linear combinations of $a_{\mu}$ and $b_{\mu}$ are identified as $a_{\mu}+b_{\mu}=\alpha_{\mu}+2 A_{\mu}$ and $a_{\mu}-b_{\mu}=\alpha_{\mu}$ where $A_{\mu}$ is the EM vector potential and $\alpha_{\mu}$ corresponds to a neutral gauge field (see also Sec.~\ref{sec:cuprateu1u1} for details). Hence, from Eq.~\eqref{eq:6da} we can see that the transition
at $T_{c}$ is an usual superconducting transition, giving mass to
the EM field $A_{\mu}$, which will account for Meissner
effect and quantization of the currents, with the usual superfluid
stiffness $\rho_{s}=2\left|z_{1}\right|^{2}/g$. Eq.~\eqref{eq:6f}
shows that the freezing of both phases $\theta$ and $\varphi$ imply
that both $\left|z_{1}\right|$ and $\left|z_{2}\right|$ condense
(which automatically leads to a long-range $n_{s}^{+}$, but the
reciprocal is not true), which is verified for $T<T_{c}$. In the range $T_c<T<T_{c}^{\prime}$, both $n_{s}^{+}\ne0$ and $n_{s}^{-}\ne0$, i.e., $\left|z_{1}\right|$ and $\left|z_{2}\right|$ attain uniform components. But we should note that since the relative phase fluctuates, the current-current
correlation (which gives the superfluid density \cite{Uemura89,Homes04}) will still be zero due to the lack of phase coherence \cite{bennemannbook} in $z_{1}$ and $z_{2}$. This
is not captured in the formulation of this paper as we do not intend to connect the current-current correlation to the stiffness of the gauge field. So, even if both $n_{s}^{+}\ne0$ and $n_{s}^{-}\ne0$, there would be no Meissner effect for $T_{c}<T<T_{c}^{\prime}$.

\subsection{Fluctuations below $T^{*}$ in the PG phase}\label{sec:fluc}

We already indicated in the last section that the freezing of the global phase $\theta$ of the spinor leaves fluctuations in the relative
phase $\varphi$ and also the amplitudes $\left|z_{1}\right|$ and $\left|z_{2}\right|$. Now, we ask the following question: What is the form of these fluctuations below $T^{*}$?

In this formulation, phase below $T^{*}$ is characterized by $n_{s}^{+}=\psi_{0}^{\dagger}\psi_{0}\neq0$. If we set the gauge field $b_{\mu}=0$ and expand the derivative in Eq.~(\ref{eq:deriexp}) to the second order we get the corresponding contribution to the action $S_{a}=1/\left(2g\right)\int d^{d}xn_{s}^{+}\left(\left(\partial_{\mu}\theta\right)^{2}+\left(\partial_{\mu}\varphi\right)^{2}+{a}_{\mu}^{2}-2\partial_{\mu}\theta{a}_{\mu}\right)+\left|\partial_{\mu}\psi_{0}\right|^{2}+n_{s}^{-}\left(2\partial_{\mu}\theta\partial_{\mu}\varphi-2{a}_{\mu}\partial_{\mu}\varphi\right)$. After freezing the phase $\theta$ and differentiating with respect to $\mathbf{a}_{\mathbf{q}}$, we obtain for $T<T^{*}$ (details given in Appendix \ref{subsec:U(1)-U(1)-theory,}), 
\begin{align}
S_{T^{*}}^{eff} =\frac{1}{2g}\int d^{d}x & \left[\frac{4\left|z_{1}\right|^{2}\left|z_{2}\right|^{2}}{n_{s}^{+}}\left(\partial_{\mu}\varphi\right)^{2}+\left(\partial_{\mu}\left|z_{1}\right|\right)^{2} \right.\nonumber\\
&\left.+\left(\partial_{\mu}\left|z_{2}\right|\right)^{2}+V\left( \psi_0\right)\right].\label{eq:9}
\end{align}
Noticing (see Appendix \ref{subsec:Explicit-form-of}) that $a_{\mu}$ from Eq.~\eqref{eq:prel4.1} has now the form $a_{\mu}=-i\left(\psi^{\dagger}\tau_{3}\partial_{\mu}\psi\right)/\left| \psi_0 \right|^2$, we obtain that Eq.~\eqref{eq:9} is the form that the SU(2) chiral model takes when the mapping to the $CP^{1}$ model is taken into account; it thus describes the fluctuations below $T^{*}$. As pointed out above, an equivalent form of the fluctuations is given with the O(3) NL$\sigma$M, using the variables $m^{a}=(E^{*})^{-1}z_{\alpha}^{*}\sigma_{\alpha\beta}^{a}z_{\beta}$ as introduced in the Appendix \ref{sec:warmup}, which satisfies the constraint $\sum_{a}\left|m^{a}\right|^{2}=1$. The corresponding action is now of the O(3) NL$\sigma$M: 
\begin{align}
S & =1/2\int d^{d}x\sum_{a=1}^{3}\left(\partial_{\mu}m^{a}\right)^{2}+V\left(m^{a}\right),\nonumber \\
\mbox{with } & \sum_{a=1}^{3}\left|m^{a}\right|^{2}=1.\label{eq:9a}
\end{align}
It is not a surprise that this is similar to the $CP^{1}$ representation in Eq.~\eqref{eq:prel4}. The Higgs mechanism at $T^{*}$ has given a mass to the sum of the squares of the fields $z_{1}$ and $z_{2}$ ($\left|z_{1}\right|^{2}+\left|z_{2}\right|^{2}$), and expanding below $T^{*}$, one thus recovers the typical structure of the chiral SU(2) model in Eq.~\eqref{eq:prel4}. The potential terms in Eqs.~\eqref{eq:9} and \eqref{eq:9a} gives a massive contribution to the NL$\sigma$M such that there is no exact SU(2) symmetry at all dopings. Due to the mass contribution, $z_{1}$ and $z_{2}$ fields order at low temperatures with power law correlations in $d=2$ \cite{Efetov13}.

With the choice of the spinor $\psi=\left(z_{1},z_{2}\right)^T$, the fluctuating fields are $m^z\equiv z_1z_1^*-z_2z_2^*$, $m^+\equiv z_1z_2^*$ and $m^-\equiv z_2z_1^*$. For cuprate superconductors, the fluctuating fields $m^{+}$ and $m^{-}$ take the form of PDW operators $\eta$ and $m^{z}$ takes the form of fluctuating densities on sites (details of this identification is given in Appendix \ref{sec:flucchirala}). These PDW operators involve the fluctuations in the amplitudes ($\left|z_1\right|$ and $\left|z_2\right|$) and the relative phase $\varphi$. They construct the SU(2) Lie Algebra corresponding to the O(3) NL$\sigma$M. The structure of the PDW fluctuations in the case of SU(2) chiral model and a contrast with the SU(2) emergent symmetry models is discussed in details in Appendix \ref{sec:PDWPGoperator}. Note that the O(3) NL$\sigma$M admits chiral structures also called skyrmions, in the fluctuation space ($\eta$ space). These local structures might account for the recent observation of huge thermal Hall constant in these materials \cite{Grissonnanche19}, in addition to the already existing proposals based on proximity to a quantum critical point of a `semion' topological ordered state \cite{Chatterjee19}, presence of spin-dependent next nearest neighbor hopping in the $\pi$-flux phase \cite{Han19} or presence of large loops of currents \cite{Varma19}.

The freezing of the global phase at $T^*$ results into the constraint in the NL$\sigma$M and thus opens up a regime of strong fluctuations in the amplitudes of both $z_1$ and $z_2$. Just below $T^*$, the amplitudes of $z_1$ and $z_2$ do not acquire uniform components. To illustrate this we can parametrize the amplitudes as,
\begin{equation}
\left| z_1 \right|=E^{*}\left|\sin{\varrho}\right|~~\mbox{and}~~\left| z_2 \right|=E^{*}\left|\cos{\varrho}\right|,
\label{eq:parrelamp}
\end{equation} 
where $\varrho$ is not a phase and just a parameter that quantifies the relative amplitude such that the constraint, $\left|z_1\right|^{2}+\left|z_2\right|^{2}={E^*}^2$, is satisfied. Remember that the constraint is written in real space and applicable for all bonds. If we do a spatial average of the square of the amplitudes, we get
\begin{align}
&\left\langle \left| z_1 \right|^2 \right\rangle={E^{*}}^2 \left\langle \sin^2{\varrho} \right\rangle = \frac{{E^{*}}^2}{2} \nonumber \\
\mbox{and} &\left\langle \left| z_2 \right|^2 \right\rangle={E^{*}}^2 \left\langle \cos^2{\varrho} \right\rangle = \frac{{E^{*}}^2}{2},
\label{eq:averagefluc}
\end{align} 
where $\left\langle .. \right\rangle$ denotes average over all sites. So, the amplitudes $\left|z_1\right|$ and $\left|z_2\right|$ fluctuate with a mean average to satisfy the constraint. But, the amplitudes $\left|z_1\right|$ and $\left|z_2\right|$ do not have uniform components or condensed values just below $T^*$. The amplitudes of the individual fields $\left|z_1\right|$ and $\left|z_2\right|$ attain uniform mean-field components at temperatures $T_c$ and $T_{c}^{\prime}$ respectively. This can be thought of as a BEC of the amplitudes $\left|z_1\right|$ and $\left|z_2\right|$. Below these temperatures, the amplitudes of these fields still fluctuate but now with a uniform component such that
\begin{equation}
\left|z_1\right|=\left|z_1\right|_0+\delta \left|z_1\right|~~\mbox{and}~~\left|z_2\right|=\left|z_2\right|_0+\delta \left|z_2\right|,
\label{eq:bosecond}
\end{equation} 
where $\left|z_1\right|_0$ and $\left|z_2\right|_0$ are the uniform components and $\delta \left|z_1\right|$ and $\delta \left|z_2\right|$ are the fluctuating parts. $\left|z_1\right|_0$ and $\left|z_2\right|_0$ also correspond to the mean-field precursor gaps in momentum space as calculated in Sec.~\ref{sec:micro}. As one lowers the temperature, the condensed parts of the amplitudes increase and gradually eat up the fluctuating parts, still satisfying the constraint. At $T_c$, the fluctuations in the relative phase also freeze owing to a second Higgs mechanism.     

\section{Fractionalized PDW and cuprate superconductors}\label{sec:cuprate}

In Sec.~\ref{sec:Higgs}, we have introduced the formalism of the Higgs mechanism for a spinor with U(1) $\times$ U(1) gauge structure and demonstrated its consequences. Though we took a cuprate superconductor as an example to illustrate various effects, our discussion in Sec.~\ref{sec:Higgs} is much more generic and is applicable to any spinor. In this section, we explicitly use the case for underdoped cuprates and connect the outcomes of this Higgs phenomenon to experimental signatures like different energy gaps in Raman spectroscopy (Sec.~\ref{sec:micro}), the charge modulation phase coherence in scanning tunneling microscopy (Sec.~\ref{sec:STM}) and the emergence of multiple orders in the PG phase (Sec.~\ref{sec:orders}).    

The special Higgs mechanism at $T^{*}$, which freezes the global phase of the spinor, but does not quench the full entropy, has strong experimental consequences. In this section, we will focus on a few prominent experimental consequences emerging from the theory, keeping in mind that we cannot yet give credit for all the fascinating observations performed over the years, but with the hope that the experiments chosen are distinguishing enough to make our case. In order to describe the phenomenology of underdoped cuprates, in this section, we first give the form of the fields constituting the spinor defined in Sec.~\ref{sec:Higgs}, then describe the $U(1) \times U(1)$ gauge structure and give the details of the alternative picture of the fractionalization of a PDW.

\subsection{Gauge theory formalism}\label{sec:cuprateu1u1}

\subsubsection{U(1) $\times$ U(1) gauge structure}\label{sec:cupratefieldcon}

In the case of underdoped cuprates, the field operators $z_{1}$ and
$z_{2}$ are identified to be the particle-particle (or Cooper) pairing
order and ($Q$-modulated) particle-hole (or bond-excitonic) pairing
order, with 
\begin{align}
z_{1} & \rightarrow \hat{d}\sum_{\sigma}\sigma c_{j-\sigma}c_{i\sigma}\equiv\hat{\Delta}_{ij},\nonumber \\
z_{2} & \rightarrow \hat{d}\sum_{\sigma}c_{i\sigma}^{\dagger}c_{j\sigma}e^{iQ\cdot\left(r_{i}+r_{j}\right)/2}\equiv\hat{\chi}_{ij},\label{eq:20}
\end{align}
where both $z_{1}$ and $z_{2}$ are defined on nearest neighbor bonds
$\langle ij\rangle$ (see Fig.~\ref{Fig:bond}) of a square lattice
where $r_{j}=r_{i}+\delta$ with $\delta=\pm\hat{u}_{x}~\mbox{or}~\pm\hat{u}_{y}$
and $\hat{u}$ is the lattice translational operator; and $\hat{d}$
is an operator describing the d-wave structure factor. We
note immediately that, although the field $\Delta_{ij}$ directly
couples to the EM field $A_{\mu}$ through the gauge field $a_{\mu}^{\Delta}=\alpha_{\mu}+2A_{\mu}$, to the first approximation, $\chi_{ij}$ is neutral to the EM field since the corresponding gauge field $a_{\mu}^{\chi}$ couples only to the gradient of the EM field $a_{\mu}^{\chi}=\alpha_{\mu}+\hat{u}.\partial_{\hat{u}}\left(A_{\mu}\right)$, which can be approximated as $a_{\mu}^{\chi}\sim\alpha_{\mu}$. Hence the gauge field $\alpha_{\mu}$ is neutral to the EM field and is related to the phase $\theta_{\chi}$ of the complex field $\chi_{ij}$ as $\alpha_{\mu}=\partial_{\mu} \theta_{\chi}$. $\alpha_{\mu}$ corresponds to the local variation of the effective modulation wave vector of the charge modulations. Note that incommensurate charge modulations are typically associated with a local phase which is responsible for the coupling to the lattice \cite{Lee79}. The stiffness of the neutral gauge field $\alpha_{\mu}$ leads to the constraint in Eq.~\eqref{eq:3}, which also defines the energy scale of the PG phase. 

The two complex field operators $\hat{\Delta}_{ij}$ and $\hat{\chi}_{ij}$ thus transform under local gauge transformations as 
\begin{align}
\hat{\Delta}_{ij}\rightarrow & e^{i\theta_{\Delta}}\hat{\Delta}_{ij},\nonumber \\
\hat{\chi}_{ij}\rightarrow & e^{i\theta_{\chi}}\hat{\chi}_{ij}.\label{eq:delchiphase1}
\end{align}
The global phase of the spinor in Eq.~\eqref{eq:1a} is then given by $(\theta_{\Delta}+\theta_{\chi})/2$ and the relative phase is given by $(\theta_{\Delta}-\theta_{\chi})/2$. The corresponding gauge fields in Eq.~\eqref{eq:1a} for the global and the relative fields respectively are given by
\begin{align}
a_{\mu} & =A_{\mu}+\alpha_{\mu},\nonumber \\
b_{\mu} & =A_{\mu}.\label{eq:gaugeconnect}
\end{align}

It is not the first time that a neutral field with a phase is related to a neutral gauge field through a constraint. This was used in the past as an ansatz for fractionalizing the electron \cite{Baskaran88,Ioffe89,Nagaosa90,Lee92} within, for example, a U(1) gauge theory 
\begin{align}
c_{i}= & f_{i}^{\dagger}b_{i},\nonumber \\
 & f_{i}^{\dagger}f_{i}+b_{i}^{\dagger}b_{i}=1,\label{eq:fract1}
\end{align}
where $b_{i}$ represents a charged ``holon'' and $f_{i}$ represents a neutral ``spinon''. The corresponding local gauge invariance writes 
\begin{align}
f_{i} & \rightarrow e^{i\theta}f_{i}, & b_{i} & \rightarrow e^{i\theta}b_{i}.\label{eq:fract2-1}
\end{align}
which corresponds to the fluctuations of the global phase of the spinor $\psi=\left(f_{i},b_{i}\right)^{T}$.
\\
\\

\subsubsection{Fractionalization of a PDW: operator construction}\label{sec:fracpdw}

Here, the emergent neutral gauge field also corresponds to a fractionalization, not of an electron as in Eq.~\eqref{eq:fract1}, but of an order parameter field: a preformed PDW pair which fractionalizes
into preformed p-p pairs ($\Delta_{ij}$) and p-h pairs ($\chi_{ij}$). Being a particle-particle field with finite modulation wave vector, the PDW operator is given by
\begin{align}
\hat{\eta} & =\left[\hat{\Delta}_{ij},\hat{\chi}_{ij}^{\dagger}\right], & \hat{\eta}^{\dagger} & =\left[\hat{\chi}_{ij},\hat{\Delta}_{ij}^{\dagger}\right],\label{eq:decomp1}
\end{align}
to which we add a constraint in analogy with Eq.~\eqref{eq:fract1}
\begin{align}
\hat{\Delta}_{ij}^{^{\dagger}}\hat{\Delta}_{ij}+\hat{\chi}_{ij}^{\dagger}\hat{\chi}_{ij} & =1,\label{eq:PDWdef2}
\end{align}

The corresponding gauge structure is analogous to Eq.~\eqref{eq:fract2-1} with the operators $\hat{\eta},\hat{\eta}^{\dagger}$ being invariant within 
\begin{align}
\hat{\Delta}_{ij} & \rightarrow e^{i\theta}\hat{\Delta}_{ij}, & \hat{\chi}_{ij} & \rightarrow e^{i\theta}\hat{\chi}_{ij}.\label{eq:fract2-1-1}
\end{align}
In order to construct a field theory, we make the correspondence $\hat{\eta}\rightarrow\Delta_{\text{PDW}}$, $\hat{\Delta}_{ij}\rightarrow\Delta_{ij}$, $\hat{\chi}_{ij}\rightarrow\chi_{ij}$, ($\hat{\eta}^{\dagger}\rightarrow\Delta_{\text{PDW}}^{*}$, $\hat{\Delta}_{ij}^{\dagger}\rightarrow\Delta_{ij}^{*}$, $\hat{\chi}_{ij}^{\dagger}\rightarrow\chi_{ij}^{*}$). The typical field theory associated with the decomposition Eq.~\eqref{eq:decomp1} with the constraint Eq.~\eqref{eq:PDWdef2} is the rotor model (see for example Eqs.~\eqref{eq:D7} and \eqref{eq:D8} in the context of emergent SU(2) symmetry):
\begin{equation}
S=\int d^{d}x~\frac{1}{2}\sum_{a,b=1}^{2}\left|\omega_{ab}\right|^{2},\label{eq:rotor}
\end{equation}
where
\begin{align}
\omega_{ab} & =z_{a}^{*}\partial_{\mu}z_{b}-z_{b}\partial_{\mu}z_{a}^{*},\label{eq:rotor2}
\end{align}
with here $z_{1}=\Delta_{ij}$, $z_{2}=\chi_{ij}$, ($z_{1}^{*}=\Delta_{ij}^{*}$, $z_{2}^{*}=\chi_{ij}^{*}$). Expanding Eq.~\eqref{eq:rotor} and using the constraint Eq.~\eqref{eq:PDWdef2} we get
\begin{align}
S & =\int d^{d}x\left(\sum_{a}\left|\partial_{\mu}z_{a}\right|^{2}-\sum_{a,b}\left(z_{a}^{*}\partial_{\mu}z_{a}\right)\left(z_{b}\partial_{\mu}z_{b}^{*}\right)\right),\label{eq:rotor3}
\end{align}
which in this form is equivalent to the chiral model Eq.~\eqref{eq:prel2}, see also Eq.~\eqref{eq:prel3} in Appendix \ref{sec:chiralmodel}. Introducing the real gauge field
\begin{align}
\alpha_{\mu} & =\frac{i}{2}\sum_{a}\left(z_{a}^{*}\partial_{\mu}z_{a}-z_{a}\partial_{\mu}z_{a}^{*}\right),\label{eq:rotor4}
\end{align}
the action in Eq.~\eqref{eq:rotor3} can be re-cast into the $CP^{1}$ model Eq.~\eqref{eq:prel4}:
\begin{align}
S & =\int d^{d}x\left|D_{\mu}\psi\right|^{2},\label{eq:fracPDWchiral3-1}\\
\mbox{with } & D_{\mu}=\partial_{\mu}-i\alpha_{\mu}.\nonumber 
\end{align}
The gauge invariance imposed by the transformation in Eq.~\eqref{eq:fract2-1-1} corresponds to $\alpha_{\mu}\rightarrow\alpha_{\mu}-\partial_{\mu}\theta$.

\subsubsection{Fractionalization of a PDW: a simplified construction}\label{sec:fracpdw1}

Since a PDW is a particle-particle field with finite modulation wave vector, the corresponding order parameter can be described in the field theoretic picture by 
\begin{equation}
\Delta_{\text{PDW}}=\Delta_{ij}\chi_{ij}^{*},\label{eq:PDWdef1}
\end{equation}
with the constraint in Eq.~\eqref{eq:PDWdef2} reading as
\begin{align}
\Delta_{ij}^{*}\Delta_{ij}+\chi_{ij}^{*}\chi_{ij} & =1,\label{eq:PDWdef2-1}
\end{align}

In the absence of EM field, the action governing the gradient terms of the $\Delta_{\text{PDW}}$ field is typically given by 
\begin{equation}
S=\int d^{d}x~\partial_{\mu}\Delta_{\text{PDW}}^{*}\partial_{\mu}\Delta_{\text{PDW}},\label{eq:PDWgrad}
\end{equation}
which is a special case of the chiral model Eq.~\eqref{eq:prel2} for which $\varphi_{11}=\varphi_{22}=0$, and $\varphi_{12}=\varphi_{21}^{*}=\Delta_{\text{PDW}}$. After inserting the form of $\Delta_{\text{PDW}}$ Eq.~\eqref{eq:PDWdef1} into Eq.~\eqref{eq:PDWgrad}, and using the constraint Eq.~\eqref{eq:PDWdef2-1}, the action can be rewritten as (details are given in Appendix \ref{sec:CP1fromPDW})
\begin{align}
S= & \int d^{d}x~\left[\left|\left(\partial_{\mu}+i\overline{\alpha}_{\mu}\right)\Delta_{ij}\right|^{2}+\left|\left(\partial_{\mu}-i\alpha_{\mu}\right)\chi_{ij}\right|^{2}\right]\label{eq:fracPDWchiral1}\\
 & \mbox{where }~\alpha_{\mu}=-i\left(\Delta_{ij}\partial_{\mu}\Delta_{ij}^{*}-\chi_{ij}^{*}\partial_{\mu}\chi_{ij}\right).\label{eq:fracPDWchiral2}
\end{align}
Thus $\alpha_{\mu}$ appears as an emergent gauge field owing to the fractionalization of the $\Delta_{\text{PDW}}$ and the constraint in Eq.~\eqref{eq:PDWdef2-1}. The action in Eq.~\eqref{eq:fracPDWchiral1} is equivalent to the $CP^{1}$ model Eq.~\eqref{eq:prel4} with a particular choice of the spinor (for details see Appendix \ref{sec:CP1fromPDW}). The gauge field $\alpha_{\mu}=\partial_{\mu}\theta$ corresponds to the global phase of the spinor $\psi=\left(\Delta_{ij},\chi_{ij}\right)^{T}$. In the presence of EM field, $\alpha_{\mu}$ is shifted by $A_{\mu}$ and the gauge field for the global phase is given by $\alpha_{\mu}+A_{\mu}$. Note that we obtain an SU(2) symmetric form for the gradients terms in Eq.~\eqref{eq:fracPDWchiral1} and thus motivates the choice of the form of $\left|D_{\mu}\psi\right|^{2}$ in our starting action Eq.~\eqref{eq:1}.  
\\
\\
\paragraph*{An alternative route to the effective action:}~
\\
\\
The fractionalization is a first route to obtain the effective action of the problem with a given constraint. But in order to generate the constraint, it is equivalent to minimize with respect to the corresponding gauge field. In Sec.~\ref{sec:Higgs}, we took precisely this other route of considering an action (Eq.~\eqref{eq:1}) with two varying gauge fields $a_{\mu}$ (for the global phase) and $b_{\mu}$ (for
the relative phase). Minimizing the action with respect to the gauge field $a_{\mu}$ leads to the constraint $\left|\chi_{ij}\right|^{2}+\left|\Delta_{ij}\right|^{2}={E^{*}}^{2}$. As a result, $\alpha_{\mu}+A_{\mu}=0$ at $T^{*}$. Only at $T_{c}$, the EM gauge field $A_{\mu}$ gets stiff when the relative phase is also frozen.
\\
\\
\paragraph*{Connection to other fractionalization theories:}~
\\
\\
As mentioned earlier, a neutral gauge field corresponding to the fractionalization of an entity is not new in the field of cuprates. For example, a fictitious gauge field is often introduced in theories with strong electronic correlations where the electrons fractionalize into `spinons' and `holons' \cite{Baskaran88,Ioffe89,Nagaosa90,Lee92}. Minimizing the action with respect to this fictitious gauge field generates the constraint of no double occupancy on each lattice site (see e.g. Ref.[\onlinecite{Lee92}]). Finite double occupancy leads to fluctuations in the gauge field. However, since this gauge field is fictitious, there is no dynamics associated with it. Instead of fractionalizing the electron, here, we fractionalize an order parameter which is the PDW. In contrast to the electron's fractionalization, the neutral gauge field in our case is dynamical. Since the neutral gauge field corresponds to the gradient of the global phase of the spinor comprising of physical fields, a restoring force term, proportional to the square of the gauge field, appears in the effective action Eq.~\eqref{eq:fracPDWchiral1}. There are other fundamental differences that appear when we fractionalize an order parameter. At the operator level, a PDW is defined by a commutator of dSC and dBDW orders. Ideally, fractionalizing the PDW should be viewed from the operator formalism which evenetually leads to an effective field theoretic description through a $CP^1$ model Eq.~\eqref{eq:fracPDWchiral3-1}. Furthermore, the fractionalization in our case enforces the constraint which defines the PG energy scale (Eq.~\eqref{eq:3}). For $T>T^*$, the $\Delta_{\text{PDW}}$ is not fractionalized with no constraint on $\Delta_{ij}$ and $\chi_{ij}$. As a result the gauge field $\alpha_{\mu}$ is no longer fixed as it is the case in Eq.~\eqref{eq:fracPDWchiral2} for $T<T^*$. This is in sharp contrast to the conventional electron's fractionalization where the constraint is not associated with an energy scale.

There is another recent proposal of fractionalizing an order parameter where fractionalization of a spin density wave order parameter gives rise to an SU(2) gauge theory \cite{Sachdev19}. Considering that the emerging fields in cuprates are due to the presence of strong electronic correlations, it is not completely out of the scope that preformed p-h, p-p or PDW pairs can emerge. It is also conceivable to make a variational ansatz, where a PDW pair fractionalizes into p-p and p-h pairs. Alternatively, we could have chosen to fractionalize p-p pairs into p-h and PDW pairs ($\Delta_{ij}=\Delta_{\text{PDW}}\chi_{ij}^{*}$) or again p-h pairs into p-p and PDW pairs ($\chi_{ij}=\Delta_{ij}\Delta_{\text{PDW}}^{*}$). We chose to fractionalize the PDW field as it is the most fragile out of the three, i.e., the
most difficult to stabilize within any theoretical scheme \cite{Agterberg19}.

\subsection{A microscopic model for precursors in the charge and Cooper pairing
channels; application to Raman Spectroscopy}\label{sec:micro}

A recent electronic Raman spectroscopy experiment \cite{Loret19} performed on Hg-1223 revealed for the first time a precursor gap in the charge channel forming due to the p-h preformed pairs. This gap scale is characterized as the center of a broad peak in the $B_{2g}$ channel, which preferentially probes the nodal regions of the Brillouin zone. This peak is seen below $T_{co}$ and the corresponding energy scale has a doping dependence which follows $T^*$ rather than $T_{co}$. This is compared with the more conventional $B_{1g}$ Raman response (preferentially probing the AN part of the Brillouin zone) which is used to extract the value of the precursor gap due to the p-p preformed pairs as a pair-breaking peak. Through a similar doping dependence as that of $T^*$, these measurements connect the gap scales in both p-h and p-p channels to the PG phase. We note that a similar peak in the $B_{2g}$ channel was also observed in Hg-1201 \cite{Li13} earlier, but lacked interpretation in terms of charge order. 

In this section, using a simplified microscopic model, we construct the gap equations corresponding to the p-p, the p-h and the PG order parameters. Using momentum independent results, we argue that the three gap scales are identical and the PG is characterized by a single energy scale. We further give the mean-field estimates of the momentum dependent gap scales in the p-p and p-h channels. Here, we only focus on finding estimates on the values of the precursor gaps. A detailed study of the electronic Raman spectrum is left for a future work. We also note that the concept of two kinds of entangled preformed pairs constrained by the relation in Eq.~\eqref{eq:3} gives $E^*$ as the PG energy scale, which is non-trivially related to the precursor gaps in momentum space. The connection between $E^*$ and the momentum space gaps is given by,
\begin{equation}
(E^*)^2=\sum_{k}\Omega_k^2=\sum_{k}\Delta_k^2+\chi_k^2+\mbox{fluctuations},\label{eq:consmom}
\end{equation}
where $\Omega_k$ is the gap corresponding to the entangled PG state, $\Delta_k$ is the mean-field p-p gap (condensate contribution of the preformed p-p pairs) and $\chi_k$ is the mean-field p-h gap (condensate contribution of the preformed p-h pairs). Even neglecting the fluctuations in Eq.~\eqref{eq:consmom}, we see that the real space constraint (of Eq.~\eqref{eq:3}) can be satisfied by repartitioning the Fermi surface with $\Delta_k$ and $\chi_k$ prevailing in different parts. $E^*$ should not be confused with the higher energy hump in the $B_{1g}$ Raman response \cite{Loret19}. This higher energy hump can be thought of as a coupling of the fermions to a collective mode due to the fluctuations in the PG phase (for e.g., similar coupling of the fermions to a collective mode is studied in Ref.~[\onlinecite{Norman97,Onufrieva99,Eschrig:2000bf}]).

\subsubsection{The Microscopic model}\label{sec:micromodel}

In Sec.\ref{sec:Higgs}, we have presented the U(1) $\times$ U(1) gauge theory of two kinds of preformed pairs. The amplitude of the p-h pairs condense to attain a uniform component at temperatures below $T_{co}$. The corresponding temperature for the amplitude of the p-p pairs is $T^{\prime}_{c}$. Below these temperatures, the uniform component of the amplitude of the preformed pairs can be observed as precursor gaps in the fermionic spectrum. In order to understand the momentum space structure of these gaps, we consider a simplified microscopic model of electrons interacting through short-range antiferromagnetic fluctuations and an off-site density-density interaction. While a model with short-range antiferromagnetic fluctuations leads to an exact degeneracy between the d-SC and the d-BDW at the hot-spots (the k-points where the Fermi surface intersects the antiferromagnetic Brillouin zone), the additional off-site density-density interaction breaks the degeneracy (slightly) even at the hot-spots by enhancing the d-BDW amplitude. The model is treated at the mean-field level in momentum space. Even if the order parameters are taken to be complex, the self-consistent equations only fix the amplitude for each of the gaps. Phase fluctuations or amplitude fluctuations are not considered in the mean-field formalism of this section. 

As a minimal model describing quasi-degenerate particle-particle and particle-hole orders, we consider the following Hamiltonian in real space with both short-range antiferromagnetic (AF) and off-site Coulomb interactions: 
\begin{align}
H & =\sum_{i,j,\sigma}{\left(t_{ij}+\mu \ \delta_{ij}\right)\ (c_{i,\sigma}^{\dagger}c_{j,\sigma}+h.c)}\nonumber \\
 & +\sum_{ij}\left( {J_{ij}\ \bm{S}_{i}\cdot\bm{S}_{j}+V_{ij}\ n_{i}n_{j}} \right),\label{eq:11b}
\end{align}
where $c_{i,\sigma}^{\dagger}$ ($c_{i,\sigma}$) is a creation (annihilation)
operator for an electron at site $i$ with spin $\sigma$, $n_{i}=\sum_{\sigma}c_{i,\sigma}^{\dagger}c_{i,\sigma}$
is the number operator and $\bm{S}_{i}=c_{i,\alpha}^{\dagger}\bm{\sigma}_{\alpha,\beta}c_{i,\beta}$
is the spin operator at site $i$ ($\bm{\sigma}$ is the vector of
Pauli matrices). $J_{ij}$ is an effective AF coupling which comes
for example from the Anderson super-exchange mechanism. The
constraint of no double occupancy typical of the strong Coulomb onsite
interaction is implemented through the Gutzwiller approximation \cite{Anderson:2004cz} by
renormalizing the hoping parameter and the spin-spin interaction with
\begin{align}
t\left(p\right) & =g_{t}\left(p\right)t=\frac{2p}{1+p}t,\\
J\left(p\right) & =g_{J}\left(p\right)J=\frac{4}{(1+p)^{2}}J,
\end{align}
where $p$ is the hole doping and the density-density interaction does not get renormalized. We also assume that the antiferromagnetic correlations are dynamic, strongly renormalized, and short-ranged, as given by the phenomenology of neutron scattering studies for cuprates \cite{Hinkov07}and $V_{ij}$ is a residual Coulomb interaction term. In the following part of this section, we will work in momentum space.

\begin{figure}
\includegraphics[width=1.0\linewidth]{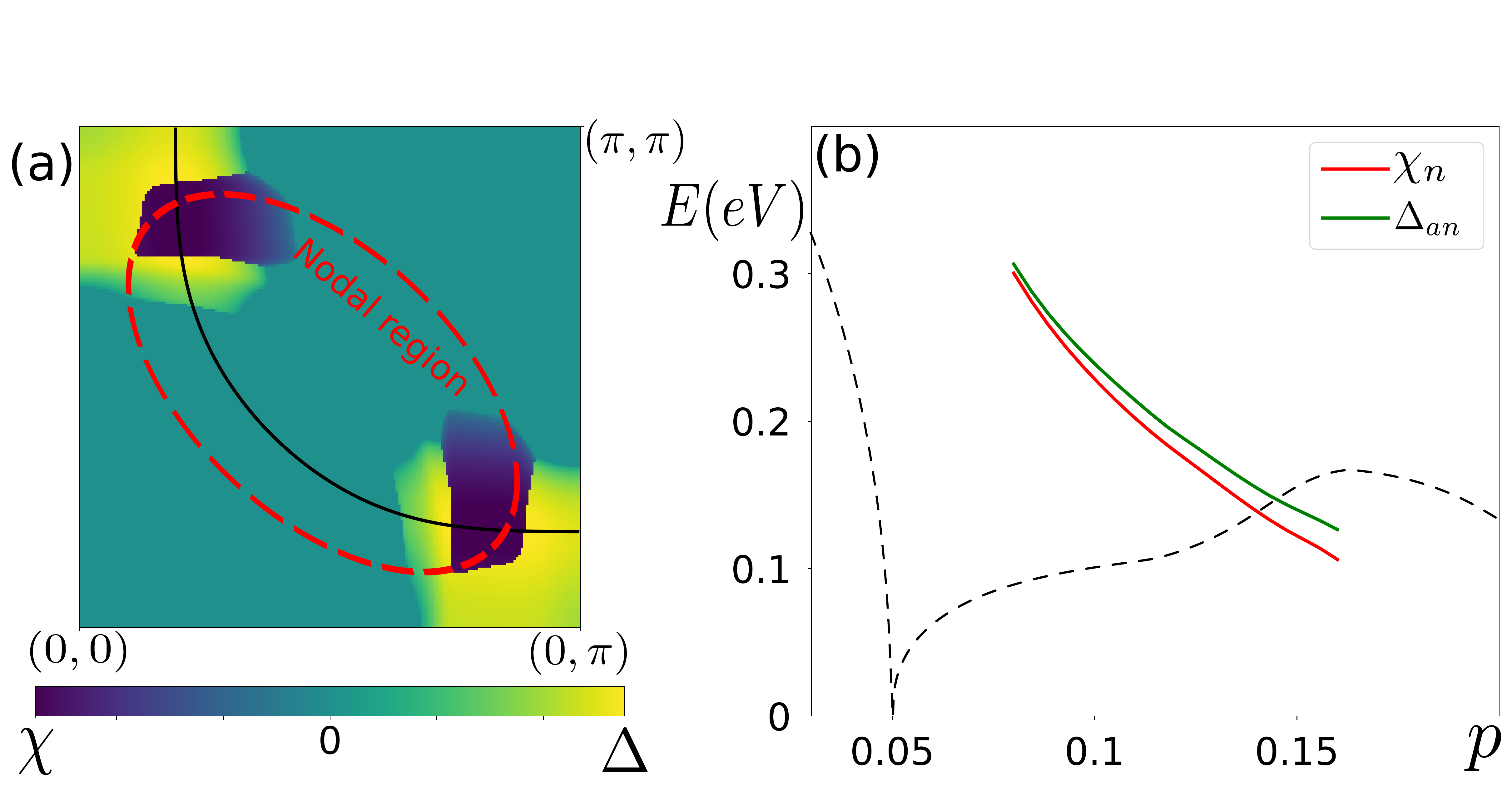} \caption{ \textbf{(a)} Gap in the particle-particle pairing channel ($\Delta$) and the particle-hole pairing channel ($\chi$) in the first quadrant of the Brillouin zone calculated using the Hamiltonian in Eq.~\eqref{eq:11b} for $p=0.12$. The particle-hole pairing is considered for an axial wave vector connecting the hot spots (also see text) in the first Brillouin zone. The black line indicates the non-interacting Fermi surface and the red dotted line indicate the nodal regions probed in $B_{2g}$ Raman response. While the particle-particle pairs gap out the AN region, the particle-hole pairs prevail the nodal region of the Fermi surface. \textbf{(b)} The doping dependence of the particle-particle gap averaged in the AN region ($\Delta_{an}$) and the particle-hole gap averaged in the nodal region ($\chi_{n}$). They both behave similarly as a function of doping in the range $0.08 >p>0.16$ with $\Delta_{an} \approx \chi_{n}$. This result fits the experimental trends \cite{Loret19} obtained in Raman spectroscopy very well. Parameters used for this plot are $J=350 \ meV$, $V=J/20$ and $\kappa_{AF}=0.1 r.l.u$. The dashed lines schematically indicate the doping region where antiferromagnetic order and the superconducting dome lies.}

\label{fig:diagram} 
\end{figure}

\subsubsection{Mean-field gap equations}\label{sec:mfgapeq}

Performing a Fourier transform and a Hubbard-Stratonovich decoupling of the interaction in Eq.~\eqref{eq:11b} in both the particle-hole and particle-particle channels, we obtain an effective fermionic action which takes the form $S_{eff}=\sum_{k,\sigma}{\Psi_{k,\sigma}^{\dagger}G^{-1}(k)\Psi_{k,\sigma}}$, where $\Psi_{k,\sigma}=(c_{k,\sigma},c_{-k,\bar{\sigma}}^{\dagger},c_{k+Q,\sigma},c_{-k+Q,\bar{\sigma}}^{\dagger})$ and 
\begin{equation}
G^{-1}(k,\omega)=\begin{pmatrix}\omega-\xi_{k} & \Delta_k & \chi_k & 0\\
\Delta^*_k & \omega+\xi_{k} & 0 & -\chi^*_k \\
\chi^*_k & 0 & \omega-\xi_{k+Q} & \Delta_{k+Q}\\
0 & -\chi_k & \Delta^*_{k+Q} & \omega+\xi_{k+Q}\\
\end{pmatrix}.\label{eq:11a}
\end{equation}
$Q$ is the modulation wave vector for the d-BDW order parameter. Motivated by experiments, in this section, we consider an axial wave-vector $Q$ relating two hot spots in the first Brillouin zone, unless otherwise stated. The issue of the leading charge instabilities in microscopic models is discussed at length in the literature. The charge order with axial wave-vector can be enhanced either by incorporating fluctuations \cite{Wang14,Meier14,Chowdhury:2014cp} or considering dynamic exchange interactions \cite{Wang15c} or an off-site Coulomb interactions \cite{Allais14c} as in our model in Eq.~\eqref{eq:11b}. Integrating the fermionic fields and minimizing the resulting action with respect to either $\Delta$ (precursor gap corresponding to the p-p pairing or d-SC order) or $\chi$ (precursor gap corresponding to the p-h pairing or d-BDW order) leads to the mean-field self-consistent gap equations. They initially form a set of coupled equations but for simplicity we will consider the decoupled equations given by:
\begin{align}
\Delta_{k,\omega} & =-\frac{1}{\beta}\sum_{q,\omega^{\prime}}\frac{J_{-}\left(q,\omega^{\prime}\right) \Delta_{k+q}}{\left(\omega+\omega^{\prime} \right)^2 - \xi_{k+q}^2 - \Delta_{k+q}^2},\label{gap_eqSC} \\
\chi_{k,\omega} & =-\frac{1}{\beta}\sum_{q,\omega^{\prime}}\frac{J_{+}\left(q,\omega^{\prime}\right) \chi_{k+q}}{\left(\omega+\omega^{\prime} - \xi_{k+q}\right)\left(\omega+\omega^{\prime} - \xi_{k+Q+q}\right)- \chi_{k+q}^2},\label{gap_eqCO}
\end{align}
with $J_{\pm} \left(q,\omega^{\prime}\right)$ being related to the original model parameter as $J_{\pm} \left(q,\omega^{\prime}\right) \sim 3J(p) \pm V$ and $\beta$ is the inverse temperature.

It is however also possible to write the action as a function of the field corresponding to the PG phase, $\Omega_k$, which is defined by the relation in Eq.~\eqref{eq:consmom}. Then we minimize the resulting action with respect to $\Omega_k$ giving the self-consistent gap equation,

{\footnotesize{\begin{align}
\Omega_k & =-\frac{1}{\beta} \sum_{q,\omega} \frac{J^* \ \left( \omega +\frac{\Delta \xi_{k+q}}{2}\right) \Omega_{k+q}}{\left(\omega^2 - \xi_{k+q}^2\right) \left(\omega-\xi_{k+q+Q}\right) -\left(\omega + \frac{\Delta \xi_{k+q}}{2} \right)\Omega_{k+q}^2},\label{gap_eqPG}
\end{align}}}
where $J^*=\frac{2 J_{+} J_{-}}{J_{+} + J_{-}}$ and $\Delta \xi_{k+q}=\xi_{k+q}-\xi_{k+q+Q}$. Minimizing with respect to $\Omega_k$ is equivalent to condensing the field $n_s^{+}$ defined in Eq.~\eqref{eq:nspdef}. While expressing the action in terms of the field $\Omega_k$, we consider that there is no condensation of $n_s^{-}$ (defined in Eq.~\eqref{eq:nsmdef}) and ignore its contribution.

In order to obtain an estimate of the energy scale associated with $\Omega_k$, we first solve the gap equations Eqs.~\eqref{gap_eqSC}-\eqref{gap_eqPG} by taking $\Delta$, $\chi$, $\Omega_k$ and $J_{\pm}$ to be momentum and frequency independent. This leads to only one energy scale corresponding to all the three gaps $\Omega_k$, $\Delta$ and $\chi$ with $J_+ \approx J_- \approx J^*$. This can also be understood if we additionally consider $\xi_{k+Q}\approx -\xi_{k}$ which gives three identical gap equations. The approximate equality $\xi_{k+Q}\approx -\xi_{k}$ is valid in the AN region for an axial $Q$ vector connecting two hot spots in the first Brillouin zone. Hence this alternative way of decoupling does not introduce a new energy scale. 

If we further ask the question: why would the system want to condense the field $n_s^{+}$ just below $T^*$ and entangle the p-p and p-h pairs instead of condensing $\Delta$ or $\chi$ separately? The answer lies in the energetics. We calculate the condensation energies of all the three possible processes, $E_{sc}$ (for condensation only in p-p pairs), $E_{co}$ (for condensation only in p-h pairs) and $E_{PG}$ (for condensation in the entangled state), given by
\begin{align}
E_{sc}=-\frac{1}{2J_{-}}\rho_0 \Delta_{k=k_F}^2, \nonumber\\
E_{co}=-\frac{1}{2J_{+}}\rho_0 \chi_{k=k_F}^2, \nonumber\\
E_{PG}=-\frac{1}{2J^*}\rho_0 \Omega_{k=k_F}^2,
\label{eq:cond_en}
\end{align}
where $\rho_0$ is the density of states at the Fermi level and $k_{F}$ is the Fermi momentum. $\Delta_{k=k_F}$, $\chi_{k=k_F}$ and $\Omega_{k=k_F}$ are the average gaps on the Fermi level obtained from the solutions of Eqs.~\eqref{gap_eqSC}-\eqref{gap_eqPG}. We find that $E_{PG}<E_{sc}\approx E_{co}$, indicating that the system maximizes the gap by choosing the entangled solution in order to gain in energy. This gives a simple argument behind the choice of our variational ansatz of entangled p-p and p-h pairs: the system choses to fractionalize the PDW pair into a p-p and p-h pairs to maximize the gap at the Fermi surface.

The real space constraint is realized by fragmenting the Fermi surface allowing the possibility of $\Delta_k$ and $\chi_k$ to exist at different places in momentum space. To get an insight into the fragmentation of the two precursor gaps in momentum space, we solve Eqs.~\eqref{gap_eqSC}-\eqref{gap_eqCO} by making a series of approximations while keeping the momentum dependence of the gaps, the assumptions are summarized here and detailed calculations are deferred to Appendix \eqref{sec:Appendix-B:-Details}. The integration over Matsubara frequency is performed analytically considering the couplings $J_{\pm}$ to be frequency independent. Then the momentum integration is performed by restricting the momentum exchange to be close to $Q_{AF}=(\pi,\pi)$ (AF wave vector) with a broadening given by $\kappa_{\text{{\scriptsize{AF}}}}$ which replicate the short-range nature of the antiferromagnetic fluctuations. In fact, this broadening can be directly related to the coherence length ($\xi_{\text{{\scriptsize{AF}}}}$) of the antiferromagnetic fluctuations with $\kappa_{\text{{\scriptsize{AF}}}} \sim \left(\xi_{\text{{\scriptsize{AF}}}}\right)^{-1}$. The restriction in the momentum integration helps us in obtaining analytical expressions for the solution of the gaps. We should note that the $\chi$ is the precursor gap and thus only represents the uniform component of the amplitude of the d-BDW order.

\subsubsection{Results}\label{sec:mfgapeqres}

We find the solutions of the gap equations for each k-point in a quadrant of the Brillouin zone independently. Owing to the competition between the two orders we only keep the solution which gives the bigger gap of the two at each k-point. We take the band parameters of Hg-1201 \cite{Loret19,Das12} with, choosing the nearest-neighbor hopping $t$ as the energy scale, $t'/t=-0.2283$, $t''/t=0.1739$, $t'''/t=-0.0435$, fix the chemical potential in order to obtain a desired doping and take $\beta=50$. The extent of the short-ranged nature of the antiferromagnetic interaction is estimated from the neutron scattering experiment and give $\kappa_{AF}\sim 0.1\frac{2\pi}{a}$. One typical result obtained for $J=350 \ meV (=0.85t)$, $V=J/20$ and $p=0.12$ is shown in Fig.~\ref{fig:diagram}(a). The p-h pairs preferentially gap the Fermi surface close to the nodal region and the p-p pairs dominate in the AN region and is in good agreement with the fact that the precursor in the charge channel has been observed in the $B_{2g}$ probing preferentially the nodal region of the Brillouin zone shown schematically by the red dotted line. Since p-h and p-p pairs prevail at different regions of the Fermi surface, it further justifies the consideration of two gap equations Eqs.~\eqref{gap_eqSC} and \eqref{gap_eqCO} as decoupled. The quasi-particle dispersion (or the excitation spectrum) can be written in a form analogous to the conventional BCS result with reconstructed bands due to the presence of a modulating order,
\begin{align}
E^2_{\pm}\left(k\right) &= \frac{1}{2} \left(\xi_k+\xi_{k+Q}\right)\pm \sqrt{\left(\xi_k-\xi_{k+Q}\right)^2+4\left|\chi_k\right|^2} + \left|\Delta_k\right|^2 \nonumber \\
&=\epsilon_{\pm}^2+\left|\Delta_k\right|^2,\label{eq:recons}
\end{align}
where $\epsilon_{\pm}$ give the form of the reconstructed bands. Eq.~\eqref{eq:recons} is a usual form of the quasi-particle dispersion in a coexisting state obtained by diagonalizing the Hamiltonian. We also show the d-BDW gap with the diagonal wave-vector in Fig.~\ref{fig:gaps}(g).

We perform the same calculation for a continuous evolution of the doping between $p=0.08$ and $p=0.16$, with axial wave-vector $Q$ which changes with doping. Note that the d-CDW wave-vector found in experiments are not exactly equal (though very close) to the value obtained this way. But the doping dependence of the experimentally observed value is similar to the one used here. In order to compare the results to recent Raman spectroscopy experiment, in Fig.~\ref{fig:diagram}(b), we look at the d-SC gap averaged in the AN region ($\Delta_{an}$) and compare it with the d-BDW gap averaged in the nodal region ($\chi_{n}$). Note that the d-BDW gap need not be in the immediate proximity of the nodal line ($k_x=k_y$) to be visible in Raman experiment as the region probed in the $B_{2g}$ symmetry also extend away from the nodal line with non-zero weights at the hot-spots as shown schematically by the red dotted line in Fig.~\ref{fig:diagram}(a). The evolution of the precursor gaps as a function of doping is depicted by green and red lines in Fig.\ref{fig:diagram}(b). We see that, for a fixed temperature, both the gaps have similar magnitudes and decrease linearly in a range of doping $p=0.08$ to $p=0.16$ similar to doping dependence of the pseudo-gap temperature $T^*$. The dome-shape of $T_{co}$ and $T_c^{\prime}$ is expected to be recovered by taking into account the effect of phase fluctuations. In a standard preformed pair scenario this effect has been included on phenomenological grounds by introducing a damping term in the electronic Green's function and a finite lifetime for the preformed pairs above $T_c$ leading to a good description of the ARPES spectra for all temperatures below $T^*$ \cite{Norman:1998va}. These results are very close to the behavior observed in Raman scattering experiment where comparison between the response in the $B_{1g}$ symmetry (which probes the AN region) and in the $B_{2g}$ symmetry (which preferentially probes the nodal region) lead to a similar conclusion of the p-h and the p-p gaps being quasi-degenerate and having the same doping dependence\cite{Loret19} as that of $T^*$. 

\begin{figure}[t]
\includegraphics[width=1.0\columnwidth]{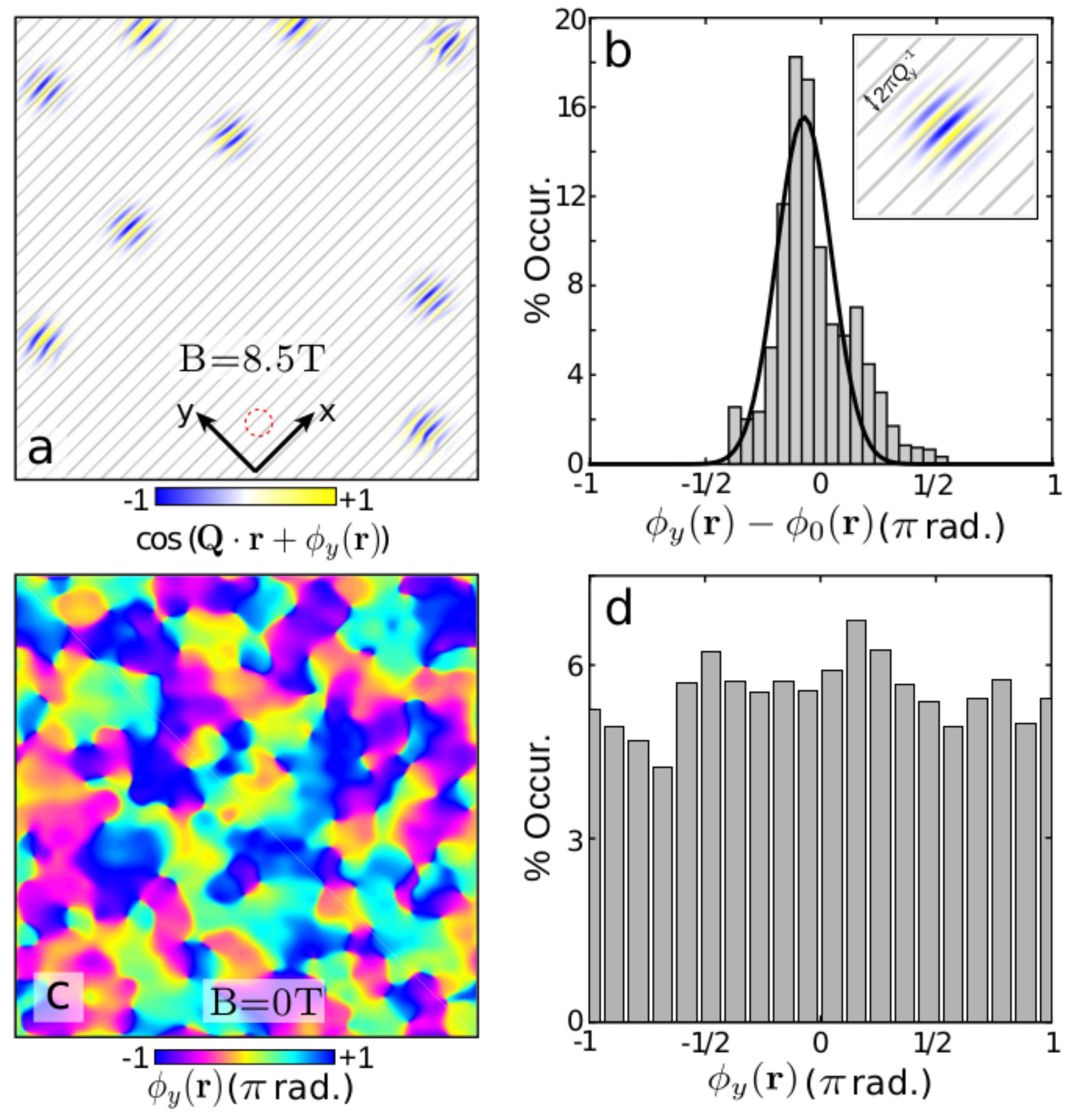} \caption{\label{fig:Global-spatial-Charge}Experimental STM data \cite{Hamidian15} showing global spatial d-CDW phase coherence inside vortex cores. \textbf{(a)} The amplitude and phase of the CDM with d-symmetry form factor with predominant wave vector $Q_{y}$ inside stable vortex cores. The vortex core has radius of 2 nm (shown by red dotted circle). The modulation is represented by blue/yellow colors which are measured with respect to an arbitrary reference phase (modulation wavelength $\lambda_{y}=2\pi/Q_{y}$), shown by the grey color. It can be seen that the measured phase inside the vortex core is relatively constant w.r.t. the reference phase. \textbf{(b)} Histogram plot of the relative phase between density wave state inside the vortex core and the reference phase. The plot shows that the relative variation of phase is mostly centralized to a single value inside the vortex core with a standard deviation of $12\%(2\pi)$. For this plot, 6-9 vortices were used because of the stability issue of vortex \cite{Hamidian15}. \textbf{(c)} The map of the spatial phase of the CDM modulating along the y-direction is shown for $B=0T$ \cite{Hamidian15a}. \textbf{(d)} Histogram plot of the spatial phase data in (c). It shows that the disorder makes the usual phase distribution of the CDM purely random with equal values in every bin. Comparison of the histograms in (b) and (d) highlights the remarkable global phase coherence of the CDM inside the vortex cores. Note that while plotting the histogram in (b), the disordered $B=0T$ part is subtracted.}
\end{figure}

\subsection{Charge modulation phase coherence on a macroscopic scale from Scanning Tunnelling Microscopy (STM)}\label{sec:STM}

The Higgs mechanism described in this paper predicts two distinct signatures in STM measurements under applied magnetic field (B). First, the Higgs mechanism at $T^*$ results into a constraint ($\left|\chi_{ij}\right|^{2}+\left|\Delta_{ij}\right|^{2}={E^*}^2$) between the d-BDW field $\chi_{ij}$ and the d-SC field $\Delta_{ij}$. In the presence of magnetic field, the amplitude of the d-SC field gets suppressed inside the halo region surrounding vortex cores. So, due to the constraint, the amplitude of the d-BDW order parameter is enhanced inside the vortex halos. Evidence for enhancement of the d-CDW (real part of the d-BDW) inside the vortex cores is already evident in STM results \cite{Hoffman02,Matsuba07,Yoshizawa13,Machida16}. This feature is also expected in theories with emergent SU(2) symmetries \cite{Einenkel14} or competing orders \cite{Kivelson:2002er}. But there is a second feature that is unique to the formalism of this paper and is expected to be captured in STM measurements. The special Higgs mechanism freezes the global phase of the spinor comprising of $\Delta_{ij}$ and $\chi_{ij}$ at $T^*$. Subsequently, the relative phase of the spinor gets frozen at a lower temperature $T_c$. Since both the global and the relative phases of the spinor gets frozen below $T_c$, $\chi_{ij}$ also acquires a spatial phase coherence along with $\Delta_{ij}$. So, we look at the real part of the d-BDW order given by
\begin{align}
Re\left(\chi_{ij}\right)=\hat{d}\left|\chi_{ij}\right|\cos\left(Q\cdot r+\phi(r)\right),
\label{eq:CDWdef}
\end{align}  
where $\phi(r)$ is the phase of $\chi_{ij}$. We remind the reader that $r=\left(r_{i}+r_{j}\right)/2$ denotes the midpoint of the bonds and $\phi(r)\equiv \theta_{\chi}(r)$. In Fig.~\ref{fig:Global-spatial-Charge}, we plot the spatial profile and the histogram of the phase $\phi(r)$ obtained in the spatial-phase resolved STM measurement of BSCCO. The details on the determination of $\phi(r)$ is given in Appendix \ref{sec:STMphase}. In the d-SC phase ($T<T_c$), the d-CDW has been observed inside the vortex halos. More recently, STM visualization \cite{Edkins18} of density-of-states modulations within the halo surrounding Bi$_{2}$Sr$_{2}$CaCu$_{2}$O$_{8}$ vortex cores reveals a complex energy dependence, with Bogoliubov quasiparticles at lower energies and two sets of particle-hole symmetric modulations occurring at energies near the gap edge. Focusing on the gap-edge modulations with $Q=(0,0.25)2\pi/a_0$ ($a_0$ is the lattice constant), they appear to exhibit spatial phase coherence between halo regions surrounding different vortices, as shown in Fig.~\ref{fig:Global-spatial-Charge}(a), with a global phase coherence length much larger than the typical size ($\sim$ 5 lattice spacings) of the vortex halo \cite{Hamidian15}. The phase coherence becomes apparent looking at the histogram of the phase $\phi(r)$ (with respect to a reference $\phi_0(r)$) in Fig.~\ref{fig:Global-spatial-Charge}(b). In contrast, the usual phase distribution of the CDM at $B=0$ is completely random due to the presence of disorder, as shown in Fig.~\ref{fig:Global-spatial-Charge}(c) and (d). This is a very unusual situation for charge ordering, which finds a natural explanation within our scenario. The idea of global phase coherence is usually discussed in the context of granular superconductors where local superconducting puddles are formed \cite{Ma85,Ghosal01,Seibold12,Dubi07}. These superconducting puddles, typically of the size of the coherence length, attain global phase coherence below a characteristic temperature \cite{Yu19}. In Fig.~\ref{fig:Global-spatial-Charge}(a), the scenario is different as the puddles are formed of the charge order and still attain phase coherence over large distances. Note that the map shown in Fig.~\ref{fig:Global-spatial-Charge}(a) is extracted by subtracting the zero field data from the data for $B=8.5$T. As a result, the d-CDW puddles in Fig.~\ref{fig:Global-spatial-Charge}(a) are only visible near the vortex cores.

\subsection{Multiple orders in the PG phase : pair density waves and loop current state}\label{sec:orders}

A remarkable outcome of the special Higgs mechanism at $T^*$ is that it can induce the formation of multiple orders in the PG phase. 

\subsubsection{Pair density waves in the vortex halos}\label{sec:PDW}

STM observation \cite{Hamidian16,Edkins18} of PDW order \cite{Agterberg19}, with both $Q$ and $Q/2$ modulations, in the halo surrounding vortex cores of BSCCO has inspired many theoretical works \cite{Wang18,Dai18,Norman18}. While some theoretical works consider the $Q/2$ PDW order as the `mother state' which drives the pseudo-gap phenomenology \cite{Dai18}, others treat the $Q/2$ PDW order as a competitor of the d-SC order \cite{Wang18}. 

Within our theory, the PDW order emerges as a composite field of the p-p and p-h pairs. From the alternative viewpoint, the PDW is a fundamental object in the theory and gets fractionalized to p-p and p-h pairs. In both the perspectives, we have the following features of the PDW order: (i) It can be observed as `short-range' PDW at low temperatures. (ii) It can be observed only in the vortex halo due to the pinning of the d-BDW order. (iii) It can have both extended s-wave and d-wave symmetric components. (iv) The modulation wave vector of the PDW order will be $Q$, which is the same as that of the d-BDW order. Within our current formulation, the $Q/2$ PDW order is not a natural outcome. However, we can in principle accommodate the $Q/2$ PDW order as one of the primary states along with d-SC and d-BDW orders by constructing a quintuplet (see Sec. \ref{sec:generalnature}). In order to analyze this possibility, we solve the mean-field gap equation (Eq.~\eqref{eq:PDWgap}) for the $Q/2$ PDW order parameter. We find that, in the nodal region, the value of the $Q/2$ PDW gap is nearly half the value of the d-BDW gap with $Q$ modulations (also see Appendix \ref{sec:PDWgap}). Hence, d-BDW with $Q$ is favored energetically as a primary state and we treat the $Q/2$ PDW as a competing order (not as a primary state) which appears only in the vortex halos.

In Sec.~\ref{sec:fracpdw}, we already indicated that the PG phase can be viewed as consisting of preformed PDW pairs which get fractionalized into p-p and p-h pairs. The PDW field locally reconfines at $T_{co}$. For $T<T_{co}$, two possible PDW fields can appear as bilinear in $\Delta_{ij}$ and $\chi_{ij}$. The one involving the relative phase is defined in Eq.~\eqref{eq:PDWdef1}. The other combination is given as
\begin{equation}
\tilde{\Delta}_{\text{PDW}}(i)=\sum_{j \in \text {n.n. of }i}\chi_{ij}\Delta_{ij},
\label{eq:pdw}
\end{equation}
where the sum is over the nearest neighbors (n.n.) of $i$. $\tilde{\Delta}_{\text{PDW}}$ carries the global phase ($\theta_{\Delta}+\theta_{\chi}$) which is the sum of the phases of $\chi_{ij}$ and $\Delta_{ij}$. Rewriting the Eq.~\eqref{eq:pdw} in terms of the amplitude and the phase, we get
\begin{align}
\tilde{\Delta}_{\text{PDW}}(i)&=\sum_{j \in \text {n.n. of }i}\left|\chi_{ij}^{Q}\right| \left|\Delta_{ij}\right|e^{i(\theta_{\Delta}+\theta_{\chi})}e^{iQ.r}\nonumber \\
&=\left|\tilde{\Delta}_{\text{PDW}}^{Q}(i)\right|e^{i(\theta_{\Delta}+\theta_{\chi})}e^{iQ.r},
\label{eq:pdwre}
\end{align}
where it should be noted that the amplitude ($\tilde{\Delta}_{\text{PDW}}^{Q}$) of the modulating PDW field also depends the value of $Q$. For $T<T^*$, $\theta_{\Delta}+\theta_{\chi}$ is frozen. But as the amplitudes $\left|\chi_{ij}^{Q}\right|$ and $\left|\Delta_{ij}\right|$ still fluctuate, they will obscure the modulations of the mean-field $\left\langle \tilde{\Delta}_{\text{PDW}}(i) \right\rangle=\left\langle \sum_{j}\left|\chi_{ij}^{Q}\right|\left|\Delta_{ij}\right|e^{iQ.r}\right\rangle$. For temperatures below $T^{\prime}_c$, the PDW order parameter will be long-ranged (in a clean system) when both the d-SC and the d-BDW fields acquire uniform mean values, though this transition is a crossover. In contrast, the other form of PDW ($\Delta_{\text{PDW}}$ as defined in Eq.~\eqref{eq:PDWdef1}) involving the relative phase acquire long-range coherence at a lower temperature $T_c$. From Eq.~\eqref{eq:pdwre}, it is evident that the PDW order parameter occurs with the same wave vector ($Q$) \cite{Hamidian16} as the d-BDW order parameter, $\chi_{ij}^{Q}$. The local amplitude of the PDW order parameter will be the maximum in regions where there is a maximum non-zero overlap of the amplitudes $\left|\Delta_{ij}\right|$ and $\left|\chi_{ij}^{Q}\right|$ on the same bond $\langle ij \rangle$ (this will be the case in the halo \cite{Edkins18,Hamidian16} region of the vortex in the presence of magnetic field). The momentum structure of the order parameter $\tilde{\Delta}_{\text{PDW}}$ will depend on the choice of the wave vector of $\chi_{ij}$. An axial modulation wave vector will give a momentum space structure of $\chi_{ij}$ with both s-wave and d-wave components. As a result, the PDW order parameter will consist of both extended s-wave and d-wave components.

\subsubsection{Loop current state}\label{sec:loopc}

Apart from the finite $Q$ orders at low temperatures, the PG also sustains $Q=0$ orders. We discuss one such $Q=0$ order, magneto-electric loop currents, which break discrete symmetries like parity and time reversal. Within our framework, the loop currents appear as an `auxiliary' or a `preemptive' order \cite{Sarkar19}.

Though the PDW order can be observed only below $T=T^{\prime}_c$, the fluctuations of the PDW field (Eq.~\eqref{eq:pdw}) in the temperature regime $T>T^{\prime}_c$ can give rise to auxiliary order parameters. With the motivation to generate a $Q=0$ (translationally invariant) emergent loop current order in the PG phase, we construct a secondary order parameter following Ref.~[\onlinecite{Agterberg:2014wf}], 
\begin{equation}
l=\left| \tilde{\Delta}_{\text{PDW}}^{Q} \right|^2-\left| \tilde{\Delta}_{\text{PDW}}^{-Q} \right|^2,
\label{eq:loop}
\end{equation}
where $\tilde{\Delta}_{\text{PDW}}^{Q}$ is the amplitude of the PDW field and its value depends on the choice of the modulation wave vector $Q$. The PDW field transforms under translation $T$, time-reversal $TR$ and parity $P$ as
\begin{align}
T(\tilde{\Delta}_{\text{PDW}}^{Q})=e^{iT.Q} \tilde{\Delta}_{\text{PDW}}^{Q};& TR(\tilde{\Delta}_{\text{PDW}}^{Q})=\left(\tilde{\Delta}_{\text{PDW}}^{-Q}\right)^{*};\nonumber\\ 
P(\tilde{\Delta}_{\text{PDW}}^{Q})&=\tilde{\Delta}_{\text{PDW}}^{-Q}.  
\label{eq:pdwsym}
\end{align} 
As $l$ is composed of terms depending on $\left(\tilde{\Delta}_{\text{PDW}}^{Q}\right)^{*}\tilde{\Delta}_{\text{PDW}}^{Q}$ and $\left(\tilde{\Delta}_{\text{PDW}}^{-Q}\right)^{*}\tilde{\Delta}_{\text{PDW}}^{-Q}$, it is a translationally invariant order parameter (under translation $T(l)=l$). The loop current order parameter $l$ also satisfies,
\begin{equation}
TR(l)=-l; P(l)=-l ~~\mbox{and}~~ TRP(l)=l.
\label{eq:tp}
\end{equation}
Thus, the loop current order parameter defined in Eq.~\eqref{eq:loop} satisfies the same symmetries as the magneto-electric loop current state proposed by Varma \cite{Varma97}, which is often used to interpret the intra unit cell magnetic order seen in polarized elastic neutron scattering measurements \cite{Fauque06}. It is important to highlight that the discrete $Z_2$ symmetries like parity or time reversal is spontaneously broken by the secondary order parameter $l$, which is composed of PDW fluctuations. So, a non-zero average value of $\langle l \rangle$ does not mean $\langle \tilde{\Delta}_{\text{PDW}}^{Q} \rangle \neq \langle \tilde{\Delta}_{\text{PDW}}^{-Q} \rangle$ (i.e. the PDW ground state does not break parity or time reversal) \cite{Agterberg:2014wf}. Possibilities of preemptive discrete $Z_2$ symmetry breaking outside the Landau paradigm \cite{Lee18} occurring due to secondary order parameters is already discussed in Refs.~[\onlinecite{Wang14}] and [\onlinecite{Agterberg:2014wf}]. Interestingly, the preemptive transition occurs at a higher temperature \cite{Wang14} than the primary order transition temperature (in our case $T^{\prime}_c$ for the PDW order), thus justifying the presence of loop current state in the $T>T^{\prime}_c$. In this paper, we only justify that the loop current state can be visible for temperatures $T>T^{\prime}_c$ and do not explicitly show that the upper temperature limit is $T^{*}$. There are also other phenomenological proposals \cite{Lee14,Scheurer18} and proposals based microscopic three orbital models \cite{Bulut:2015jt,Carvalho16} for the existence of loop current order in the PG phase. We also note that the magnetic moments derived from microscopic mean field models are usually far smaller compared to what it is found in experiments \cite{Dai18}. The preemptive transition can also give way to nematicity \cite{Sato2017,Murayama18} and the breaking of the inversion symmetry recently observed in the study of the optical second harmonic generation \cite{Zhao2016}. 

\section{Conclusions and discussions}\label{sec:conclu}

\subsection{Summary of the work}

Through out this paper, we have shed light into the `Frankenstein' nature of the PG phase of underdoped cuprates. We proposed that the PG phase is an entangled state of p-p and p-h preformed pairs. In the following, we summarize the exclusive features of this proposal:
\begin{enumerate}
\item{A special Higgs mechanism entangles the two preformed pairs at $T^*$ by freezing the global phase and a gap opens in the fermionic excitation spectrum. This entanglement results into a strong competition between p-p and p-h pairs. The relative phase and the two amplitudes of the fields corresponding to the two pairs fluctuate. The amplitude fluctuations are related by a constraint, $\left|z_1\right|^2+\left|z_2\right|^2=(E^*)^2$. This is followed by a unique sequence of events occurring as the temperature is reduced. The amplitudes of the p-h and p-p pairs get condensed at lower temperatures $T_{co}$ and $T^{\prime}_c$ respectively. A second Higgs mechanism occurs at $T=T_c$ and both the superconducting and bond-excitonic orders acquire phase coherence leading to a `super-solid' like phase. Thus, we have different temperature lines in the rich phase diagram of cuprates.}
\item{Equivalently, the pseudo-gap phase can be understood as a ``fractionalized" PDW order. Indeed at $T<T_{c}$ the system orders into ``short range'' PDW state. In the PG phase the PDW order deconfines to release two elementary components, p-p and p-h preformed pairs, which stay entangled through the constraint $\Delta_{ij}^{2}+\chi_{ij}^{2}=E^{*2}$. The corresponding variational ansatz for this entangled state is $\ket{PG}=\ket{d\text{-}SC}+\ket{d\text{-}BDW}$, which corresponds to a coherent superposition of ``dead cat'' and ``alive cat'' in the Schrödinger's thought experiment.}
\item{For the first time, this theory relates the PG phase to both p-p and p-h instabilities, without being restricted to particular parameter regimes. Using a simplified microscopic model, we obtain the doping dependence of mean-field precursor gaps arising out of both these instabilities and the gap repartition in the Brillouin zone. These results show close resemblance to the Raman \cite{Loret19} spectroscopy findings.}
\item{The two stage Higgs mechanism has distinct experimental consequences at low temperatures. For temperatures $T<T_c$, both d-SC and d-BDW show phase coherence. A distinguishing feature occurs with the application of magnetic field. In the presence of small magnetic field, superconducting vortices appear with suppressed superconducting order parameter inside the halo region surrounding vortex cores. The competing d-BDW order is enhanced inside the vortex halos. Remarkably, STM measurements see a locking of the phase slips of the charge modulations in a much larger region of space than the typical size of a vortex halo. Our theory can explain this unusual feature seen in STM measurements.}
\item{Other unique nature of two states d-SC and d-BDW forming a doublet is the emergence of multiple orders like PDW or loop currents, which are higher order combinations of the primary state. From the perspective of `fractionaized' PDW, the PDW reconfines at low temperatures. Thus this formalism not only accommodates finite $Q$ orders like d-CDW and PDW at lower temperatures, it also gives possibilities of $Q=0$ orders at higher temperatures.}
\end{enumerate}

\subsection{Generic nature of the model} \label{sec:generalnature}

In this section, we outline the generic nature of the model proposed in this paper and discuss the possibilities of accommodating multiple orders as primary states.

The spinor in this work consists of the d-SC field and the d-BDW field with $Q$ modulations. A PDW operator rotates one constituent of the spinor to another. This structure of the spinor is chosen with motivations from experiments in underdoped cuprates. Some of them include: (i) Ubiquity of CDM with $Q$ modulations. (ii) Competition between the d-CDW order and the d-SC order. (iii) Signatures of near degeneracy of these two orders in the underdoped regime of the phase diagram. Even with this form of the spinor, the first generic aspect is the choice of the $Q$ vector for the d-BDW field. To add to this, we can also consider d-BDW fields with multiple wave vectors. As an example, the case for two wave vectors is already shown in the Appendix \ref{sec:su3cp2}.

One of the challenges in obtaining a generic model for cuprates is the presence of plethora of non-superconducting orders \cite{Fradkin:2015ch,Keimer:2015vy,Chakravarty01}. Theoretically, these multiple orders are often treated as competing or intertwined with superconductivity \cite{Zaanen89,Zhang97,Fradkin:2015ch,Keimer:2015vy,Chakravarty01,Berg09,Moon09,Zhang:2002hz,Benfatto00,Caprara:2016vh}. The skeleton of the theory presented in this paper leaves room for multiple components in the spinor to accommodate many primary states. As an example, here we show how to incorporate the recently observed $Q/2$ PDW order \cite{Hamidian16,Edkins18} as one of the primary states. If we consider d-BDW with two wave vectors $\mathbf{Q}_x$ and $\mathbf{Q}_y$ and d-wave PDW with two wave vectors $\mathbf{Q}_x/2$ and $\mathbf{Q}_y/2$, the spinor is given as a quintuplet,
\begin{equation}
\psi=\left(\begin{array}{c}
z_{1}\\
z_{2}\\
z_{3}\\
z_{4}\\
z_{5}
\end{array}\right)=\left(\begin{array}{c}
\chi_{Q_{x}}\\
\chi_{Q_{y}}\\
\Delta\\
\Delta_{Q_{x}/2}\\
\Delta_{Q_{y}/2}
\end{array}\right),\label{eq:quintuplet}
\end{equation}
where (for $\left\langle i,j\right\rangle $ site indices on a bond and $\sigma$ the spin index) $\chi_{Q_{x}}=\hat{d}\sum_{\sigma}c_{i\sigma}^{\dagger}c_{j\sigma}e^{i\theta_{1}}$, $\chi_{Q_{y}}=\hat{d}\sum_{\sigma}c_{i\sigma}^{\dagger}c_{j\sigma}e^{i\theta_{2}}$, $\Delta=\hat{d}\sum_{\sigma}\sigma c_{j-\sigma} c_{i\sigma}e^{i\theta_{3}}$, $\Delta_{Q_{x}/2}=\hat{d}\sum_{\sigma}\sigma c_{j-\sigma} c_{i\sigma}e^{i\theta_{4}}$ and $\Delta_{Q_{y}/2}=\hat{d}\sum_{\sigma}\sigma c_{j-\sigma} c_{i\sigma}e^{i\theta_{5}}$ with $\theta_{1}=\mathbf{Q_x}\cdot\mathbf{r}+\tilde{\theta}_{1}$, $\theta_{2}=\mathbf{Q_y}\cdot\mathbf{r}+\tilde{\theta}_{2}$, $\theta_{4}=\mathbf{Q_x}\cdot\mathbf{r}/2+\tilde{\theta}_{4}$, $\theta_{5}=\mathbf{Q_y}\cdot\mathbf{r}/2+\tilde{\theta}_{5}$ and $\mathbf{r}=\left(\mathbf{r}_i+\mathbf{r}_j\right)/2$. The global phase of the quintuplet is frozen at $T^*$ which will give the constraint $\sum_{a=1}^5 z_a^{*}z_a=(E^*)^2$. The fluctuations in the PG phase will be governed by an SU(5) chiral model or equivalently a $CP^4$ model. The corresponding collective modes will be $\eta$ modes with charge 2 and spin 0; and density modes (similar to $\eta_z$ in Sec.~\ref{sec:flucchiral}) with charge 0 and spin 0. Note that in this case, we can have PDW $\eta$ modes with different wave vectors.          

To illustrate the power of the concept, let us try to infer what happens when oxygen doping is lowered, below $p=0.06$. We are then in a regime closer to the Mott insulator, hence it is legitimate to guess that the superconducting modes will be absent whereas the AF and charge modes can be strengthened. We can construct the SU(2) spinor made of incommensurate AF and charge fluctuations 
\begin{align}
\psi= & \left(\begin{array}{c}
z_{1}\\
z_{2}
\end{array}\right)=\left(\begin{array}{c}
\varphi_{Q_{1}}^{AF}\\
\chi_{Q_{2}}
\end{array}\right),\label{eq:stripy}
\end{align}
where (for $\left\langle i,j\right\rangle $ site indices on a bond and $\sigma$ the spin index) $\varphi_{Q_{1}}^{AF}=\sum_{\sigma}\left(c_{i\sigma}^{\dagger}c_{i-\sigma}-c_{j\sigma}^{\dagger}c_{j-\sigma}\right)e^{i\theta_{1}}$, $\chi_{Q_{2}}=\hat{d}\sum_{\sigma}c_{i\sigma}^{\dagger}c_{j\sigma}e^{i\theta_{2}}$
and with $\theta_{1}=\mathbf{Q}_{1}\cdot\mathbf{r}+\tilde{\theta}_{1}$
, $\mathbf{Q}_{1}\simeq\left(\pi,\pi\right)+\delta$ and $\theta_{2}=\mathbf{Q}_{2}\cdot\mathbf{r}+\tilde{\theta}_{2}$,
$\mathbf{r}=\left(\mathbf{r}_i+\mathbf{r}_j\right)/2$. A gap can then open at $T^{*}$ due to the
constraint $\sum_{a=1}^{2}z_{a}^{*}z_{a}=(E^*)^2$, made of a superposition of short-range AF fluctuations ($z_{1}$)
and short patches of charge modulations ($z_{2}$). At lower temperatures,
the quantum superposition of those two modes will form ``stripes", a feature which is ubiquitous in La-compounds \cite{Tranquada95,Tranquada97,Hucker11,Zaanen:1998cl,Emery99}.

As a final illustration, we show the possibility of including the d-SC, AF, d-BDW with two wave vectors $\mathbf{Q}_x$ and $\mathbf{Q}_y$ and PDW with two wave vectors $\mathbf{Q}_x/2$ and $\mathbf{Q}_y/2$, all as primary states in the spinor. The associated spinor is given as
\begin{equation}
\psi=\left(\begin{array}{c}
z_{1}\\
z_{2}\\
z_{3}\\
z_{4}\\
z_{5}\\
z_{6}
\end{array}\right)=\left(\begin{array}{c}
\varphi_{Q_{1}}^{AF}\\
\chi_{Q_{x}}\\
\chi_{Q_{y}}\\
\Delta\\
\Delta_{Q_{x}/2}\\
\Delta_{Q_{y}/2}
\end{array}\right),\label{eq:SU6spinor}
\end{equation}
where $\varphi_{Q_{1}}^{AF}$ is defined in the same way as in Eq.~\eqref{eq:stripy} and the other components are defined as in Eq.~\eqref{eq:quintuplet}. The constraint in this case will be given as $\sum_{a=1}^{6}z_{a}^{*}z_{a}=(E^*)^2$. The corresponding fluctuations in the PG phase will be governed by a SU(6) chiral model or a $CP^5$ model.

\subsection{Links with previous works}

\subsubsection{Competing order scenarios}

The ubiquitous observation of charge order and the evidences of its competition with superconductivity led to several works based on the competing order scenario \cite{Kivelson:2002er,Caprara:2016gs,Caprara17,Benfatto00,Yu19}. There is substantial evidence that the competition between the d-SC and d-CDW is not of the usual Ginzburg Landau type with two independent energy scales \cite{Chakraborty18}. Experiments also indicate a near degeneracy of the two orders through: (a) Similarity of $T_{co}^{3D}$ (transition temperature of high field 3D uniaxial long-range charge order \cite{Gerber:2015gx,Chang16,LeBoeuf13}) and $T_{c}$. (b) Closeness of the pair breaking peaks in $B_{2g}$ and $B_{1g}$ Raman response \cite{Loret19}. Our theory is motivated from this near degeneracy. The entanglement between the p-p and p-h pairs results into a strong competition between them and their energy scales are constrained by the PG energy scale, which makes it different from the usual Ginzburg Landau approach. In addition, our theory is based on the presence of preformed pairs in contrast to the competing order scenarios.

\subsubsection{Fractionalization of an order parameter}

The idea to associate the $T^{*}$ line of the pseudo-gap to a Higgs phenomenon has received a huge amount of attention recently, in the special case of SU(2) gauge theories where the spin density wave order fractionalizes into Higgs fields \cite{Sachdev19} with spinons being an integral part of the theory. Our model has in common the Higgs mechanism at $T^{*}$, but in the context of a U(1)$\times$ U(1) gauge theory. Though in both approaches the electron is not fractionalized, we fractionalize a PDW field for the first time. In particular, our theory does not require spinons to be an essential ingredient for the PG phase, but rather to have a quasi-degenerate doublet of preformed pairs, which then undergo the Higgs mechanism at $T^{*}$. One might wonder about the role of magnetism in the whole picture. In the simplified microscopic model (Sec.~\ref{sec:micro}) developed in order to extract the precursor gaps, magnetism is the ``glue'' for the formation of both the precursors. Dynamic, short-range, antiferromagnetic correlations are at the core of the formation of both order parameters \cite{Montiel16}. Note that the spin fluctuations are also a key in the emergent SU(2) theories. Within this theory, the experimentally observed spin excitation spectrum \cite{Chan16,Chan:2016vk} has already been discussed for the compound Hg-1201 \cite{Montiel17}. We expect these results to remain similar within our approach.

\subsubsection{SU(2) fluctuations}

Now, we would make links with previous works based on SU(2) fluctuations. The Higgs mechanism at $T^*$ proposed in this paper is a new idea, which supports a scenario where there is no fractionalization of the electron above $6\%$ of doping, but a complex class of SU(2) fluctuations emerges. The fluctuations below $T^*$ can be described by an SU(2) chiral model. The real space chiral models have the tendency of resulting into phase separation, which was described in a previous work with the image of droplet formation \cite{Kloss15a}. The competition between the d-CDW and the d-SC order revealed by the magnetic field-temperature phase diagram can be described within an O(3) NL$\sigma$M analogous to the previous works \cite{Chakraborty18}. The PDW ladder operators $\eta$ and $\eta^{\dagger}$ of the SU(2) fluctuations can form a collective mode, which is a signature of the O(3) fluctuations below $T^{*}$ \cite{Morice18b}. The new concept introduced in this paper give some similar phenomenology as that of the emergent SU(2) symmetry picture. 
\\
\\

Historically, the PG state was either discussed as a crossover due to the formation of preformed Copper pairs or a phase transition induced by a competing p-h instability. Here the two approaches are not opposed anymore, but are amalgamated into a single model: the PG state involves a true phase transition with two kinds of entangled preformed pairs. The model is a perfect synthesis of earlier debates.

The Higgs mechanism involving a spinor is a novel theoretical idea. To the best of our knowledge, this concept is unique not only in the field of condensed matter physics, but an analog is also absent in other areas of theoretical physics. Connections between different fields of theoretical physics is not unusual. For example, the pioneer work of Anderson \cite{Anderson:1963vi} in the context of superconductivity inspired the remarkable discovery of its relativistic counterpart in the form of the ``Higgs particle" \cite{Higgs64} in particle physics. We believe our theory can also find its applications in diverse fields of physics motivating future theoretical and experimental discoveries. For instance, the spinor Higgs mechanism will likely lead to emergence of new states of matter in condensed matter physics like in graphene, Weyl semimetals, topological superconductors; or even in particle physics like in quantum chromodynamics.

\begin{acknowledgments}

We thank C. Bena, P. Bourges, A. Ferraz, H. Freire, T. Jolicoeur, M.-H. Julien, X. Montiel, M. R. Norman, I. Paul, C. Proust, A. Sacuto, S. Sarkar and S. Verret for valuable discussions and critical reading of our manuscript. This work has received financial support from the ERC, under grant agreement AdG-694651-CHAMPAGNE. J.C.S.D. acknowledges support from Science Foundation Ireland under Award SFI 17/RP/5445 and from the European Research Council (ERC) under Award DLV-788932. The authors also like to thank the IIP (Natal, Brazil), where the inspiration for this work came from and where parts of this work were done, for hospitality.

\end{acknowledgments}

\appendix

\section{Analogy of spinor Higgs mechanism to the conventional one}\label{subsec:Higgsanalogy}

We give a brief overview of the special Higgs mechanism of a spinor to describe the PG phase of underdoped cuprates in the form of table (Table.~\ref{tab:table1}). While identifying different features of the Higgs mechanism, we also give the corresponding analogy to the Higgs mechanism in a conventional superconductor.

\begin{table*}[t]
\begin{center}
 \begin{tabular}{|c | c | c|}
 \hline
  & \bf{Higgs mechanism} & \bf{Higgs mechanism at} $\mathbf{T^*}$ \\
   & \bf{for a conventional superconductor} &  \\
 \hline\hline
 Higgs field & Superconducting order parameter & Field $n_{s}^{+}=\psi^{\dagger}\psi=\left|z_1\right|^2+\left|z_2\right|^2$ \\
 \hline
 Broken Symmetry & Charge U(1) & Global U(1) phase of the spinor $\psi=\left(\begin{array}{c}{z}_{1}\\{z}_{2}\end{array}\right)$ \\
  &  & (Results in entangled preformed pairs) \\
 \hline
 Gauge Field & $A_{\mu}$ & $a_{\mu}$ \\
  & (EM vector potential) & (Defined in Eq.~\eqref{eq:1a}) \\
 \hline
 Gap in the excitation spectrum & Superconducting Gap & Pseudo-gap energy scale $E^*=\sqrt{\left|z_1\right|^2+\left|z_2\right|^2}$ \\
 (Mass of the Higgs boson) &  &  \\
 \hline
 Experimental signature & Meissner effect & No Meissner effect \\
  &  & (Signatures at lower temperatures like charge modulation phase coherence) \\
 \hline
\end{tabular}
\caption{Analogy of the special Higgs mechanism to describe the PG phase of underdoped cuprates and the Higgs mechanism of a conventional superconductor. The term ``Meissner effect" is used to identify the expulsion of the EM field. No Meissner effect in the case of the Higgs mechanism at $T^*$ means that the EM field will not be expelled. \label{tab:table1}}
\end{center}
\end{table*}

\section{\label{subsec:Pseudo-spin-Higgs-mechanism} Spinor Higgs mechanism}

\subsection{The standard Higgs mechanism}

Let us recall in this section how the standard Higgs mechanism is working.
We start with an action

\begin{align}
S_{a} & =\frac{1}{2g}\sum_{\mathbf{q}}\psi^{\dagger}\left(\mathbf{q}\theta_{q}+\mathbf{a}_{\mathbf{q}}\right)^{2}\psi+\frac{1}{4}F_{\mu\nu}F^{\mu\nu}.\label{eq:B1}
\end{align}

The goal is to integrate out the Goldstone mode $\theta_{q}$ and
for this we complete the square in $\theta$ in Eq. (\ref{eq:B1}),
which leads to (assuming condensation of the field $\psi^{\dagger}\psi=\left|\psi_{0}\right|^{2}$),
{\small{}{ 
\begin{align}
S_{a} & =\sum_{\mathbf{q}}\left[\frac{\left|\psi_{0}\right|^{2}}{2g}\left(q^{2}\left(\theta_{q}+\frac{\mathbf{q}\cdot\mathbf{a}_{\mathbf{q}}}{q}\right)^{2}+\left(a_{q}^{\perp}\right)^{2}\right)+\frac{q^{2}}{2}\left(a_{q}^{\perp}\right)^{2}\right],\label{eq:B1a}\\
\mbox{with } & \mathbf{a}_{\mathbf{q}}^{\perp}=\mathbf{a}_{\mathbf{q}}-\mathbf{q}\left(\mathbf{q}\cdot\mathbf{a_{q}}\right)/q^{2},\nonumber \\
\mbox{and } & \left(a_{q}^{\perp}\right)^{2}=\mathbf{a}_{\mathbf{q}}^{\perp}\cdot \mathbf{a}_{-\mathbf{q}}^{\perp}.\nonumber 
\end{align}
}}{\small \par}

The integration over $\theta$ is now straightforward and leads to
an effective action 
\begin{align*}
S_{a}^{eff} & =\sum_{\mathbf{q}}\left(\frac{\left|\psi_{0}\right|^{2}}{2g}+\frac{q^{2}}{2}\right)\left(a_{q}^{\perp}\right)^{2}.
\end{align*}

\subsection{\label{subsec:U(1)-U(1)-theory,}U(1) $\times$U(1) theory, the special ``Higgs mechanism''}

We now treat the first Higgs mechanism, starting with the action

\begin{align*}
S_{a,b} & =\sum_{\mathbf{q}}\frac{1}{2g}\psi^{\dagger}\left(\mathbf{q}\theta_{q}+\tau_{3}\mathbf{q}\varphi_{q}+\mathbf{a}_{\mathbf{q}}+\tau_{3}\mathbf{b}_{\mathbf{q}}\right)^{2}\psi+\frac{1}{4}F_{\mu\nu}F^{\mu\nu},
\end{align*}

where $\psi$ is the spinor defined in Eq. (\ref{eq:1}) and we assume
that $\psi^{\dagger}\psi$ condenses so that $n_{s}^{+}=\left|\psi_{0}\right|^{2}$
is a constant. We want to integrate out the phase $\theta$ and for
this we complete the square in $\theta_{q}$, leading to (dropping
the $F_{\mu\nu}F^{\mu\nu}$ terms for a while) {\small{}{ 
\begin{align}
S_{a,b} & =\frac{1}{2g}\sum_{\mathbf{q}}[\left|\psi_{0}\right|^{2}q^{2}\left(\theta_{q}+\frac{\mathbf{q}}{\left|\psi_{0}\right|^{2}q^{2}}\cdot\psi^{\dagger}\left(\mathbf{q}\tau_{3}\varphi_{q}+\tau_{3}\mathbf{b}_{\mathbf{q}}+\mathbf{a}_{\mathbf{q}}\right)\psi\right)^{2}\nonumber \\
 & -\frac{\left(\mathbf{q}\cdot\psi^{\dagger}\left(\mathbf{q}\tau_{3}\varphi_{q}+\tau_{3}\mathbf{b}_{\mathbf{q}}+\mathbf{a}_{\mathbf{q}}\right)\psi\right)^{2}}{\left|\psi_{0}\right|^{2}q^{2}}\nonumber \\
 & +\left.\psi^{\dagger}\left(\mathbf{q}\tau_{3}\varphi_{q}+\tau_{3}\mathbf{b}_{\mathbf{q}}+\mathbf{a}_{\mathbf{q}}\right)^{2}\psi\right].\label{eq:B1b}
\end{align}
}}{\small \par}

Integrating out the phase $\theta_{q}$ in Eq. \eqref{eq:B1b} leads
to the effective action 
\begin{align}
S_{\varphi,a,b}^{eff} & =\frac{1}{2g}\sum_{\mathbf{q}}[\psi^{\dagger}\left(\mathbf{q}\tau_{3}\varphi_{q}+\tau_{3}\mathbf{b}_{\mathbf{q}}+\mathbf{a}_{\mathbf{q}}\right)^{2}\psi\nonumber \\
 & -\frac{\left(\psi^{\dagger}\mathbf{q}\cdot\left(\mathbf{q}\tau_{3}\varphi_{q}+\tau_{3}\mathbf{b}_{\mathbf{q}}+\mathbf{a}_{\mathbf{q}}\right)\psi\right)^{2}}{q^{2}\left|\psi_{0}\right|^{2}}].\label{eq:B1c}
\end{align}

\subsubsection{Differentiation with respect to $\mathbf{a}_{\mathbf{q}}$}

The first mean field equation comes from the constraint
$\partial S_{\varphi,a,b}^{eff}/\partial \mathbf{a}_{\mathbf{q}}=0$ which gives 
\begin{align}
0= & \psi^{\dagger}\left(\mathbf{q}\tau_{3}\varphi_q+\tau_{3}\mathbf{b}_{\mathbf{q}}+\mathbf{a}_{\mathbf{q}}\right)\psi\label{eq:B1c.1}\\
 & -\frac{\psi^{\dagger}\mathbf{q}\psi}{q^{2}\left|\psi_{0}\right|^{2}}\psi^{\dagger}\mathbf{q}\cdot\left(\mathbf{q}\tau_{3}\varphi_q+\tau_{3}\mathbf{b}_{\mathbf{q}}+\mathbf{a}_{\mathbf{q}}\right)\psi,\nonumber 
\end{align}

which finally leads to 
\begin{align}
\left|\psi_{0}\right|^{2}\mathbf{a}_{\mathbf{q}}^{\perp} & =-\psi^{\dagger}\tau_{3}\psi \mathbf{b}_{\mathbf{q}}^{\perp},\label{eq:B1c.2}
\end{align}

with $\mathbf{a}^{\perp}$ and $\mathbf{b}^{\perp}$ defined as in Eq.~\eqref{eq:B1a}. Eq.~\eqref{eq:B1c.2} is important since it tells us that the freezing of the phase $\theta$ and the condensation of the Higgs boson $\left|\psi_{0}\right|^{2}$ could not provoke the expulsion of the field $\mathbf{a}^{\perp}$ until $\psi^{\dagger}\tau_{3}\psi$ in Eq.~\eqref{eq:B1c.2} is condensed.

Adding back the $F_{\mu\nu}F^{\mu\nu}$ and $\tilde{F}_{\mu\nu}\tilde{F}^{\mu\nu}$ terms leads to

\begin{align}
\left(\frac{\left|\psi_{0}\right|}{2g}^{2}+q^{2}\right)\mathbf{a}_{\mathbf{q}}^{\perp} & =-\left(\frac{\psi^{\dagger}\tau_{3}\psi}{2g}+q^{2}\right)\mathbf{b}_{\mathbf{q}}^{\perp},\label{eq:B1c.3}
\end{align}

\subsubsection{Effective action at $T^{*}$} \label{sec:flucderiv}

The system is invariant with respect to the second
U(1), which means that we can always re-absorb the phase $\varphi$
into a re-definition of $\mathbf{b}_{\mu}^{\parallel}\rightarrow \mathbf{b}_{\mu}^{\parallel}-\partial_{\mu}\varphi$.
We can choose the gauge such that, for example, $\mathbf{b}^{\parallel}=0$.

Writing the fields $\mathbf{a}$ and $\mathbf{b}$ in terms of the longitudinal and transverse components as $\mathbf{a}=\mathbf{a}^{\perp}+\mathbf{a}^{\parallel}$ and $\mathbf{b}=\mathbf{b}^{\perp}$ and putting back in Eq.~(\ref{eq:B1c}), we get
\begin{align}
& S_{\varphi,a,b}^{eff} =\frac{1}{2g}\sum_{\mathbf{q}}[\psi^{\dagger}\left( \left(\mathbf{q}\tau_{3}\varphi_{q}+\mathbf{a}_{\mathbf{q}}^{\parallel}\right)^2+\left(\mathbf{a}_{\mathbf{q}}^{\perp}+\tau_{3}\mathbf{b}_{\mathbf{q}}^{\perp}\right)^{2}\right)\psi\nonumber \\
& -\frac{\left(\psi^{\dagger}\tau_3\psi q^2\varphi_{q}+\mathbf{q}\cdot \mathbf{a}_{\mathbf{q}}^{\parallel}\psi^{\dagger}\psi+\mathbf{q}\cdot\left( \psi^{\dagger}\tau_3\psi \mathbf{b}_{\mathbf{q}}^{\perp}+\psi^{\dagger}\psi \mathbf{a}_{\mathbf{q}}^{\perp}\right)\right)^2}{q^{2}\left|\psi_{0}\right|^{2}}].\label{eq:B1c.4}
\end{align} 
From Eq.~(\ref{eq:B1c.4}) and using Eq.~\eqref{eq:B1c.2}, we obtain

\begin{align}
S_{\varphi,b}^{eff} & =\frac{1}{2g}\sum_{\mathbf{q}}\left(\left|\psi_{0}\right|^{2}-\frac{\left(\psi^{\dagger}\tau_{3}\psi\right)^{2}}{\left|\psi_{0}\right|^{2}}\right)\left(q^{2}\varphi_{q}^{2}+\left(b_{q}^{\perp}\right)^2\right).\label{eq:B1e}
\end{align} 
Using $n_{s}^{+}=\psi^{\dagger}\psi$ and $n_{s}^{-}=\psi^{\dagger}\tau_{3}\psi$ we finally get
\begin{align}
S_{\varphi,b}^{eff} & =\frac{1}{2g}\sum_{\mathbf{q}}\frac{4\left|z_{1}\right|^{2}\left|z_{2}\right|^{2}}{n_{s}^{+}}\left(q^{2}\varphi_{q}^{2}+\left(b_{q}^{\perp}\right)^2\right).\label{eq:B1e.1}
\end{align}
We recover the $\left(\partial_{\mu}\varphi\right)^{2}$ in Eq.~(\ref{eq:9}), hence proving that the fluctuations below $T^{*}$ are described by the SU(2) chiral model, itself equivalent to the
$CP^{1}$ model, or also the O(3) NL$\sigma$M with fluctuating $\eta$-fields.

\subsection{\label{subsec:Derivation-of-the} Integration of all the Goldstone modes: Derivation of the effective action $S_{a,b}^{eff}$}

In this appendix, we formally integrate out the Goldstone modes $\theta$
and $\varphi$ in the action in Eq.~\eqref{eq:1} to arrive at an
effective action in Eq.~(\ref{eq:6}). As mentioned in the main text,
the field derivative term in Eq.~(\ref{eq:1}) has a quadratic form
$\left|D_{\mu}\psi\right|^{2}=$ $\left|\partial_{\mu}\psi_{0}\right|^{2}+\psi^{\dagger}\left(\partial_{\mu}\theta+\tau_{3}\partial_{\mu}\varphi-a_{\mu}-\tau_{3}b_{\mu}\right)^{2}\psi$.
Ignoring the amplitude fluctuations ($\left|\partial_{\mu}\psi_{0}\right|^{2}=0$),
the action in the momentum space reads as: 
\begin{align}
{\cal S}_{a,b}[\theta,\varphi] & =\sum_{\mathbf{q}}\left[\frac{1}{2g}\psi^{\dagger}\left(\mathbf{q}\theta_{q}+\tau_{3}\mathbf{q}\varphi_{q}+\mathbf{a}_{\mathbf{q}}+\tau_{3}\mathbf{b}_{\mathbf{q}}\right)^{2}\psi\right.\nonumber \\
 & \left.+V\left(\psi_{0}\right)+\frac{1}{2}\left(\mathbf{q}\times \mathbf{a}_{\mathbf{q}}\right)^{2}+\frac{1}{2}\left(\mathbf{q}\times \mathbf{b}_{\mathbf{q}}\right)^{2}\right].\label{eq:A1}
\end{align}
Note that $V\left(\psi\right)$ has been replaced by
$V\left(\psi_{0}\right)$ in Eq.~(\ref{eq:A1}) as it is independent
of the Goldstone modes. We have used $\frac{1}{4}F_{\mu\nu}F^{\mu\nu}=\frac{1}{2}\left(\mathbf{q}\times \mathbf{a}_{\mathbf{q}}\right)^{2}$
and $\frac{1}{4}$$\tilde{F}_{\mu\nu}\tilde{F}^{\mu\nu}=\frac{1}{2}\left(\mathbf{q}\times \mathbf{b}_{\mathbf{q}}\right)^{2}$.
Expanding the terms in Eq.~(\ref{eq:A1}) and using the spinor structure
of $\psi$ in Eq.~(\ref{eq:1.0}), we get, 
\begin{align}
{\cal S}_{a,b}[\theta,\varphi] & =\frac{1}{2g}\sum_{\mathbf{q}}\left\{ n_{s}^{+}q^{2}\theta_{q}^{2}+2\theta_{q}\left(n_{s}^{-}q^{2}\varphi_{q}+n_{s}^{+}\mathbf{q}\cdot \mathbf{a}_{\mathbf{q}}\right.\right.\nonumber \\
 & \left.\left.+n_{s}^{-}\mathbf{q}\cdot \mathbf{b}_{\mathbf{q}}\right)\right\} +\frac{1}{2g}\psi^{\dagger}\left(\tau_{3}\mathbf{q}\varphi_{q}+\tilde{\mathbf{a}}_{\mathbf{q}}+\tau_{3}\mathbf{A}_{\mathbf{q}}\right)^{2}\psi\nonumber \\
 & +V\left(\psi_{0}\right)+\frac{1}{4}F_{\mu\nu}F^{\mu\nu}+\frac{1}{4}\tilde{F}_{\mu\nu}\tilde{F}^{\mu\nu}.\label{eq:A2}\\
\mbox{where } & n_{s}^{+}=\left|z_{1}\right|^{2}+\left|z_{2}\right|^{2},\nonumber \\
 & n_{s}^{-}=\left|z_{1}\right|^{2}-\left|z_{2}\right|^{2}.\label{eq:A2a}
\end{align}

We use the definition

\begin{align}
T_{1} & :=\frac{1}{2g}\sum_{\mathbf{q}}\psi^{\dagger}\left(\tau_{3}\mathbf{q}\varphi_{q}+\mathbf{a}_{\mathbf{q}}+\tau_{3}\mathbf{b}_{\mathbf{q}}\right)^{2}\psi\nonumber \\
 & =\frac{1}{2g}\left[n_{s}^{+}q^{2}\varphi_{q}^{2}+2\varphi_{q}\left\{ n_{s}^{-}\mathbf{q}\cdot \mathbf{a}_{\mathbf{q}}+n_{s}^{+}\mathbf{q}\cdot \mathbf{b}_{\mathbf{q}}\right\} \right.\nonumber \\
 & \left.+n_{s}^{+}a_{q}^{2}+n_{s}^{+}b_{q}^{2}+2n_{s}^{-}\mathbf{a}_{\mathbf{q}}\cdot \mathbf{b}_{\mathbf{q}}\right]\nonumber \\
 & =\frac{1}{2g}\left[n_{s}^{+}q^{2}\varphi_{q}^{2}+2\varphi_{q}\left\{ n_{s}^{-}\mathbf{q}\cdot \mathbf{a}_{\mathbf{q}}+n_{s}^{+}\mathbf{q}\cdot \mathbf{b}_{\mathbf{q}}\right\} \right.\nonumber \\
 & \left.+n_{s}^{+}\left(\mathbf{a}_{\mathbf{q}}+\frac{n_{s}^{-}}{n_{s}^{+}}\mathbf{b}_{\mathbf{q}}\right)^{2}+n_{s}^{+}\left(1-\left(\frac{n_{s}^{-}}{n_{s}^{+}}\right)^{2}\right)b_{q}^{2}\right].\label{eq:A3}
\end{align}
Completing the square in $\theta_{q}$ in Eq.~(\ref{eq:A2}), and
neglecting $V\left(\psi_{0}\right)$ and the potential strength $F_{\mu\nu}F^{\mu\nu}$
and $\tilde{F}_{\mu\nu}\tilde{F}^{\mu\nu}$ we obtain {\small{}{
\begin{align}
\delta\tilde{{\cal S}}_{a,b}[\theta,\varphi] & =\frac{n_{s}^{+}q^{2}}{2g}\sum_{\mathbf{q}}\left\{ \theta_{q}+\left(\frac{n_{s}^{-}q^{2}\varphi_{q}+n_{s}^{+}\mathbf{q}\cdot \mathbf{a}_{\mathbf{q}}+n_{s}^{-}\mathbf{q}\cdot \mathbf{b}_{\mathbf{q}}}{n_{s}^{+}q^{2}}\right)\right\} \nonumber \\
 & -\frac{n_{s}^{+}q^{2}}{2g}\left(\frac{n_{s}^{-}q^{2}\varphi_{q}+n_{s}^{+}\mathbf{q}\cdot \mathbf{a}_{\mathbf{q}}+n_{s}^{-}\mathbf{q}\cdot \mathbf{b}_{\mathbf{q}}}{n_{s}^{+}q^{2}}\right)^{2}+T_{1}.\label{eq:A4-1}
\end{align}
}} First, we integrate over $\theta_{q}$ which results into an effective
action $\delta\tilde{{\cal S}}_{a,b}^{eff}[\varphi]$ where $e^{-\delta\tilde{{\cal S}}_{a,b}^{eff}[\varphi]}\equiv\int{\cal D}\theta~{\cal \delta\tilde{S}}_{a,b}[\theta,\varphi]$
with, {\small{}{ 
\begin{align}
\delta\tilde{{\cal S}}_{a,b}^{eff}[\varphi] & =-\sum_{\mathbf{q}}\frac{n_{s}^{+}q^{2}}{2g}\left(\frac{n_{s}^{-}q^{2}\varphi_{q}+n_{s}^{+}\mathbf{q}\cdot \mathbf{a}_{\mathbf{q}}+n_{s}^{-}\mathbf{q}\cdot \mathbf{b}_{\mathbf{q}}}{n_{s}^{+}q^{2}}\right)^{2}+T_{1}\nonumber \\
 & =-\sum_{q}[\frac{n_{s}^{+}q^{2}}{2g}\left(\frac{n_{s}^{-}}{n_{s}^{+}}\right)^{2}\varphi_{q}^{2}+\frac{n_{s}^{-}}{g}\varphi_{q}\left(\mathbf{q}\cdot \mathbf{a}_{\mathbf{q}}+\frac{n_{s}^{-}}{n_{s}^{+}}\mathbf{q}\cdot \mathbf{b}_{\mathbf{q}}\right)\nonumber \\
 & +\frac{n_{s}^{+}}{2gq^{2}}\left(\mathbf{q}\cdot \mathbf{a}_{\mathbf{q}}+\frac{n_{s}^{-}}{n_{s}^{+}}\mathbf{q}\cdot \mathbf{b}_{\mathbf{q}}\right)^{2}]+T_{1}.\label{eq:A5}
\end{align}
}}{\small \par}

Using the form of $T_{1}$ given in Eq.~(\ref{eq:A3}) and after
a bit of algebraic manipulations, the action in Eq.~(\ref{eq:A5})
can be written as, 
\begin{align}
\delta\tilde{{\cal S}}_{a,b}^{eff}[\varphi] & =\sum_{\mathbf{q}}[\frac{n_{s}^{\perp}q^{2}}{2g}\left(\left(\varphi_{q}+\frac{\mathbf{q}\cdot \mathbf{b}_{\mathbf{q}}}{q^{2}}\right)^{2}-\left(\frac{\mathbf{q}\cdot \mathbf{b}_{\mathbf{q}}}{q^{2}}\right)^{2}\right)\nonumber \\
 & +\frac{n_{s}^{\perp}}{2g}b_{q}^{2}+\frac{n_{s}^{+}}{2g}\left(\mathbf{a}_{\mathbf{q}}^{\perp}+\frac{n_{s}^{-}}{n_{s}^{+}}\mathbf{b}_{\mathbf{q}}^{\perp}\right)^{2}],\label{eq:A6}\\
\mbox{with } & n_{s}^{\perp}=n_{s}^{+}\left(1-\left(\frac{n_{s}^{-}}{n_{s}^{+}}\right)^{2}\right),\label{eq:A6b}
\end{align}
with the notation $\mathbf{a}^{\perp}=\mathbf{a}-\mathbf{q}\left(\mathbf{q}\cdot\mathbf{a}\right)/q^{2}$
and $a_{q}^{2}=\mathbf{a}_{q}\cdot\mathbf{a}_{-q}$ (idem for $\mathbf{b}$).
We now integrate over $\varphi_{q}$ which results into an effective
action $\delta{\cal S}_{a,b}^{eff}$ where $e^{-\delta{\cal S}_{a,b}^{eff}}\equiv\int{\cal D}\varphi~\delta\tilde{{\cal S}}_{a,b}^{eff}[\varphi]$
with, 
\begin{align}
\delta{\cal S}_{a,b}^{eff} & =\sum_{\mathbf{q}}\left[\frac{n_{s}^{\perp}}{2g}\left(b_{q}^{\perp}\right)^{2}+\frac{n_{s}^{+}}{2g}\left(\mathbf{a}_{\mathbf{q}}^{\perp}+\frac{n_{s}^{-}}{n_{s}^{+}}\mathbf{b}_{\mathbf{q}}^{\perp}\right)^{2}\right].\label{eq:A7}
\end{align}
Simplifying Eq.~(\ref{eq:A7}), and putting back the potential terms,
we obtain Eq.~(\ref{eq:6}) of the main text, 
\begin{align*}
{\cal S}_{a,b}^{eff} & =\sum_{\mathbf{q}}\frac{n_{s}^{+}}{2g}\left(\left(a_{q}^{\perp}\right)^{2}+\left(b_{q}^{\perp}\right)^{2}\right)+\frac{2n_{s}^{-}}{2g}\mathbf{a}_{\mathbf{q}}^{\perp}\cdot \mathbf{b}_{\mathbf{q}}^{\perp}\\
 & +\frac{q^{2}}{2}\left(a_{q}^{\perp}\right)^{2}+\frac{q^{2}}{2}\left(b_{q}^{\perp}\right)^{2}+V\left(\psi_{0}\right).
\end{align*}


\section{Derivation of the $CP^1$ model from fractionalization of PDW} \label{sec:CP1fromPDW}

In this appendix, we derive the effective action due to the fractionalization of the preformed PDW order and show its similarity with the $CP^1$ model. The action governing the gradients of the PDW field $\Delta_{\text{PDW}}$ is given, in the absence of EM field, as
\begin{equation}
S=\int d^{d}x ~ \partial_{\mu}\Delta_{\text{PDW}}^{*}\partial_{\mu}\Delta_{\text{PDW}}.
\label{eq:PDWgradapp}
\end{equation}

\subsection{Proof with $\Delta_{\text{PDW}}=\Delta_{ij} \chi_{ij}^{*}$}

We fractionalize the $\Delta_{\text{PDW}}$ field into p-p and p-h pairs as
\begin{align}
\Delta_{\text{PDW}}=\Delta_{ij} \chi_{ij}^{*} \equiv z_1 z_2^*, \label{eq:PDWdef1app} \\
\mbox{with } z_1 \equiv \Delta_{ij} ~\mbox{and}~ z_2 \equiv \chi_{ij}.\nonumber
\end{align}
Using the constraint $\left| z_1 \right|^2+\left| z_2 \right|^2=1$ and substituting Eq.~\eqref{eq:PDWdef1app} in Eq.~\eqref{eq:PDWgradapp}, we get
\begin{align}
S=&\int d^{d}x ~ \left[ \sum_{a=1}^{2} \partial_{\mu}z_a^{*}\partial_{\mu}z_a \right. \nonumber\\
&\left. -\left( z_1 \partial_{\mu}z_1^{*}-z_2^* \partial_{\mu}z_2 \right)\left( z_1^* \partial_{\mu}z_1-z_2 \partial_{\mu}z_2^* \right) \right].
\label{eq:chiralderi1}
\end{align}
Lets consider the gauge fields 
\begin{align*}
\alpha_{\mu} & =-i\left(z_{1}\partial_{\mu}z_{1}^{*}-z_{2}^{*}\partial_{\mu}z_{2}\right);\\
\overline{\alpha}_{\mu} & =i\left(z_{1}^{*}\partial_{\mu}z_{1}-z_{2}\partial_{\mu}z_{2}^{*}\right).
\end{align*}
Then Eq.~\eqref{eq:chiralderi1} can be re-cast into 
\begin{align}
S & =\int d^{d}x~\sum_{a=1}^{2}\left|D_{\mu}^{a}z_{a}\right|^{2},\label{eq:inter1}\\
D_{\mu}^{1} & =\partial_{\mu}+i\overline{\alpha}_{\mu,}\nonumber \\
D_{\mu}^{2} & =\overline{D}_{\mu}^{1}=\partial_{\mu}-i\alpha_{\mu}.\nonumber 
\end{align}

In order to connect this action with a $CP^1$ model, we choose a spinor 
\begin{equation}
\psi  =\left(\begin{array}{c} z_{1}^{*}\\ z_{2} \end{array}\right),
\label{eq:spinorcp1app}
\end{equation}
and write down the corresponding action in the form of a $CP^1$ model
\begin{align}
S_{a} & =\int d^{d}x\left|D_{\mu}\psi\right|^{2},\label{eq:pdwcp1app} \\
\mbox{with } & D_{\mu}=\partial_{\mu}-i\tau_3 \alpha_{\mu}~\mbox{and }~\psi^{\dagger} \psi  =1,
\end{align}
where the gauge field $\alpha_{\mu}$ corresponding to the relative phase of the spinor in Eq.~\eqref{eq:spinorcp1app} is defined by the condition $\partial S_{a}/\partial \alpha=0$. This gives
\begin{equation}
\alpha_{\mu}=-i\psi^{\dagger} \tau_3 \partial_{\mu} \psi.
\label{eq:alphavalue}
\end{equation}
Putting the value of $\alpha_{\mu}$ from Eq.~\eqref{eq:alphavalue} in Eq.~\eqref{eq:pdwcp1app} and using the form of the spinor in Eq.~\eqref{eq:spinorcp1app}, we can write the action as 
\begin{align}
S=&\int d^{d}x ~ \left[ \sum_{a=1}^{2} \partial_{\mu}z_a^{*}\partial_{\mu}z_a - \bar{\alpha}_{\mu} \alpha_{\mu}\right] \label{eq:chiralderi2.0}\\
& \mbox{where }~\alpha_{\mu}=-i\left( z_1 \partial_{\mu}z_1^{*}-z_2^* \partial_{\mu}z_2 \right).
\label{eq:chiralderi2}
\end{align}
This $CP^1$ model is same as the action in Eq.~\eqref{eq:chiralderi1} obtained by fractionalizing the PDW field. Due to the fractionalization of the PDW field defined in Eq.~\eqref{eq:PDWdef1app}, the $CP^1$ model involves the gauge field corresponding to the relative phase of spinor in Eq.~\eqref{eq:spinorcp1app} or equivalently the global phase of the spinor $\psi=\left( \Delta_{ij},\chi_{ij} \right)^{T}$. 

If we parameterize $z_1$ and $z_2$ in terms of the relative and the global phases as
\begin{equation}
z_1=\left|z_1\right| e^{i(\theta+\varphi)};~z_2=\left|z_2\right| e^{i(\theta-\varphi)},
\label{eq:z1z2def}
\end{equation} 
we can rewrite the $CP^1$ model in Eq.~\eqref{eq:chiralderi2.0} as
\begin{align}
S&=\int d^{d}x \left( 4\left|z_1\right|^2\left|z_2\right|^2 \left( \partial_{\mu}\varphi \right)^2 +\left|z_2\right|^2 \left( \partial_{\mu}\left|z_1\right| \right)^2 \right.\nonumber\\
&\left.+\left|z_1\right|^2 \left( \partial_{\mu}\left|z_2\right| \right)^2 + 2\left|z_1\right| \left|z_2\right|\left( \partial_{\mu}\left|z_1\right| \right)\left( \partial_{\mu}\left|z_2\right| \right) \right). 
\label{eq:actionwph}
\end{align}
The effective action depends only on the fluctuations of the relative phase $\varphi$, the amplitude fluctuations and is independent of the global phase $\theta$. The form of the action is similar to the action used in Eq.~\eqref{eq:9} but with renormalized coefficients in the amplitude fluctuation terms.

\subsection{Proof with $\tilde{\Delta}_{\text{PDW}}=\Delta_{ij} \chi_{ij}$}

In principle, we could have also fractionalized a PDW field with the definition
\begin{align}
\tilde{\Delta}_{\text{PDW}}=\Delta_{ij} \chi_{ij} \equiv z_1 z_2. \label{eq:PDWdef1appalt}
\end{align}
The action governing the gradients of this PDW will be 
\begin{equation}
S=\int d^{d}x ~ \partial_{\mu}\tilde{\Delta}_{\text{PDW}}^{*}\partial_{\mu}\tilde{\Delta}_{\text{PDW}}.
\label{eq:PDWgradapp1}
\end{equation}
Again using the constraint $\left| z_1 \right|^2+\left| z_2 \right|^2=1$ and substituting Eq.~\eqref{eq:PDWdef1appalt} in Eq.~\eqref{eq:PDWgradapp1}, we get
\begin{align}
S=&\int d^{d}x ~ \left[ \sum_{a=1}^{2} \partial_{\mu}z_a^{*}\partial_{\mu}z_a \right. \nonumber\\
&\left. -\left( z_1 \partial_{\mu}z_1^{*}-z_2 \partial_{\mu}z_2^* \right)\left( z_1^* \partial_{\mu}z_1-z_2^* \partial_{\mu}z_2 \right) \right].
\label{eq:chiralderi1a}
\end{align}
As previously, lets consider the gauge fields
\begin{align*}
\alpha_{\mu} & =-i\left(z_{1}^{*}\partial_{\mu}z_{1}-z_{2}^{*}\partial_{\mu}z_{2}\right);\\
\overline{\alpha}_{\mu} & =i\left(z_{1}\partial_{\mu}z_{1}^{*}-z_{2}\partial_{\mu}z_{2}^{*}\right).
\end{align*}
The Eq.(\ref{eq:chiralderi1}) can then be re-cast into 
\begin{align}
S & =\int d^{d}x~\sum_{a=1}^{2}\left|D_{\mu}^{a}z_{a}\right|^{2},\label{eq:inter1-1}\\
D_{\mu}^{1} & =\partial_{\mu}+i\alpha_{\mu,}\nonumber \\
D_{\mu}^{2} & =\partial_{\mu}-i\alpha_{\mu}.\nonumber 
\end{align}

With a choice of a spinor
\begin{equation}
\psi  =\left(\begin{array}{c} z_{1}^{*}\\ z_{2}^{*} \end{array}\right),
\label{eq:spinorcp1app1}
\end{equation}
the action Eq.~\eqref{eq:inter1-1} in the form of a $CP^{1}$ model is the same as previously Eq.~\eqref{eq:pdwcp1app}, with
\begin{align}
S_{a} & =\int d^{d}x\left|D_{\mu}\psi\right|^{2},\label{eq:pdwcp1app-1}\\
\mbox{with } & D_{\mu}=\partial_{\mu}-i\tau_{3}\alpha_{\mu}~\mbox{and }~\psi^{\dagger}\psi=1,
\end{align}
where the gauge field $\alpha_{\mu}$ corresponding to the relative phase of the spinor in Eq.~\eqref{eq:spinorcp1app1} is defined by the condition $\partial S_{a}/\partial\alpha=0$. This gives 
\begin{align}
\alpha_{\mu} & =-i\psi^{\dagger}\tau_{3}\partial_{\mu}\psi.\label{eq:alphavalue-1}\\
 & =-i\left(z_{1}\partial_{\mu}z_{1}^{*}-z_{2}\partial_{\mu}z_{2}^{*}\right)\nonumber 
\end{align}

Using the parametrization as in Eq.~\eqref{eq:z1z2def}, the effective action can be written as
\begin{align}
S&=\int d^{d}x \left( 4\left|z_1\right|^2\left|z_2\right|^2 \left( \partial_{\mu}\theta \right)^2 +\left|z_2\right|^2 \left( \partial_{\mu}\left|z_1\right| \right)^2 \right.\nonumber\\
&\left.+\left|z_1\right|^2 \left( \partial_{\mu}\left|z_2\right| \right)^2 + 2\left|z_1\right| \left|z_2\right|\left( \partial_{\mu}\left|z_1\right| \right)\left( \partial_{\mu}\left|z_2\right| \right) \right). 
\label{eq:actionwph1s}
\end{align}
Thus, fractionalizing the PDW of the form given in Eq.~\eqref{eq:PDWdef1appalt}, we obtain an effective action with fluctuating global phase. So, fractionalizing the PDW field as in Eq.~\eqref{eq:PDWdef1appalt} would result to freezing of the relative phase $\varphi$ at $T^{*}$ and the effective action in the PG phase involves only the fluctuation of the global phase.

\section{The chiral SU(n+1) model}

\label{sec:chiralmodel}

\subsection{$CP^{n}$ representation of a chiral SU(n+1) model}

\label{sec:warmup}

In this appendix, we give the details of the $CP^{n}$ representation
of a chiral SU(n+1) model for a generic $n$. Let us consider the SU(n+1)
invariant chiral model \cite{Perelomov81}. A generic field $\varphi$
belonging to the Lie algebra of the group SU(n+1) can be cast into
the form 
\begin{align}
\varphi_{ab} & =\frac{\delta_{ab}}{n+1}-z_{a}z_{b}^{*},\label{eq:prel1}
\end{align}
where $z_{a}$ is a set of n+1 complex numbers verifying the constraint
\begin{align}
\sum_{a=1}^{n+1}z_{a}^{*}z_{a} & =1.\label{eq:prel1a}
\end{align}
The action for this model is 
\begin{align}
S=\frac{1}{2}\int d^{d}x & Tr[\partial_{\mu}\varphi^{\dagger}\partial_{\mu}\varphi],\label{eq:aprel2}
\end{align}
which using the constraint can be put into the form ($a=1\cdots n+1$)
\begin{align}
S=\int d^{d}x & \left[\sum_{a}\partial_{\mu}z_{a}^{*}\partial_{\mu}z_{a}-\sum_{a,b}\left(z_{a}^{*}\partial_{\mu}z_{a}\right)\left(z_{b}\partial_{\mu}z_{b}^{*}\right)\right].\label{eq:prel3}
\end{align}
Eq.~(\ref{eq:aprel2}) can be recast to the action 
\begin{align}
S_{a} & =\int d^{d}x\left|D_{\mu}z\right|^{2},\label{eq:aprel4}\\
\mbox{with } & \sum_{a=1}^{n+1}\left|z_{a}\right|^{2}=1,\nonumber \\
 & D_{\mu}=\partial_{\mu}-ia_{\mu},\nonumber \\
\mbox{and } & a_{\mu}=-i\sum_{a}z_{a}^{*}\partial_{\mu}z_{a},\nonumber 
\end{align}
where $z$ is a short hand notation for the multiplet $z=\left(z_{1,}z_{2},\cdots z_{n+1}\right)$,
$D_{\mu}=\partial_{\mu}-ia_{\mu}$, and $a_{\mu}=-i/2\sum_{a}\left(z_{a}^{*}\partial_{\mu}z_{a}-z_{a}\partial_{\mu}z_{a}^{*}\right)=-i\sum_{a}z_{a}^{*}\partial_{\mu}z_{a}$. One can convince oneself of this equivalence by solving for the mean value of the gauge field $\delta S_{a}/\delta a_{\mu}=0$ which leads
to the definition of the gauge field $a_{\mu}$ in Eq.~(\ref{eq:aprel4}),
and then reporting it into $S_{a}$ in Eq. \eqref{eq:aprel4} leads
to Eq.~(\ref{eq:prel3}). The model defined in Eq.~(\ref{eq:aprel4})
is called the $CP^{n}$ model. It is remarkable that it is invariant
under the gauge transformation $z_{a}\rightarrow e^{i\theta}z_{a}$,
$a_{\mu}\rightarrow a_{\mu}+\partial_{\mu}\theta$. The gauge structure
enforced by the gauge field $a_{\mu}$ reflects the topological character
of the $CP^{n}$ model, with $\pi_{2}\left(CP^{n}\right)=Z$. Said
in simpler words, n+1 complex fields verifying the constraint Eq.~(\ref{eq:prel1a})
are not purely independent, but lead to a field theory of n independent
fluctuating fields subjected to the action Eq.~(\ref{eq:aprel4}).

For the specific case of SU(2), the $CP^{1}$ model is equivalent
to the O(3) NL$\sigma$M. To see this, it is convenient to take a
representation of the fields in terms of Pauli matrices 
\begin{align}
m^{a} & =z_{\alpha}^{*}\sigma_{\alpha\beta}^{a}z_{\beta}, & a=1,3\label{eq:aprel5}
\end{align}
which satisfies the constraint 
\begin{align}
\sum_{a=1}^{3}\left|m^{a}\right|^{2} & =1.\label{eq:aprel6}
\end{align}
The corresponding action reads 
\begin{align}
S & =1/2\int d^{d}x\sum_{a=1}^{3}\left(\partial_{\mu}m^{a}\right)^{2}.\label{eq:aprel7}
\end{align}

The action Eq. \eqref{eq:aprel7} is typical of an O(3) NL$\sigma$M.
This equivalence between $CP^{1}\sim O(3)$ NL$\sigma$M is not generically
valid for all n. In particular the $O(n+1)$ NL$\sigma$M does not
have topological defects for $n\geq3$ since $\pi_{2}(S^{n})=0$ for
$n\geq3$, whereas the $CP^{n}$ model does with $\pi_{2}\left(CP^{n}\right)=Z$
for all $n$. The topological charge can be written
as\cite{Perelomov81}
\begin{align}
Q & =\int d^{2}x\epsilon_{\mu\nu}\left(\partial_{\mu}z^{*}\partial_{\nu}z\right),\label{eq:aprel8}\\
 & =\int d^{2}x\sum_{a}\epsilon_{\mu\nu}\left(\partial_{\mu}z_{a}^{*}\partial_{\nu}z_{a}\right),\nonumber 
\end{align}

where $\epsilon_{\mu\nu}$ is the totally anti-symmetric
tensor. In terms of the gauge field the topological charge writes 
\begin{align}
Q & =\frac{1}{2\pi}\int d^{2}x\epsilon_{\mu\nu}\partial_{\mu}a_{\nu}.\label{eq:aprel9}
\end{align}

\subsection{\label{subsec:Explicit-form-of}Explicit forms of the $SU(n+1)\rightarrow CP^{n}$ mapping}

\subsubsection{\label{subsec:The-case-with}The case with two fields: $SU(2)\rightarrow CP^{1}$}

In the case for example where the d-BDW order has only one wave vector, two fields $z_{1}$ and $z_{2}$ form the spinor
$\psi$ in Eq. (\ref{eq:1.0}). We take the form of the spinor in Eq. (\ref{eq:5a}) and assume that we are below $T^{*}$ so that the phase $\theta$ is frozen. Since the upper energy scale is $E^{*}=\left|\psi_{0}\right|$, the constraint in Eq. (\ref{eq:prel1a}) writes 
\begin{align}
\sum_{a=1}^{2}z_{a}^{*}z_{a} & =\left|\psi_{0}\right|^{2}\label{eq:AC1}
\end{align}
From Eqs. (\ref{eq:prel3}) and (\ref{eq:prel4}), we see that the SU(2) chiral model can be written as 
\begin{align}
S=\int d^{2}x & \left[\sum_{a}\partial_{\mu}z_{a}^{*}\partial_{\mu}z_{a}-\bar{a}_{\mu}a_{\mu}\left|\psi_{0}\right|^{2}\right],\label{eq:AC1.1}\\
\mbox{with } & a_{\mu}=\frac{-i}{2\left|\psi_{0}\right|^2}\sum_{a}\left(z_{a}^{*}\partial_{\mu}z_{a}-z_{a}\partial_{\mu}z_{a}^{*}\right).\nonumber 
\end{align}
Reporting the explicit form of the spinor in term of the phase $\varphi$ leads to 
\begin{align}
S= & \int d^{2}x\left|\psi_{0}\right|^{2}\left(\partial_{\mu}\varphi\right)^{2}+\left|\partial_{\mu}\psi_{0}\right|^{2}\nonumber \\
 & -\frac{\left(\psi^{\dagger}\tau_{3}\psi\right)^{2}}{\left|\psi_{0}\right|^{2}}\left(\partial_{\mu}\varphi\right)^{2},\label{eq:AC1.2}
\end{align}
which finally gives after a recombination of terms
\begin{align}
S= & \int d^{2}x\frac{4\left|z_{1}\right|^{2}\left|z_{2}\right|^{2}}{n_{s}^{+}}\left(\partial_{\mu}\varphi\right)^{2}+\left(\partial_{\mu}\left|z_{1}\right|\right)^{2}+\left(\partial_{\mu}\left|z_{2}\right|\right)^{2},\label{eq:AC1.3}
\end{align}
where the standard notations $n_{s}^{+}=\left|\psi_{0}\right|^{2}=\left|z_{1}\right|^{2}+\left|z_{2}\right|^{2}$ have been used. We see that Eq. (\ref{eq:AC1.3}) is identical to Eq. (\ref{eq:9}) and to Eq. (\ref{eq:B1e.1}) which makes the point that the chiral model describes the fluctuations below $T^{*}$.

\subsubsection{The case for three fields: $SU(3)\rightarrow CP^{2}$} \label{sec:su3cp2}

In the case, for example, where the d-BDW has two wave vectors $\mathbf{Q}_{x}$ and $\mathbf{Q}_{y}$, which is the
most typical case for cuprates, we have three fields $z_{1}$, $z_{2}$ and $z_{3}$ which form the spinor
\begin{align}
\psi^{\dagger}=\left(z_{1}^{*},z_{2}^{*},z_{3}^{*}\right) & \ \ \!\ \ \ \ \psi=\left(\begin{array}{c}
z_{1}\\
z_{2}\\
z_{3}
\end{array}\right).\label{eq:AC2.0}
\end{align}
We take the following parametrization of $\psi$
with 
\begin{align}
\psi & =e^{i\theta}e^{i\hat{\delta}_{2}\varphi_{2}}e^{i\hat{\delta}_{3}\varphi_{3}}\psi_{0},\label{eq:AC2.1}\\
\mbox{with } & \hat{\delta}_{2}=\left(\begin{array}{ccc}
1\\
 & -1\\
 &  & 1
\end{array}\right),\nonumber \\
\mbox{and } & \hat{\delta}_{3}=\left(\begin{array}{ccc}
1\\
 & 1\\
 &  & -1
\end{array}\right),\nonumber 
\end{align}
and where $\psi_{0}$ is parametrized as in Eq. (\ref{eq:5a}). We now expand the action Eq.(\ref{eq:prel3}) in this basis, assuming that the phase $\theta$ is frozen at $T^{*}$ and that as in Sec.~\ref{subsec:The-case-with}, the constraint in Eq. (\ref{eq:AC1}) is extended as $\sum_{a=1}^{3}z_{a}^{*}z_{a}=\left|\psi_{0}\right|^{2}$.
We get
\begin{align}
S & =\int d^{2}x\left[\left|\partial_{\mu}\psi_{0}\right|^{2}+\left(\left|\psi_{0}\right|^{2}-\frac{\left(\psi^{\dagger}\hat{\delta}_{2}\psi\right)^{2}}{\left|\psi_{0}\right|^{2}}\right)\left(\partial_{\mu}\varphi_{2}\right)^{2}\right.\label{eq:AC2.2}\\
 & +\left(\left|\psi_{0}\right|^{2}-\frac{\left(\psi^{\dagger}\hat{\delta}_{3}\psi\right)^{2}}{\left|\psi_{0}\right|^{2}}\right)\left(\partial_{\mu}\varphi_{3}\right)^{2}\nonumber \\
 & \left.+2\left(\psi^{\dagger}\hat{\delta}_{2}\hat{\delta}_{3}\psi-\frac{\left(\psi^{\dagger}\hat{\delta}_{2}\psi\right)\left(\psi^{\dagger}\hat{\delta}_{3}\psi\right)}{\left|\psi_{0}\right|^{2}}\right)\left(\partial_{\mu}\varphi_{2}\right)\left(\partial_{\mu}\varphi_{3}\right)\right].\nonumber 
\end{align}
Reducing in terms of the components of the fields yields 
\begin{align}
S & =\int d^{2}x\left[\frac{4\left(\left|z_{1}\right|^{2}+\left|z_{3}\right|^{2}\right)\left|z_{2}\right|^{2}}{\left|\psi_{0}\right|^{2}}\left(\partial_{\mu}\varphi_{2}\right)^{2}\right.\nonumber \\
&\left.+\frac{4\left(\left|z_{1}\right|^{2}+\left|z_{2}\right|^{2}\right)\left|z_{3}\right|^{2}}{\left|\psi_{0}\right|^{2}}\left(\partial_{\mu}\varphi_{3}\right)^{2}\right.\label{eq:AC2.3}\\
 & \left.-\frac{4\left|z_{2}\right|^{2}\left|z_{3}\right|^{2}}{\left|\psi_{0}\right|^{2}}\left(\partial_{\mu}\varphi_{2}\right)\left(\partial_{\mu}\varphi_{3}\right)+\sum_{a=1}^{3}\left(\partial_{\mu}\left|z_{a}\right|\right)^{2}\right].\nonumber 
\end{align}


\section{PDW ($\eta$)-fluctuations below $T^*$: Operator formalism} \label{sec:PDWPGoperator}

As already highlighted in Sec.~\ref{sec:fluc} and detailed in Sec.~\ref{sec:flucderiv}, the fluctuations in the PG phase can be described by an SU(2) chiral model. In this appendix, we will show that the fluctuations will take the form of PDW operators when the fields forming the spinor are written in terms of electronic operators. We also compare the structure of fluctuations in the SU(2) chiral model with earlier works related to the idea of SU(2) emergent symmetry (see e.g. Ref. [\onlinecite{Efetov13}]). We further construct an effective quantum rotor model within the operator formalism.


\subsection{The SU(2) chiral model}

\label{sec:flucchiral}

In order to obtain the form of the SU(2) fluctuations (Eq. \eqref{eq:9a})
in the case of cuprate superconductors, it is worth going back to the chiral model in Eq.~\eqref{eq:aprel2} and make the following
identifications. In the operator formalism, we define 
\begin{align}
z_{1} & \rightarrow \hat{\Delta}_{ij} \equiv \hat{d}\sum_{\sigma}\sigma c_{j-\sigma}c_{i\sigma}e^{-i\theta_{\Delta}},\nonumber \\
z_{2} & \rightarrow \hat{\chi}_{ij} \equiv \hat{d}\sum_{\sigma}c_{i\sigma}^{\dagger}c_{j\sigma}e^{iQ\cdot r+i\theta_{\chi}},\nonumber \\
\overline{z}_{1} & \rightarrow \hat{\Delta}_{ij}^{\dagger} \equiv \hat{d}\sum_{\sigma}\sigma c_{i\sigma}^{\dagger}c_{j-\sigma}^{\dagger}e^{i\theta_{\Delta}},\nonumber \\
\overline{z}_{2} & \rightarrow \hat{\chi}_{ij}^{\dagger} \equiv \hat{d}\sum_{\sigma}c_{j\sigma}^{\dagger}c_{i\sigma}e^{-iQ\cdot r-i\theta_{\chi}},\label{eq:D2a}
\end{align}
with $r=\left(r_{i}+r_{j}\right)/2$ the bond midpoint. We can now write the commutators (using $\hat{d}^{2}=1$) 
\begin{align}
\frac{1}{2}\left[{z}_{1},\overline{z}_{1}\right] & =-\frac{1}{2}\left(\hat{n}_{i}+\hat{n}_{j}\right)+1\equiv-\hat{\eta}_{z},\nonumber \\
\frac{1}{2}\left[{z}_{2},\overline{z}_{2}\right] & =\frac{1}{2}\left(\hat{n}_{i}-\hat{n}_{j}\right)\equiv \hat{\eta}_{0},\label{eq:D2b}\\
\frac{1}{2}\left[z_{1},\overline{z}_{2}\right] & =-\frac{1}{2}\sum_{\sigma}\sigma e^{-iQ\cdot r-i\left(\theta_{\Delta}+\theta_{\chi}\right)}c_{i-\sigma}c_{i\sigma}\equiv \hat{\eta}_{i,Q},\nonumber \\
\frac{1}{2}\left[z_{1},z_{2}\right] & =-\frac{1}{2}\sum_{\sigma}\sigma e^{iQ\cdot r-i\left(\theta_{\Delta}-\theta_{\chi}\right)}c_{j-\sigma}c_{j\sigma}\equiv \hat{\eta}_{j,-Q},\nonumber 
\end{align}
where $\hat{n}_{i}=\sum_{\sigma}c_{i\sigma}^{\dagger}c_{i\sigma}$. With the help of the PDW operators in Eqs.~\eqref{eq:D2b} one can form the raising and lowering operators as 
\begin{align}
\hat{\eta}_{+} & =\frac{1}{\sqrt{AB}}\left(A\hat{\eta}_{i,Q}^{\dagger}+B\hat{\eta}_{j,-Q}^{\dagger}\right),\nonumber \\
\hat{\eta}_{-} & =\frac{1}{\sqrt{AB}}\left(B\hat{\eta}_{i,Q}+A\hat{\eta}_{j,-Q}\right).\label{eq:D2c}
\end{align}

At first look it sounds that we have six generators $\hat{\eta}_{z}$, $\hat{\eta}_{0}$,
$\hat{\eta}_{i,Q}$, $\hat{\eta}_{j,-Q}$, $\hat{\eta}_{i,Q}^{\dagger}$, $\hat{\eta}_{j,-Q}^{\dagger}$,
but they are not independent. We have two copies of the SU(2) field
theory corresponding to two choices of the spinors. In the following
two subsections, we give the choices of the spinors and identify the
corresponding fluctuations.

\begin{lrbox}{\mybox} $\psi=\left(\begin{array}{c}
z_{1}\\
z_{2}
\end{array}\right)$ \end{lrbox}

\subsubsection{\usebox{\mybox}}

\label{sec:flucchirala}

It is the form of the spinor that we chose to start with, the generic fields in Eq.~\eqref{eq:prel1} that constitute
the chiral SU(2) model are given in a matrix form (using expressions
in Eqs.~\eqref{eq:D2b}), 
\begin{align}
\hat{\varphi}=\left(\begin{array}{cc}
-\hat{\eta}_{z} & \hat{\eta}_{i,Q}\\
\hat{\eta}_{i,Q}^{\dagger} & \hat{\eta}_{0}
\end{array}\right),\label{eq:D2d}
\end{align}
where we have used $\varphi_{ab}=\frac{1}{2}\left[z_{a},\overline{z}_{b}\right]$.

We can now write the first copy of the O(3) NL$\sigma$M in Eq.~\eqref{eq:9a},
with the identification 
\begin{align}
\hat{m}^{z} & \equiv\frac{1}{4}\sigma_{ab}^{z}\left[z_{a},\overline{z}_{b}\right]=\frac{1}{2}\left(-\hat{\eta}_{z}-\hat{\eta}_{0}\right)=\frac{1}{2}\left(1-\hat{n}_{i}\right),\nonumber \\
\hat{m}^{+} & \equiv\frac{1}{2}\sigma_{ab}^{+}\left[z_{a},\overline{z}_{b}\right]=\hat{\eta}_{i,Q},\nonumber \\
\hat{m}^{-} & \equiv\frac{1}{2}\sigma_{ab}^{-}\left[z_{a},\overline{z}_{b}\right]=\hat{\eta}_{i,Q}^{\dagger},\label{eq:D2e}
\end{align}
and the constraint 
\begin{align}
\left|\hat{m}^{z}\right|^{2}+\hat{m}^{+}\hat{m}^{-}+\hat{m}^{-}\hat{m}^{+} & =1.\label{eq:D2f}
\end{align}

We note that the SU(2) algebra formed by the operators
Eq.~\eqref{eq:D2e} is self adjoint: namely the $l=1$ representation
associated to it is itself. With the notation 
\begin{align*}
\hat{\Delta}_{1} & =\frac{-1}{\sqrt{2}}\hat{m}^{+},\\
\hat{\Delta}_{0} & =\hat{m}^{z},\\
\hat{\Delta}_{-1} & =\frac{1}{\sqrt{2}}\hat{m}^{-},\label{eq:10b-2-1}
\end{align*}
we get 
\begin{align}
\left[\hat{m}^{\pm},\hat{\Delta}_{m}\right] & =\sqrt{l\left(l+1\right)-m\left(m\pm1\right)}\hat{\Delta}_{m\pm1},\nonumber \\
\left[\hat{m}^{z},\hat{\Delta}_{m}\right] & =m\hat{\Delta}_{m}.
\end{align}

The form of the fluctuations coming from the chiral
model Eq.~\eqref{eq:prel3} and corresponding to the O(3) NL$\sigma$M
Eq.~\eqref{eq:aprel7} thus consists of three types of $\eta$-fields
forming an SU(2) algebra acting on the $\eta$-fields themselves. With the appropriate rotation of the basis 
\begin{align}
\hat{n}_{1} & =\frac{1}{2}\left(\hat{\Delta}_{1}+\hat{\Delta}_{-1}\right),\nonumber \\
\hat{n}_{2} & =\hat{\Delta}_{0},\label{eq:D2g}\\
\hat{n}_{3} & =\frac{-i}{2}\left(\hat{\Delta}_{1}-\hat{\Delta}_{-1}\right),\nonumber 
\end{align}

the Lie algebra writes 
\begin{align}
\hat{L}= & \left[\begin{array}{ccc}
0 & * & *\\
-i\frac{\hat{m}^{+}-\hat{m}^{-}}{2} & 0 & *\\
-\hat{m}^{z} & \frac{\hat{m}^{+}+\hat{m}^{-}}{2} & 0
\end{array}\right].\label{eq:D1-1}
\end{align}

\begin{lrbox}{\mybox} $\psi_{2}=\left(\begin{array}{c}
z_{1}\\
\overline{z}_{2}
\end{array}\right)$ \end{lrbox}

\subsubsection{\usebox{\mybox}}

\label{sec:flucchiralb}

We could have chosen a second form for the spinor, the generic fields in Eq.~\ref{eq:prel1} that constitute the chiral
SU(2) model are given in a matrix form (using expressions in Eqs.~\eqref{eq:D2b}),
\begin{align}
\hat{\varphi}_{2}=\left(\begin{array}{cc}
-\hat{\eta}_{z} & \hat{\eta}_{j,-Q}\\
\hat{\eta}_{j,-Q}^{\dagger} & -\hat{\eta}_{0}
\end{array}\right).\label{eq:D2d}
\end{align}

We can now write the second copy of the O(3) NL$\sigma$M in Eq.~\eqref{eq:9a},
with the identification 
\begin{align}
\hat{m}_{2}^{z} & \equiv\frac{1}{4}\sigma_{ab}^{z}\left[z_{a},\overline{z}_{b}\right]=\frac{1}{2}\left(-\hat{\eta}_{z}+\hat{\eta}_{0}\right)=\frac{1}{2}\left(1-\hat{n}_{j}\right),\nonumber \\
\hat{m}_{2}^{+} & \equiv\frac{1}{2}\sigma_{ab}^{+}\left[z_{a},\overline{z}_{b}\right]=\hat{\eta}_{j,-Q},\nonumber \\
\hat{m}_{2}^{-} & \equiv\frac{1}{2}\sigma_{ab}^{-}\left[z_{a},\overline{z}_{b}\right]=\hat{\eta}_{j,-Q}^{\dagger},\label{eq:D2e-1}
\end{align}
and the constraint is unchanged. As for Eq.(\ref{eq:D2e}) the algebra of Eq.(\ref{eq:D2e-1}) is self-adjoint.

It is interesting to note that the first copy of the O(3) NL$\sigma$M
corresponds to fluctuations living on site `i' of the bond $\langle ij\rangle$
and the second copy lives on site `j'. Thus, by constructing the preformed
pairs on bonds with the definition of the fields given in Eq.~\eqref{eq:D2a},
we have duplicated O(3) NL$\sigma$M into two copies living on different
sites of a bond. The angular fluctuations are given by $\hat{m}^{\pm}$
which correspond to PDW fields. On the other hand, $\hat{m}^{z}$ correspond
to the fluctuations in the density. At $T^{*}$, if we freeze the
global phase by choosing the spinor as in first copy $\psi$,
the PDW operators corresponding to the second copy acquires phase
coherence as they involve global phase. So, the special Higgs mechanism
at $T^{*}$ restricts the fluctuation space to only one copy of O(3)
NL$\sigma$M.


\subsection{SU(2) emergent symmetry}

After showing that the fluctuations in SU(2) chiral model are governed by the PDW operators, we now review the framework of SU(2) emergent symmetry and give the set of SU(2) operators, which rotates a particle-particle pairing field on a lattice bond, to a particle-hole pairing field sitting as well on a bond. We show that these operators resemble the form of a PDW operator.

We work in real space and introduce the following $l=1$ representation
of the SU(2) algebra in terms of the operators $\hat{\Delta}_{-1}=1/\sqrt{2}\hat{\Delta}_{ij}$,
$\hat{\Delta}_{1}=-1/\sqrt{2}\hat{\Delta}_{ij}^{\dagger}$, and $\hat{\Delta}_{0}$,
a linear combination of $\hat{\chi}_{ij}$ and $\hat{\chi}_{ij}^{\dagger}$: 
\begin{align}
\hat{\Delta}_{1} & =\frac{-1}{\sqrt{2}}\hat{d}\sum_{\sigma}\sigma c_{i\sigma}^{\dagger}c_{j-\sigma}^{\dagger}e^{i\theta_{\Delta}},\nonumber \\
\hat{\Delta}_{0} & =\frac{1}{2\sqrt{AB}}\hat{d}\sum_{\sigma}[Ac_{i\sigma}^{\dagger}c_{j\sigma}e^{iQ\cdot\left(r_{i}+r_{j}\right)/2}e^{i\theta_{\chi}}\nonumber \\
 & +Bc_{j-\sigma}^{\dagger}c_{i-\sigma}e^{-iQ\cdot\left(r_{i}+r_{j}\right)/2}e^{-i\theta_{\chi}},\nonumber \\
\hat{\Delta}_{-1} & =\frac{1}{\sqrt{2}}\hat{d}\sum_{\sigma}\sigma c_{j-\sigma}c_{i\sigma}e^{-i\theta_{\Delta}},\label{eq:8a}
\end{align}
where $A$ and $B$ are generic complex numbers, $\theta_{\Delta}$ and
$\theta_{\chi}$ are phases. $\hat{\Delta}_{-1}$ and $\hat{\Delta}_{1}=-\hat{\Delta}_{-1}^{\dagger}$
are proportional to the d-SC field and its conjugate, whereas $\hat{\Delta}_{0}$
is a modulated bond particle-hole operator. The phases of the various
operators are independent from each other. The representation in Eq.~(\ref{eq:8a})
has a large degree of generality. It supports a complex bond-excitonic
field $\hat{\chi}$ carrying both an amplitude and a phase, thus able to
host d-currentDW as well as d-CDW. The modulation vector ${\bf Q}$
does not need to be commensurate with the lattice. The SU(2) ladder
pseudo-spin operators are defined in the following way: 
\begin{align}
\hat{\eta}_{+} & =\frac{1}{2\sqrt{AB}}\sum_{\sigma}\sigma[Ac_{i\sigma}^{\dagger}c_{i-\sigma}^{\dagger}e^{iQ\cdot\left(r_{i}+r_{j}\right)/2}e^{i\left(\theta_{\Delta}+\theta_{\chi}\right)}\nonumber \\
 & +Bc_{j\sigma}^{\dagger}c_{j-\sigma}^{\dagger}e^{-iQ\cdot\left(r_{i}+r_{j}\right)/2}e^{i\left(\theta_{\Delta}-\theta_{\chi}\right)}],\nonumber \\
\hat{\eta}_{-} & =\frac{1}{2\sqrt{AB}}\sum_{\sigma}\sigma[Bc_{i-\sigma}c_{i\sigma}e^{-iQ\cdot\left(r_{i}+r_{j}\right)/2}e^{-i\left(\theta_{\Delta}+\theta_{\chi}\right)}\nonumber \\
 & +Ac_{j-\sigma}c_{j\sigma}e^{iQ\cdot\left(r_{i}+r_{j}\right)/2}e^{-i\left(\theta_{\Delta}-\theta_{\chi}\right)}],\nonumber \\
\hat{\eta}_{z} & =\frac{1}{2}\left[\hat{\eta}_{+},\hat{\eta}_{-}\right],\nonumber \\
 & =\frac{1}{2}\sum_{\sigma}\left(\hat{n}_{i\sigma}+\hat{n}_{j\sigma}-1\right),\label{eq:8b}
\end{align}
where $\hat{n}_{i\sigma}=c_{i\sigma}^{\dagger}c_{i\sigma}$. With
these definitions, the three $\eta$ operators form a closed SU(2)
algebra 
\begin{align}
\left[\hat{\eta}_{z},\hat{\eta}_{\pm}\right] & =\pm\hat{\eta}_{\pm},\nonumber \\
\left[\hat{\eta}_{+},\hat{\eta}_{-}\right] & =2\hat{\eta}_{z},\nonumber \\
\left[\hat{\eta}^{2},\hat{\eta}_{a}\right] & =0~~~\mbox{with}\quad a=(+,-,z),
\end{align}
where $\hat{\eta}^{2}\equiv\hat{\eta}_{+}\hat{\eta}_{-}+\hat{\eta}_{z}^{2}-\hat{\eta}_{z}$ is the
Casimir operator commuting with all the generators. The $\hat{\Delta}_{m}$
operators then form a $l=1$ representation under this algebra, satisfying
the commutation relations: 
\begin{align}
\left[\hat{\eta}_{\pm},\hat{\Delta}_{m}\right] & =\sqrt{l\left(l+1\right)-m\left(m\pm1\right)}\hat{\Delta}_{m\pm1},\nonumber \\
\left[\hat{\eta}_{z},\hat{\Delta}_{m}\right] & =m\hat{\Delta}_{m},\nonumber \\
\left[\hat{\eta}^{2},\hat{\Delta}_{m}\right] & =l(l+1)\hat{\Delta}_{m}.
\end{align}
The operators $(\hat{\eta}_{+},\hat{\eta}_{-})$ have the form of a particle-particle
pairing order with finite center of mass momentum (which is equal to the modulation wave vector of the d-BDW) and thus define a PDW operator. 

In writing the $l=1$ representation in Eq. (\ref{eq:8a}), we have considered both the d-SC and the d-BDW fields to be complex. This is necessary for the Hopf fibration discussed in Sec.~\ref{sec:subhopf}. On the contrary, the case of emergent SU(2) symmetries works well with a purely real d-BDW or a purely imaginary d-BDW. While Eqs. (\ref{eq:8a}) and (\ref{eq:8b}) are completely generic, in the following, we show two special $l=1$ representations corresponding to $A=B=1$ and $A=-B=i$.

\subsubsection{For $A=B=1$}

\begin{align}
\hat{\Delta}_{1} & =\frac{-1}{\sqrt{2}}\hat{d}\sum_{\sigma}\sigma c_{i\sigma}^{\dagger}c_{j-\sigma}^{\dagger}e^{i\theta_{\Delta}},\nonumber \\
\hat{\Delta}_{0}^{a} & =\frac{1}{2}\hat{d}\sum_{\sigma}[c_{i\sigma}^{\dagger}c_{j\sigma}e^{iQ\cdot\left(r_{i}+r_{j}\right)/2}e^{i\theta_{\chi}}\nonumber \\
 & +c_{j\sigma}^{\dagger}c_{i\sigma}e^{-iQ\cdot\left(r_{i}+r_{j}\right)/2}e^{-i\theta_{\chi}},\nonumber \\
\hat{\Delta}_{-1} & =\frac{1}{\sqrt{2}}\hat{d}\sum_{\sigma}\sigma c_{j-\sigma}c_{i\sigma}e^{-i\theta_{\Delta}}.\label{eq:10a}
\end{align}
Note that in this case $\hat{\Delta}_{0}^{a}$ corresponds to the real part of the excitonic order $\hat{\Delta}_{0}^{a}=\hat{\chi}+\hat{\chi}^{\dagger}$ , \emph{i.e.} to the charge modulations. We can construct the PDW operators

\begin{align}
\hat{\eta}_{+}^{a} & =\frac{1}{2}\sum_{\sigma}\sigma[c_{i\sigma}^{\dagger}c_{i-\sigma}^{\dagger}e^{iQ\cdot\left(r_{i}+r_{j}\right)/2}e^{i\left(\theta_{\Delta}+\theta_{\chi}\right)}\nonumber \\
 & +c_{j\sigma}^{\dagger}c_{j-\sigma}^{\dagger}e^{-iQ\cdot\left(r_{i}+r_{j}\right)/2}e^{i\left(\theta_{\Delta}-\theta_{\chi}\right)}],\nonumber \\
\hat{\eta}_{-}^{a} & =\frac{1}{2}\sum_{\sigma}\sigma[c_{i-\sigma}c_{i\sigma}e^{-iQ\cdot\left(r_{i}+r_{j}\right)/2}e^{-i\left(\theta_{\Delta}+\theta_{\chi}\right)}\nonumber \\
 & +c_{j-\sigma}c_{j\sigma}e^{iQ\cdot\left(r_{i}+r_{j}\right)/2}e^{-i\left(\theta_{\Delta}-\theta_{\chi}\right)}],\nonumber \\
\hat{\eta}_{z} & =\frac{1}{2}\left[\hat{\eta}_{+},\hat{\eta}_{-}\right],\nonumber \\
 & =\frac{1}{2}\sum_{\sigma}\left(\hat{n}_{i\sigma}+\hat{n}_{j\sigma}-1\right),\label{eq:10b}
\end{align}

and 
\begin{align}
\left[\hat{\eta}_{\pm},\hat{\Delta}_{m}\right] & =\sqrt{l\left(l+1\right)-m\left(m\pm1\right)}\hat{\Delta}_{m\pm1},\nonumber \\
\left[\hat{\eta}_{z},\hat{\Delta}_{m}\right] & =m\hat{\Delta}_{m}.\label{eq:10b-2}
\end{align}
Note that in all cases $\hat{\eta}_{z}$ is real, with $\hat{\eta}^{\dagger}_{z}=\hat{\eta}_{z}$. In this case we have $\hat{\eta}_{+}=\hat{\eta}_{-}^{\dagger}$ ( the subscript $\left(*\right)^{a,b}$
has been dropped in the commutation relations for clarity).

\subsubsection{For $A=-i$ and $B=i$ }

\begin{align}
\hat{\Delta}_{1} & =\frac{-1}{\sqrt{2}}\hat{d}\sum_{\sigma}\sigma c_{i\sigma}^{\dagger}c_{j-\sigma}^{\dagger}e^{i\theta_{\Delta}},\nonumber \\
\hat{\Delta}_{0}^{b} & =\frac{i}{2}\hat{d}\sum_{\sigma}[-c_{i\sigma}^{\dagger}c_{j\sigma}e^{iQ\cdot\left(r_{i}+r_{j}\right)/2}e^{i\theta_{\chi}}\nonumber \\
 & +c_{j\sigma}^{\dagger}c_{i\sigma}e^{-iQ\cdot\left(r_{i}+r_{j}\right)/2}e^{-i\theta_{\chi}},\nonumber \\
\hat{\Delta}_{-1} & =\frac{1}{\sqrt{2}}\hat{d}\sum_{\sigma}\sigma c_{j-\sigma}c_{i\sigma}e^{-i\theta_{\Delta}}.\label{eq:10a-1}
\end{align}

Note that in this case $\hat{\Delta}_{0}^{b}$ corresponds to the imaginary
part of the excitonic order $\hat{\Delta}_{0}^{b}=-i\hat{\chi}+i\hat{\chi}^{*}$ , \emph{i.e.}
to the current modulations. We can construct the PDW operators

\begin{align}
\hat{\eta}_{+}^{b} & =\frac{i}{2}\sum_{\sigma}\sigma[-c_{i\sigma}^{\dagger}c_{i-\sigma}^{\dagger}e^{iQ\cdot\left(r_{i}+r_{j}\right)/2}e^{i\left(\theta_{\Delta}+\theta_{\chi}\right)}\nonumber \\
 & +c_{j\sigma}^{\dagger}c_{j-\sigma}^{\dagger}e^{-iQ\cdot\left(r_{i}+r_{j}\right)/2}e^{i\left(\theta_{\Delta}-\theta_{\chi}\right)}],\nonumber \\
\hat{\eta}_{-}^{b} & =\frac{-i}{2}\sum_{\sigma}\sigma[c_{i-\sigma}c_{i\sigma}e^{-iQ\cdot\left(r_{i}+r_{j}\right)/2}e^{-i\left(\theta_{\Delta}+\theta_{\chi}\right)}\nonumber \\
 & -c_{j-\sigma}c_{j\sigma}e^{iQ\cdot\left(r_{i}+r_{j}\right)/2}e^{-i\left(\theta_{\Delta}-\theta_{\chi}\right)}],\nonumber \\
\hat{\eta}_{z} & =\frac{1}{2}\left[\hat{\eta}_{+},\hat{\eta}_{-}\right],\nonumber \\
 & =\frac{1}{2}\sum_{\sigma}\left(\hat{n}_{i\sigma}+\hat{n}_{j\sigma}-1\right).\label{eq:10b-1}
\end{align}

As above, $\hat{\eta}_{z}$ is real, with $\hat{\eta}_{z}^{\dagger}=\hat{\eta}_{z}$
and $\hat{\eta}_{+}=-\hat{\eta}_{-}^{\dagger}$.

\subsection{Lie algebra and quantum rotor model}

In the operator formalism, it is interesting to reformulate the discussion on fluctuations with the help of Lie algebra. For this we introduce a basis of a four vector $n=\left(n_{1},n_{2},n_{3},n_{4}\right)$ such that, 
\begin{align}
n_{1} & =\frac{1}{2}\left(\hat{\Delta}+\hat{\Delta}^{\dagger}\right),\nonumber \\
n_{2} & =\frac{1}{2}\left(\hat{\chi}+\hat{\chi}^{\dagger}\right),\nonumber \\
n_{3} & =\frac{-i}{2}\left(\hat{\Delta}^{\dagger}-\hat{\Delta}\right),\nonumber \\
n_{4} & =\frac{-i}{2}\left(\hat{\chi}^{\dagger}-\hat{\chi}\right).\label{eq:Dlie1}
\end{align}

Using the definitions of $\hat{\eta}_{\pm}^{a}$ in Eq.~\eqref{eq:10b}
and $\hat{\eta}_{\pm}^{b}$ in Eq.~\eqref{eq:10b-1}, we can form the SO(4)
Lie algebra with the anti-symmetric operator $\hat{L}$ such that
$L_{ab}=-L_{ba}$, 
\begin{align}
\hat{L}= & \left[\begin{array}{cccc}
0 & * & * & *\\
-i\frac{\hat{\eta}_{+}^{a}-\hat{\eta}_{-}^{a}}{2} & 0 & * & *\\
-\hat{\eta}_{z} & \frac{\hat{\eta}_{+}^{a}+\hat{\eta}_{-}^{a}}{2} & 0 & *\\
i\frac{\hat{\eta}_{+}^{b}-\hat{\eta}_{-}^{b}}{2} & \hat{\eta}_{0} & -\frac{\hat{\eta}_{+}^{b}+\hat{\eta}_{-}^{b}}{2} & 0
\end{array}\right],\label{eq:D1}
\end{align}

where $*$ means the anti-symmetric component of $L_{ab}$. The basis
works for general groups SO(n) with $n$ vectors $n_{a}$ and the
constraint $\sum_{a}n_{a}^{2}=1$ is implicit. Here we have SO(4),
which has six such components. One can check the relations- valid
as well for the SO(n) algebra, 
\begin{align}
\left[L_{ab},n_{c}\right] & =i\delta_{ac}n_{b}-i\delta_{bc}n_{a},\nonumber \\
\left[L_{ab},L_{cd}\right] & =i\delta_{ad}L_{cb}+i\delta_{ac}L_{bd}+i\delta_{bc}L_{da}+i\delta_{bd}L_{ac}.\label{eq:D2}
\end{align}

We can form the Hamiltonian 
\begin{align}
\hat{H} & =\frac{1}{2\zeta}\hat{L}^{2}+\frac{\rho}{2}\hat{v}^{2}+U(n),\label{eq:D3}
\end{align}

where $\zeta$ is a susceptibility, $\rho$ is the stiffness, $\hat{v}_{ab}=n_{a}\nabla n_{b}-n_{b}\nabla n_{a}$,
and $U(n)$ is a potential term. We can take formally the Legendre
transform by introducing the conjugate momentum to the quantum rotor
$\hat{L}$, 
\begin{align}
\hat{\omega} & =\frac{\partial\hat{H}}{\partial\hat{L}}=\frac{\hat{L}}{\zeta}.\label{eq:D4}
\end{align}

The corresponding Lagrangian thus writes 
\begin{align}
{\cal L} & =\frac{\zeta}{2}\hat{\omega}^{2}-\frac{\rho}{2}\hat{v}^{2}-U(n).\label{eq:D5}
\end{align}

The form of the momentum $\hat{\omega}$ can be obtained by taking
the operator representation for $\hat{L}$ with 
\begin{align}
L_{ab} & =\hat{n}_{a}\hat{p}_{b}-\hat{p}_{a}\hat{n}_{b},\label{eq:D6}
\end{align}

with $\hat{p}$ the momentum conjugate to $\hat{n},$ with $\left[\hat{p}_{a},\hat{n}_{b}\right]=i\delta_{ab}$.
Using the Hamilton equation $\dot{n}_{a}=\partial H/\partial p_{a}$,
and making use of the constraint $\sum_{a}n_{a}^{2}=1$, we get the
following expression 
\begin{align}
\omega_{ab} & =n_{a}\dot{n}_{b}-n_{b}\dot{n}_{a}.\label{eq:D7}
\end{align}
Noticing that, by differentiating the constraint, one has the relation
$\sum_{a}n_{a}\dot{n}_{a}=0$, one obtains $\sum_{a,b}\omega_{ab}^{2}=\sum_{a}\dot{n}_{a}^{2}$,
thus the Lagrangian in Eq.~(\ref{eq:D5}) can be cast into the form
\begin{align}
{\cal L} & =\sum_{a=1}^{4}\frac{\zeta}{2}\dot{n}_{a}^{2}-\frac{\rho}{2}\left(\nabla n_{a}\right)^{2}-U(n),\nonumber \\
\mbox{ with } & \sum_{a=1}^{4}n_{a}^{2}=1.\label{eq:D8}
\end{align}

Eq.~(\ref{eq:D8}) is typical of the SO(4) NL$\sigma$M. As mentioned
before, since in the model invoked for cuprates, the form of $\hat{L}$
in Eq.~(\ref{eq:D1}) involves only $\eta$-operators, one can conclude
that the fluctuations below $T^{*}$ are made of PDW modes.

The $\hat{L}$ matrix in Eq.~(\ref{eq:D1}) can be constructed by
the generators of the two copies of the chiral SU(2) model of Sec.~\ref{sec:flucchiral}.
The SO(4) group has 6 generators and the two copies of SU(2) also
consists of 3+3=6 generators. In order to see this analogy, we write
the anti-symmetric operator $\hat{L}$ in terms of the $m$ field operators in Eqs.~\eqref{eq:D2e} and \eqref{eq:D2e-1}, 
\begin{align}
\hat{L}= & \left[\begin{array}{cccc}
0 & * & * & *\\
-\hat{m}^{y}-\hat{m}_{2}^{y} & 0 & * & *\\
\hat{m}^{z}+\hat{m}_{2}^{z} & \hat{m}^{x}+\hat{m}_{2}^{x} & 0 & *\\
\hat{m}^{x}-\hat{m}_{2}^{x} & \hat{m}_{2}^{z}-\hat{m}^{z} & \hat{m}^{y}-\hat{m}_{2}^{y} & 0
\end{array}\right],\label{eq:Lchir}
\end{align}
where $*$ means the anti-symmetric component of $L_{ab}$ (with $L_{ba}=-L_{ab}$
for $a,b\in[1,2,3,4]$) and 
\begin{align}
\hat{m}^{x} & =\frac{1}{2}\left(\hat{m}^{+}+\hat{m}^{-}\right);~\hat{m}^{y}=\frac{1}{2i}\left(\hat{m}^{+}-\hat{m}^{-}\right),\nonumber \\
\hat{m}_{2}^{x} & =\frac{1}{2}\left(\hat{m}_{2}^{+}+\hat{m}_{2}^{-}\right);~\hat{m}_{2}^{y}=\frac{1}{2i}\left(\hat{m}_{2}^{+}-\hat{m}_{2}^{-}\right),\label{eq:mdef}
\end{align}
with $\hat{m}^{\pm}$, $\hat{m}^{z}$ defined in Eqs.~\eqref{eq:D2e}
and $\hat{m}_{2}^{\pm}$, $\hat{m}_{2}^{z}$ defined in Eqs.~\eqref{eq:D2e-1}.
$\hat{m}^{x}$, $\hat{m}^{y}$, $\hat{m}^{z}$ are the generators of the
first chiral SU(2) copy in Sec.~\ref{sec:flucchirala} and $\hat{m}_{2}^{x}$,
$\hat{m}_{2}^{y}$, $\hat{m}_{2}^{z}$ are the generators of the second chiral
SU(2) copy in Sec.~\ref{sec:flucchiralb}.

\section{\label{sec:Appendix-B:-Details} Details of the microscopic model}

\subsection{Solving the gap equations}\label{subsec:gapeq}

We explicit here the solution of the two gap equations given in
Eq.~\eqref{gap_eqSC} and Eq. \eqref{gap_eqCO}. 
\begin{align}
\Delta_{k,\omega} & =-T\sum_{q,\omega^{\prime}}\frac{J_{-}\left(q,\omega^{\prime}\right)\Delta_{k+q}}{\left(\omega+\omega^{\prime}\right)^{2}-\xi_{k+q}^{2}-\Delta_{k+q}^{2}},\nonumber \\
\chi_{k,\omega} & =-T\sum_{q,\omega^{\prime}}\frac{J_{+}\left(q,\omega^{\prime}\right)\chi_{k+q}}{\left(\omega+\omega^{\prime}-\xi_{k+q}\right)\left(\omega+\omega^{\prime}-\xi_{k+Q+q}\right)-\chi_{k+q}^{2}}.\label{eq:B2}
\end{align}

\begin{figure}
\centering \includegraphics[scale=0.2]{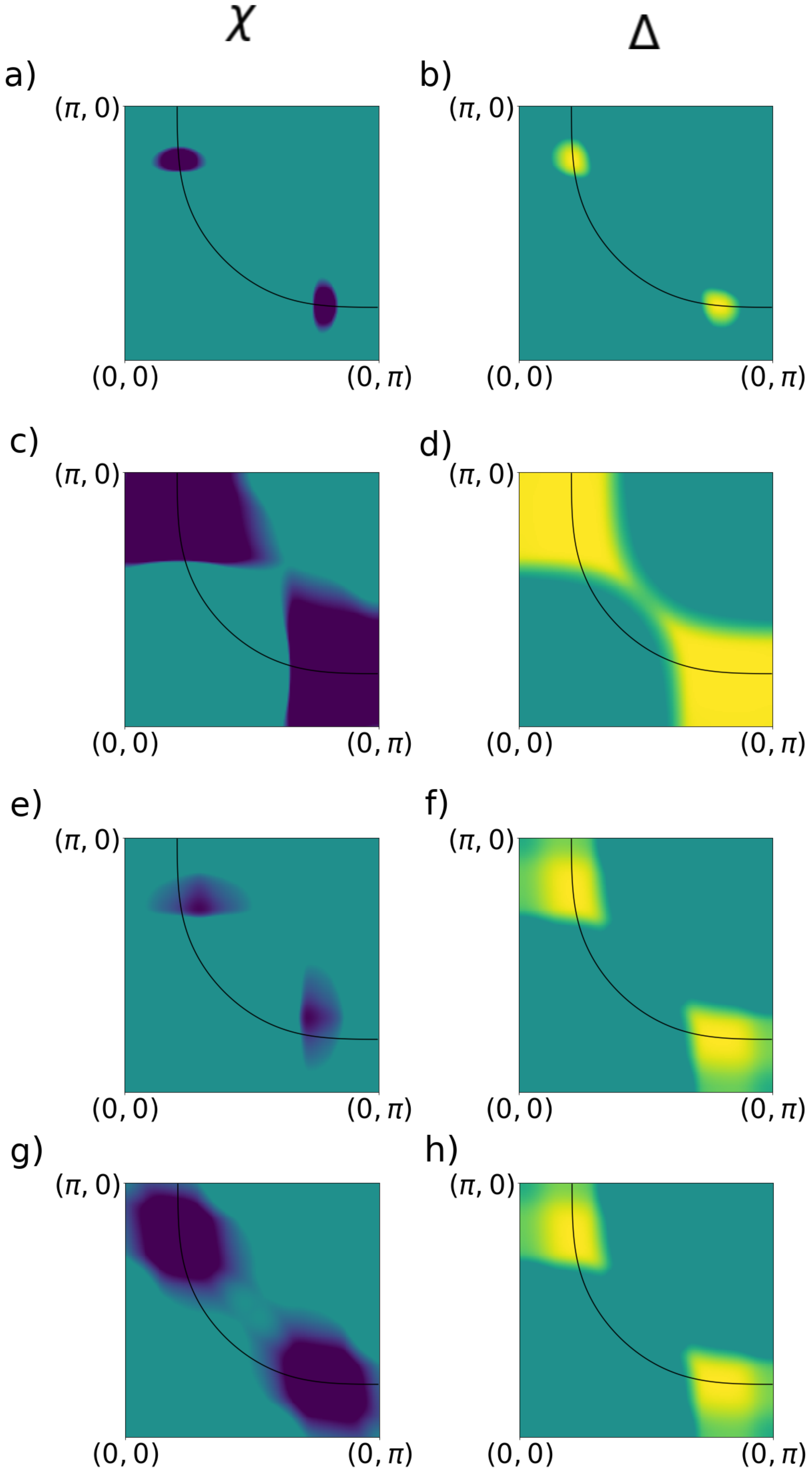} \caption{\textbf{(a,c,e,g)} Gap in the particle-hole pairing channel ($\chi$) for different choices of parameters. \textbf{(b,d,f,h)} Corresponding gaps
in the particle-particle pairing channel ($\Delta$). For all figures
$J=0.6\ eV$, $V=J/3$. We compare the gaps with $\kappa_{AF}=0.05\ r.l.u$ (a and b) and $\kappa_{AF}=0.16\ r.l.u$ (c and d) for $\chi$ with an axial $Q$. We also show solutions for $\Delta$ and $\chi$ with $\kappa_{AF}=0.1\ r.l.u$ and $Q=(0,\pm0.3)\ r.l.u$ (e and f) which is close to the experimentally observed value. Solutions of $\chi$ with a diagonal wave-vector linking hot-spots $Q=(\pm Q_{x},\pm Q_{y})$ for $\kappa_{AF}=0.1\ r.l.u$ is shown in (g) with its equivalent comparison for $\Delta$ in (h).}
\label{fig:gaps} 
\end{figure}

We start by assuming $J_{\pm}$, $\Delta_{k,\omega}$ and $\chi_{k,\omega}$ to be frequency independent. We also take the interaction to be maximal for $q=Q_{AF}$. An example is to take $J_{\pm}(q)$ to be of the form 
\begin{equation}
J_{\pm}(q)=\frac{J_{\pm}}{\left(q-Q_{AF}\right)^{2}+\kappa_{\text{{\scriptsize{AF}}}}^{2}},
\label{eq:Jdef}
\end{equation}
where the constant part $J_{\pm}=3J\left(p\right)\pm V$ is related
to the original parameter of the real space model and $\kappa_{\text{{\scriptsize{AF}}}}$ is a mass that translate the short-range nature of the fluctuations. We use the fact that $J_{\pm}$ is peaked around the antiferromagnetic wave-vector to restrict the momentum sum in Eq.~\eqref{eq:B2} to a small region around $Q_{AF}$ of size $\kappa_{\text{{\scriptsize{AF}}}}$ in which $J_{\pm}$, $\Delta$ and $\chi$ are taken to be constant. In order not to make any assumption on the relation between $\chi_{k+Q_{AF}}$ and $\chi_{k}$ or $\Delta_{k+Q_{AF}}$ and $\Delta_{k}$
we use equation Eq. \eqref{eq:B2} to express $\chi_{k+Q_{AF}}$
and $\Delta_{k+Q_{AF}}$ respectively, leading to {\small{}{}{ 
\begin{align}
\chi_{k} & =(J_{+})^{2}\sum_{q,\omega^{\prime}}\frac{1}{(\omega+\omega^{\prime}-\tilde{\xi}_{k+Q+q})(\omega+\omega^{\prime}-\tilde{\xi}_{k+q})-(\tilde{\chi}_{k})^{2}}\nonumber \\
 & \times\sum_{q^{\prime},\omega^{\prime}}\frac{\chi_{k}}{(\omega+\omega^{\prime}-\xi_{k+Q+q^{\prime}})(\omega+\omega^{\prime}-\xi_{k+q^{\prime}})-(\chi_{k})^{2}},\label{eq:B4}\\
\Delta_{k} & =(J_{-})^{2}\sum_{q,\omega^{\prime}}\frac{1}{(\omega+\omega^{\prime}-\tilde{\xi}_{k+q})(\omega+\omega^{\prime}+\tilde{\xi}_{-k-q})-(\tilde{\Delta}_{k})^{2}}\nonumber \\
 & \times\sum_{q^{\prime},\omega^{\prime}}\frac{\Delta_{k}}{(\omega+\omega^{\prime}-\xi_{k+q^{\prime}})(\omega+\omega^{\prime}+\xi_{-k-q^{\prime}})-(\Delta_{k})^{2}},\label{eq:B5}
\end{align}
}} where we used the notation $\tilde{\xi}_{k}=\xi_{k+Q_{AF}}$,
$\tilde{\chi}_{k}=\chi_{k+Q_{AF}}$ and $\tilde{\Delta}_{k}=\Delta_{k+Q_{AF}}$.
We can now simplify the equations on both side and perform analytically the two summations over the Matsubara frequencies which leads to the same result starting from either Eq.~\eqref{eq:B4} or Eq. \eqref{eq:B5}: 
\begin{equation}
1=(J_{\pm})^{2}\sum_{q}\frac{n_{f}\left(\tilde{\omega}_{+}\right)-n_{f}\left(\tilde{\omega}_{-}\right)}{\tilde{\omega}_{+}-\tilde{\omega}_{-}}\sum_{q^{\prime}}\frac{n_{f}\left(\omega_{+}\right)-n_{f}\left(\omega_{-}\right)}{\omega_{+}-\omega_{-}}\label{eq:B6}
\end{equation}
with 
\begin{align}
\omega_{\pm}^{\chi} & =\frac{1}{2}\left(\xi_{k+q}+\xi_{k+Q+q}\pm\sqrt{\left(\xi_{k+Q+q}-\xi_{k+q}\right)^{2}+4 \chi_k^{2}}\right),\label{eq:B7}\\
\omega_{\pm}^{\Delta} & =\pm\sqrt{\xi_{k+q}^{2}+\Delta_k^{2}}\label{eq:B8}
\end{align}
and $\tilde{\omega}$ have the same expression with all momenta shifted
by $Q_{AF}$, \emph{i.e.} $k\rightarrow k+Q_{AF}$. Neglecting the
$q$-dependence of $\left(\omega_{+}-\omega_{-}\right)$ in Eq. \eqref{eq:B6}
we can also perform analytically the momentum summation. This is done
by linearization of $\omega_{\pm}$ around $k$ and by limiting the
integration to the direction parallel to the Fermi velocity in a range $\kappa_{\text{{\scriptsize{AF}}}}$, 
\begin{equation}
\sum_{q}\rightarrow\int{\frac{dq^{n}}{\left(2\pi\right)^{n}}}=\int_{-\kappa_{\text{{\scriptsize{AF}}}}/2}^{\kappa_{\text{{\scriptsize{AF}}}}/2}{\frac{dq_{||}}{2\pi}}.
\end{equation}
This leads us to the following expression: 
\begin{equation}
\sum_{q}n_{f}\left(\omega\right)=-\frac{1}{\beta v_{f}}\log{\frac{1+e^{\beta\left(\omega+v_{f}\kappa_{\text{{\scriptsize{AF}}}}/2\right)}}{1+e^{\beta\left(\omega-v_{f}\kappa_{\text{{\scriptsize{AF}}}}/2\right)}}}.
\end{equation}
Finally we start by neglecting $\tilde{\chi}$ and $\tilde{\Delta}$
so that we can solve the implicit equations: {\small{}{ 
\begin{align}
4\ \chi_k^{2} & =\left(J_{+}\right)^{4}\left(\frac{\Delta n_{f}\left(\tilde{\omega}_{+},\tilde{\omega}_{-}\right)}{\tilde{\xi}_{k+Q}-\tilde{\xi}_{k}}\right)^{2}\ \Delta n_{f}\left(\omega_{+},\omega_{-}\right)^{2}-\left(\xi_{k+Q}-\xi_{k}\right)^{2}\\
\Delta_k^{2} & =\frac{\left(J_{-}\right)^{4}}{4}\left(\frac{\Delta n_{f}\left(\tilde{\omega}_{+},\tilde{\omega}_{-}\right)}{2\tilde{\xi}_{k}}\right)^{2}\Delta n_{f}\left(\omega_{+},\omega_{-}\right)^{2}-\xi_{k}^{2}
\end{align}
}} Note that the right hand sides still depend on $\chi_k$ or $\Delta_k$
through $\omega_{\pm}$ in Eq. \eqref{eq:B7}. However, these equations can be solve independently for all k in the first Brillouin zone. As a second step, we use the previous solution as the input value of $\tilde{\chi}_k$ and $\tilde{\Delta}_k$ and compute the modification it implies on $\chi$ and ${\Delta}$ respectively. This procedure
converge to a stable solution within a few iterations. Having used the same set of approximations for the computation of the particle-particle gap and the particle-hole gap, allows us to have a direct comparison of their amplitudes. We finally consider that only one gap open at each k-point, the one that is the larger of the two.

\subsection{Exploration of the parameter space}

The solution of the gap equation depends on the choice of different parameters, namely the spin-spin interaction $J$, the density-density interaction $V$, the d-BDW ordering wave vector $Q$ and the range
of the AF coupling $\kappa_{\text{{\scriptsize{AF}}}}$. We present in Fig. \ref{fig:gaps}
the solutions for certain choices of parameters. We start by two results
obtained from the same value of interactions as in the main text (Fig.
\ref{fig:diagram}) but with larger and smaller value of $\kappa_{\text{{\scriptsize{AF}}}}$ than the one obtained form experiments. As expected, the resulting solutions are limited to a region closer to the hot-spots when $\kappa_{\text{{\scriptsize{AF}}}}$ is reduced while we obtain a non-zero solution in an extended part of the Brillouin zone for large $\kappa_{\text{{\scriptsize{AF}}}}$. Solutions obtained when we change the d-BDW ordering wave vector, show that the gap opening on the Fermi surface is the largest with an axial wave vector connecting the hot spots in the first Brillouin zone and hence will be favored. The diagonal wave-vector for d-BDW leads to a solution which is degenerate with the superconducting gap or with the axial d-BDW at the hot-spots but will have a larger overlap with the d-SC gap away from them.

\subsection{Mean-field solution for PDW gap}\label{sec:PDWgap}
\begin{figure}[t]
\centering
{\includegraphics[width=0.8\linewidth]{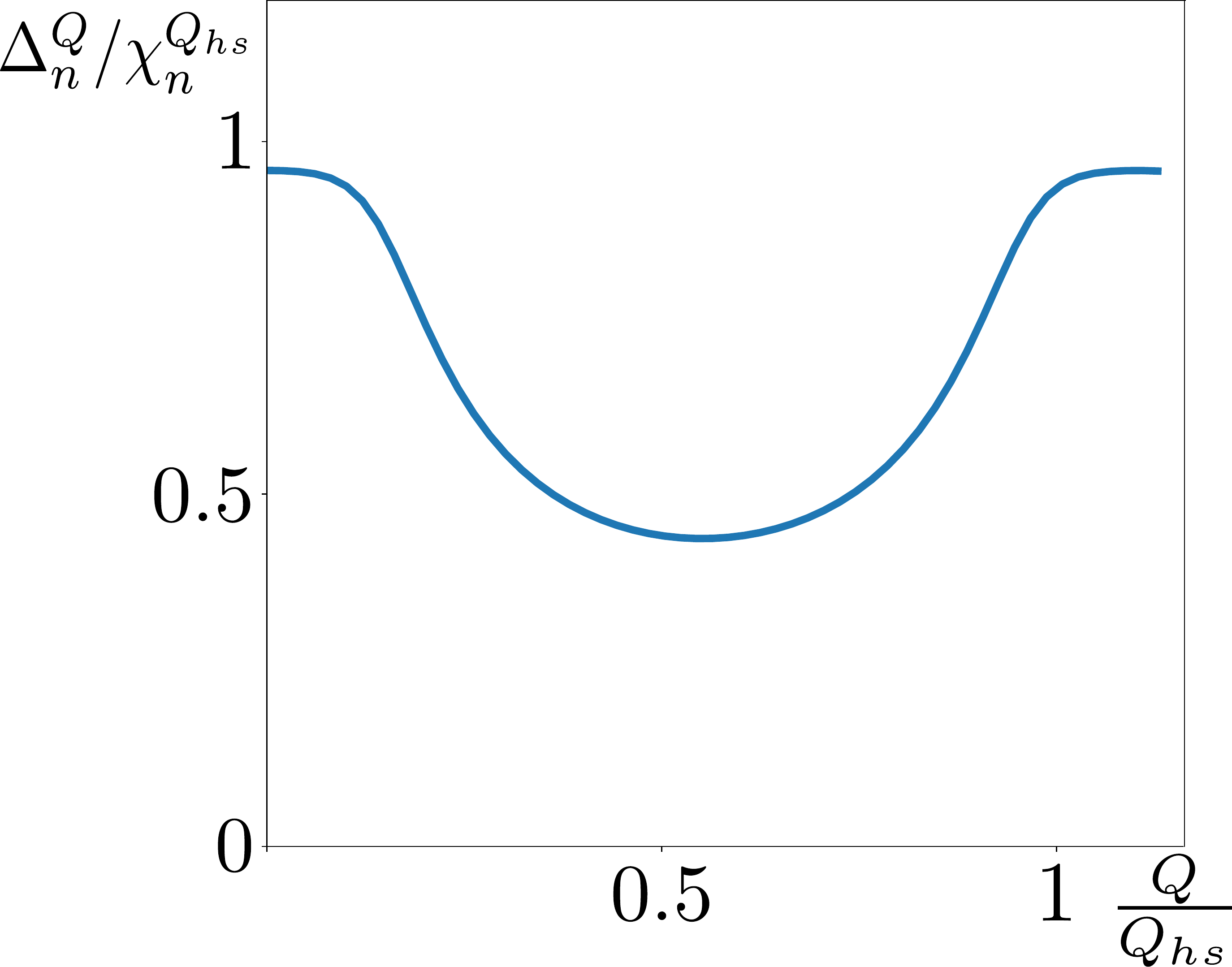}
} 
\caption{PDW gap averaged in the nodal region ($\Delta_{n}^{Q}$) scaled by $\chi^{Q_{hs}}_n$ as a function of $Q/Q_{hs}$. $\Delta_{n}^{Q_{hs}/2}/\chi^{Q_{hs}}_n$ is nearly 0.5. Thus the $Q_{hs}$ d-BDW order is energetically favored compared to the $Q_{hs}/2$ PDW as a choice for a primary state. The parameters for the short AF interactions are chosen as $\kappa_{AF}=0.05\ \text{r.l.u}$ and $J=0.5\ eV$. We used the same doping $p=0.12$, inverse temperature $\beta=50$ and ratio $J/V = 20$ as in Fig. \ref{fig:diagram}.}
\label{Fig:PDWgap} 
\end{figure}

We now look for finite momentum superconducting order arising from the microscopic model in Eq.~\eqref{eq:11b} at the mean-field level. The gap equation for the PDW gap ($\Delta^Q$) is given by 
{\small{
\begin{equation}
\Delta^Q_{k,\omega} =-T\sum_{q,\omega^{\prime}}\frac{J_{-}\left(q,\omega^{\prime}\right)\Delta^Q_{k+q}}{\left(\omega+\omega^{\prime}-\xi_{k+q}\right)\left(\omega+\omega^{\prime}+\xi_{k+Q+q}\right)-\left(\Delta^Q_{k+q}\right)^{2}}.\label{eq:PDWgap}
\end{equation}}}
We solve this equation in the same way as described previously in Sec.~\ref{subsec:gapeq} for a modulation wave-vector varying between $Q=0$ and $Q=Q_{hs}$. $Q_{hs}$ is the axial wave-vector relating two hot-spots in the first Brillouin zone. We then compare the solution with the d-BDW gap with wave-vector $Q=Q_{hs}$ as obtained in the main text. The result for the values averaged in the nodal region is shown in Fig.\ref{Fig:PDWgap}. We first note that $\Delta^{Q=0}_n$ (d-SC gap) is slightly smaller than $\chi^{Q_{hs}}_n$ as we have the density-density interaction included in our model. We see that the PDW gap for $Q=Q_{hs}/2$ is approximately half of the d-BDW gap with $Q=Q_{hs}$ in the nodal region. We checked that the PDW gap with $Q=Q_{hs}/2$ averaged in the AN region ($\Delta_{an}^{Q_{hs}/2}$) also gives $\Delta_{an}^{Q_{hs}/2} \approx 1/2 \chi^{Q_{hs}}_n$. Thus energetically, the choice of the d-BDW with $Q_{hs}$ as a primary state over $Q_{hs}/2$ PDW is justified.

\section{\label{sec:STMphase} Determination of the phase of the density modulations in STM}

We outline the procedure to extract the phase of the charge density modulations in STM measurements. The sublattice segregation method described in Ref.~\onlinecite{Hamidian15} generates $\tilde{D}^{\delta Z}(\mathbf{q},E)$, which measures the spectral weight of the field induced d-CDW at an energy $E=eV$ ($V$ is the bias voltage). Note that $\delta Z(r,E)=Z(r,E,8.5T)-Z(r,E,0T)$ with $Z(r,E)=g(r,V)/g(r,-V)$ where $g$ is the differential tunneling conductance. We hypothesize that $\tilde{D}^{\delta Z}$ is proportional to the amplitude of the d-CDW ($Re\left(\chi(r)\right)$). The phase is obtained using the following procedure:
\begin{enumerate}

\item{$\tilde{D}^{\delta Z}(\mathbf{q},E)$ is filtered individually around $\mathbf{Q}_x=(0.25,0)2\pi/a_0$ and $\mathbf{Q}_y=(0,0.25)2\pi/a_0$ with a filter of width $\Lambda$ to obtain 
\begin{align}
\tilde{D}_{x,y}(\mathbf{q},E)=2\tilde{D}^{\delta Z}(\mathbf{q},E)e^{-\frac{\left(\mathbf{q}-\mathbf{Q}_{x,y}\right)^2}{2\Lambda^2}}.
\end{align} 
}

\item{An inverse Fourier transform of $\tilde{D}_{x,y}(\mathbf{q})$ gives 
\begin{align}
D_{x,y}(r)&=ReD_{x,y}(r)+iImD_{x,y}(r)\nonumber\\
&=\frac{1}{2\pi^2}\int d\mathbf{q}e^{i\mathbf{q}.r}\tilde{D}_{x,y}(\mathbf{q}). 
\end{align}
Since $\tilde{D}_{x,y}(\mathbf{q})$ is not a perfect Gaussian, the real space structure $D_{x,y}(r)$ contains information about both spatially varying phase and amplitude.}

\item{The spatial-phase map is generated using 
\begin{align}
\phi_{x,y}(r)=tan^{-1}\left[ ImD_{x,y}(r)/ReD_{x,y}(r)\right].
\end{align}
}

\item{In Fig.~\ref{fig:Global-spatial-Charge}(a), we plot $Cos(\phi_{x,y}(r))$ masked around the vortex regions.}

\item{A reference modulation is constructed with $Cos(\phi_{ref}(r))$ where $\phi_{ref}(r)=\mathbf{Q}_y\cdot r +\phi_0$. The grey lines in Fig.~\ref{fig:Global-spatial-Charge}(a) represents the points where the reference phase function is 0 mod 2$\pi$.}

\item{The histogram of $\phi_y(r)-\phi_{ref}(r)$ is shown in Fig.~\ref{fig:Global-spatial-Charge}(b) at only the vortex regions considering 6-9 vortex cores. A common radius of $2nm$ around each vortex core is used.}

\end{enumerate}

 \bibliographystyle{apsrev4-1}
\bibliography{Cuprates}

\end{document}